\documentclass[11pt,a4paper]{article}

\usepackage[spacing=compact,font=jhep,nonatbib]{nuclthpreprint}
\usepackage{amsmath}
\usepackage{amssymb}
\usepackage{amsthm}
\usepackage{amscd}
\usepackage{amsfonts}
\usepackage{graphicx}
\usepackage{fancyhdr}
\usepackage{hyperref}
\usepackage{geometry}
\usepackage{cite}
\usepackage{setspace}
\usepackage{xcolor}
\usepackage{authblk}

\shorttitle{}

\title{Hydrodynamic attractors}

\author[1]{Tilman Enss\email{enss@thphys.uni-heidelberg.de}}
\author[2,3]{Michal P.\ Heller\email{michal.p.heller@ugent.be}}
\author[2]{Alexandre Serantes\email{alexandre.serantesrubianes@ugent.be}}
\author[2]{Clemens Werthmann\email{clemens.werthmann@ugent.be}}

\affil[1]{Institut f{\"u}r Theoretische Physik, Universit{\"a}t Heidelberg, 69120 Heidelberg, Germany}
\affil[2]{Department of Physics and Astronomy, Ghent University, 9000 Ghent, Belgium}
\affil[3]{Institute of Theoretical Physics and Mark Kac Center for Complex Systems Research, Jagiellonian University, 30-348 Cracow, Poland}

\renewcommand{\d}{\mathrm{d}}
\renewcommand{\P}{\mathcal{P}}
\newcommand{\E}{\mathcal{E}}

\begin{document}

\maketitle

\begin{abstract}
Attractors are effective low‑dimensional structures, defined in a chosen set of observables, that trajectories from different initial states approach; hydrodynamic attractors are those on which the late‑time evolution is governed by hydrodynamic constitutive relations. Motivated by nuclear collisions and ultracold atomic gases, this chapter distinguishes attractorization (the loss of sensitivity to some directions in the space of initial states) from hydrodynamization (the onset of validity of hydrodynamic constitutive relations). Using mainly conformal, boost‑invariant Bjorken flow, we compare Müller--Israel-–Stewart-type theories, kinetic theory, holography, and classical Yang–Mills fields. Forward attraction, produced by the decay of non-hydrodynamic modes, is distinguished from pullback attraction, which can select a unique regular solution, and from expansion‑driven attraction, which can suppress initial‑state sensitivity before microscopic relaxation, with or without a subsequent fluid regime. Divergent gradient expansions and their transseries completions, the state‑space picture, and adiabatic hydrodynamization provide complementary descriptions. Applications to nuclear collisions include particle production, transverse energy and flow, the initialization of and linear response around attracting backgrounds, jet quenching, and attractor‑informed modifications of hydrodynamic models. For ultracold Fermi gases, we review theoretical proposals for bulk‑channel attractors under scattering‑length drives and distinguish them from recent measurements of short‑time contact and momentum‑distribution dynamics.
\end{abstract}

\section{Introduction and summary}
\label{sec:introduction}

High-energy nuclear collisions create matter under conditions that seem particularly unfavorable for hydrodynamics: the system is small, expands rapidly, can have substantially different pressures along and perpendicular to the collision axis and can exhibit a significant inhomogeneity in the transverse plane. Nevertheless, viscous fluid-dynamical models successfully describe much of the collective behavior of the resulting QGP~\cite{Niemi:2015qia,Romatschke:2017ejr}. Why can a description involving so few macroscopic variables work under such extreme conditions? Microscopic calculations sharpen the question. In holography, which relates certain strongly coupled quantum field theories to gravitational dynamics, viscous hydrodynamics can reproduce the stress-energy tensor while pressure anisotropies remain large~\cite{Chesler:2009cy,Chesler:2010bi,Heller:2011ju,Casalderrey-Solana:2013aba,Casalderrey-Solana:2013sxa,Jankowski:2014lna,Chesler:2015bba,Chesler:2015wra,Chesler:2015fpa,Chesler:2016ceu}. Kinetic descriptions of weakly coupled quantum chromodynamics (QCD) also exhibit the onset of hydrodynamic behavior before complete equilibration~\cite{Kurkela:2018wud,Kurkela:2018oqw}. This onset, in distinction to the subsequent approach to isotropy or equilibrium, is called \emph{hydrodynamization}.

Ultracold atomic gases offer a complementary setting for the same question. Their connection to nuclear matter through nearly perfect fluidity and shear viscosity is long established~\cite{schaefer2009,cao2011}. They are not replicas of QCD: their microscopic constituents, symmetries, and experimental protocols differ. Their advantage is control over interactions and preparation, together with access to observables at selected stages of the evolution~\cite{bloch2008,chin2010}. By repeating an experiment with different initial states and measurement times, one can test directly how preparation affects transient dynamics. This complements nuclear collisions, where the early stage is inferred principally from particles detected after the evolution. Studies of non-hydrodynamic modes and interaction-driven dissipation have developed this connection beyond near-equilibrium transport~\cite{brewer2015,fujii2018}.

\paragraph{Constitutive relations and their dynamical implementation.}
From a theoretical standpoint, hydrodynamization concerns the validity of \emph{constitutive relations}. Conservation laws do not by themselves close the evolution equations: stresses and currents must also be related to local densities, fluid velocity, and their derivatives. The equation of state determines the ideal-fluid contribution, while viscosity and higher-order transport coefficients characterize derivative corrections. Hydrodynamization occurs when these relations, truncated at low or optimal order, reproduce the relevant microscopic observables to a specified accuracy. Viscous stresses need not yet be negligible relative to the equilibrium pressure. Equally, a large pressure anisotropy alone neither establishes nor excludes hydrodynamic behavior.

The experimental test is less direct. Nuclear-collision phenomenology usually evolves dissipative stresses with relaxation equations of M\"uller--Israel--Stewart (MIS) type, including formulations derived from kinetic theory, and compares predicted final-state observables with data~\cite{1967ZPhy..198..329M,Israel:1979wp,Denicol:2012cn,Niemi:2015qia}. Such equations reproduce specified constitutive relations in a slowly-varying regime but also contain additional transient dynamics. Matching a finite number of transport coefficients does not uniquely determine their behavior far from that regime. Thus, the phenomenological success of a relaxation model must be distinguished from a direct microscopic test of its underlying constitutive approximation~\cite{Baier:2007ix,Romatschke:2017ejr}.

\paragraph{From constitutive relations to attractors.}
The first encounter of a hydrodynamic attractor in~\cite{Heller:2015dha} studied a conformal MIS-type model with the nonlinear second-order structure of Baier--Romatschke--Son--Starinets--Stephanov (BRSSS) theory~\cite{Baier:2007ix}. Its setting was Bjorken flow: longitudinally boost-invariant expansion with transverse homogeneity and rotational symmetry~\cite{Bjorken:1982qr}. These assumptions reduce the dynamics to dependence on proper time $\tau$. With $T$ the effective temperature inferred from the energy density, the natural dimensionless clock is $w=\tau T$. The chapter uses $\mathrm{Kn}=1/w$ as a dimensionless measure of expansion strength; a definition using a specific microscopic relaxation time differs by a theory-dependent factor. For comparisons across interaction strengths, $\tilde w=\tau T/(4\pi\eta/s)$ is useful, where $\eta$ is the shear viscosity and $s$ the equilibrium entropy density~\cite{Kurkela:2018vqr}.

In this model, solutions with different initial dissipative stresses approach a common relation between pressure anisotropy and $w$. At large $w$, this relation reproduces the hydrodynamic gradient expansion; at smaller $w$, it can remain predictive when low-order truncations fail~\cite{Heller:2015dha}. It is therefore natural to interpret the attractor as an \emph{effective constitutive relation} extending beyond the vicinity of local thermal equilibrium. Related attracting behavior appears in kinetic theory in the relaxation time approximation (RTA), QCD kinetic theory, and holography~\cite{Heller:2016rtz,Kurkela:2018wud,Du:2020zqg,Romatschke:2017vte,Spalinski:2018mqg}. Its onset and precision depend on the microscopic description and initial states. Holographic hydrodynamization commonly occurs at $w$ of order unity, while kinetic pre-equilibrium calculations can support matching to hydrodynamics around $1\,\mathrm{fm}/c$ for phenomenologically motivated parameters; neither statement defines a universal hydrodynamization time~\cite{Heller:2011ju,Kurkela:2018wud,Kurkela:2018vqr}.

An all-order relation in a symmetric flow is not, however, a constitutive law for arbitrary space-time dependence. Distinct gradient structures become indistinguishable or vanish under Bjorken symmetry. Moreover, the higher-order coefficients generated by a finite-parameter MIS-type model are predictions of that completion, not independently matched microscopic transport data. The attractor extends the model's constitutive description along the flow studied; its wider applicability must be tested~\cite{Baier:2007ix,Heller:2015dha,An:2023yfq}.

\paragraph{What becomes universal?}
We use \emph{attractorization} to mean reduced sensitivity to some directions in the space of initial conditions, so that evolution approaches an effectively lower-dimensional structure in the chosen observables~\cite{Heller:2020anv,Spalinski:2025ngd}.This is distinct from hydrodynamization. The attracting set need not be a single curve, and energy scales or conserved densities can remain as coordinates along it~\cite{Du:2020zqg}. Different moments of a kinetic distribution can also approach their attracting behavior at different rates~\cite{Strickland:2018ayk}. Information loss here means suppression of observable sensitivity, not erasure of all information from an underlying quantum state.

Two limiting procedures make this distinction particularly clear. \emph{Forward attraction} compares solutions initialized at a fixed time as the observation time becomes late. In the examples studied here, decay of non-hydrodynamic excitations produces this late-time contraction. \emph{Pullback attraction} instead keeps the observation time fixed while moving the initialization earlier, toward the singular early-time endpoint in Bjorken flow. In conformal BRSSS theory, this procedure selects a distinguished regular solution, whereas forward contraction alone does not select a unique representative of the common late-time behavior~\cite{Heller:2015dha,Heller:2020anv}.

The corresponding mechanisms need not act on the same timescale. In MIS-type and relaxation-time kinetic models, rapid longitudinal expansion can suppress initial-state differences algebraically while microscopic relaxation is still ineffective~\cite{Kurkela:2019set,Blaizot:2019scw,Blaizot:2020gql}. We call this \emph{expansion-driven attraction}. It can precede hydrodynamization, retain information in other observables, or occur without a subsequent fluid regime. The reduced classical Yang--Mills example discussed in this chapter demonstrates the last possibility of partial recovery of previously suppressed sensitivity~\cite{Werthmann:2026zso}.

The mathematical connection to hydrodynamics comes from the large-order behavior of its derivative expansion. In holographic Bjorken flow, conformal BRSSS theory, and relaxation-time kinetic theory, the late-time gradient series exhibits factorial growth~\cite{Heller:2013fn,Heller:2015dha,Heller:2016rtz}. This statement concerns those expansions, not every series used in hydrodynamics. An asymptotic series may be accurate at low order even though adding terms indefinitely does not converge. Resummation and resurgence relate its large-order behavior to transient contributions that are exponentially suppressed in the hydrodynamic limit. Together they form a \emph{transseries}: a derivative expansion supplemented by sectors invisible at every finite derivative order. These sectors are nonperturbative with respect to gradients, not necessarily to the interaction coupling. Their amplitudes retain initial-state information; canceling resummation ambiguities does not by itself fix those amplitudes or select the pullback attractor~\cite{Heller:2015dha}.

Other universal regimes need not resemble a fluid. Near a \emph{nonthermal fixed point}, distributions or correlation functions can approach a self-similar form that becomes stationary after appropriate rescaling, without being thermal~\cite{Micha:2002ey,Berges:2008wm,Berges:2013eia,Berges:2013fga,AbraaoYork:2014hbk,Kurkela:2015qoa}, see ~\cite{Berges:2020fwq,Mikheev:2023juq} as well as the dedicated chapter of this volume for reviews. Adiabatic hydrodynamization offers a possible common framework: in adapted variables, rapidly damped modes become subdominant to a slowly evolving sector, provided spectral separation is accompanied by sufficiently weak mixing between the sectors~\cite{Brewer:2019oha,Rajagopal:2024lou}. Scaling can simplify this construction, but is neither necessary nor sufficient for attraction~\cite{Rajagopal:2025nca}. The explicit connection between this framework and quasinormal modes---the characteristic linearized relaxation modes---of a nonthermal attractor provides a concrete bridge between these perspectives~\cite{DeLescluze:2025gaa}.

\paragraph{Scope, emphasis, and main conclusions.}
Our principal relativistic examples are conformal, transversely homogeneous Bjorken flows. We use them as controlled starting points, not as complete descriptions of nuclear collisions. Non-conformal and finite-density examples, spatial perturbations, and the bulk-response protocols in cold atoms delineate important extensions~\cite{Romatschke:2017acs,Du:2020zqg,Du:2025hyk,fujii2018}. The broader hydrodynamic and cold-atom context is developed in Refs.~\cite{Romatschke:2017ejr,schaefer2009,enss2025quantum}. Within this scope, the chapter gives particular attention to three aspects.

First, we develop the phenomenological implications in detail. The energy-density attractor connects early energy deposition to subsequent cooling and entropy production, yielding estimates of final multiplicity under stated assumptions~\cite{Giacalone:2019ldn,Andronic:2025ylc}. Applied locally, it also constrains transverse energy, evolving geometry, and the initial buildup of transverse flow~\cite{Ambrus:2021fej,Ambrus:2022koq}. These results inform consistent hydrodynamic initialization and the linear response of transverse perturbations around an attracting background, as implemented in K\o MP\o ST~\cite{Kurkela:2020wwb,Kurkela:2018wud}. We also discuss photons, dileptons, jet quenching, and attempts to build improved macroscopic descriptions around anisotropic or attracting distributions~\cite{Garcia-Montero:2023lrd,Garcia-Montero:2024lbl,Pablos:2025cli,Strickland:2017kux,DuPlessis:2026qjy}.

Second, we distinguish different levels of universality across theories. Early expansion-driven attraction is explicit in the simple relaxation models; QCD kinetic evolution introduces additional dependence on coupling, observable, and the timescale used to compare trajectories~\cite{Kurkela:2019set,Boguslavski:2023jvg}. In holography, late-time attraction is well supported~\cite{Spalinski:2018mqg,Romatschke:2017vte,Kurkela:2019set} and we clarify some aspects of the preceding early-time regime using the upcoming results of~\cite{hha:2026}.

Third, we examine proposed attractors in driven ultracold gases. The main protocols vary the scattering length of a spatially uniform, normal Fermi gas near unitarity and probe the \emph{bulk channel}, absent in exactly conformal Bjorken hydrodynamics~\cite{fujii2018,fujii2024}. Monotonic driving, periodic driving, and early expansion-driven contraction test different dynamical regimes~\cite{fujii2024,mazeliauskas2026,heller2025early}. These remain theoretical attractor constructions. Rapid contact measurements in Fermi gases, contact dynamics in a Bose gas, and momentum-distribution dynamics in quenched Fermi gases provide complementary experimental advances, not interchangeable demonstrations of those constructions~\cite{xie2026,journeaux2026,yi2025quantum}. In particular, short-time few-body scaling must be distinguished from many-body hydrodynamic attraction~\cite{sharell2026fractional}.

The main conclusions are therefore that attraction is relative to specified observables and initial states; its mechanism and onset need not coincide with hydrodynamization; and its phenomenological usefulness does not establish a universal constitutive theory beyond the flow and response that have been tested. Section~\ref{sec:hydrodynamic_attractors} develops the concrete constructions. Section~\ref{sec:general_definition} presents the state-space and adiabatic perspectives; Secs.~\ref{sec:phenomenology} and~\ref{sec:cold_atoms} develop the nuclear and atomic applications. Section~\ref{sec:outlook} turns these conclusions into theoretical and experimental questions.

Finally, we would like to acknowledge two earlier review articles dedicated to hydrodynamic attractors~\cite{Soloviev:2021lhs,Jankowski:2023fdz}. The main differences with these earlier attempts to summarize the state of the art of hydrodynamic attractors originate from progress that occurred since 2023, in particular, establishing the connection with cold atoms, significant advances in adiabatic hydrodynamization and progress on understanding early-time hydrodynamic attractors in holography.

\noindent \textbf{Note on conventions.} In the review we use the mostly minus metric signature and adopt natural units, i.e. $c = k_{B} = \hbar = 1$. $\hbar$ is restored in Sec.~\ref{sec:cold_atoms}. Unless stated otherwise, we assume that the relativistic theories we work with are in four-dimensional Minkowski spacetime.

\section{Hydrodynamic attractors}
\label{sec:hydrodynamic_attractors}

This section develops the notion of hydrodynamic attractors through a sequence of increasingly general perspectives. Sec.~\ref{sec:hydrodynamization} first distinguishes hydrodynamization from local equilibration and attractorization, and relates late-time forward attraction to the divergent gradient expansion, transseries, and the decay of nonhydrodynamic excitations. Sec.~\ref{sec:BRSSS_attractors} then makes these ideas concrete in conformal BRSSS Bjorken flow, where forward attraction, pullback selection, early expansion-driven contraction, and late-time transient decay can be distinguished explicitly. We then examine how these ideas are realized, and where their interpretation becomes more subtle, in microscopic theories, focusing on kinetic theory in Sec.~\ref{sec:kt_attractors} and on holography in Sec.~\ref{sec:holography_attractors}. Finally, Sec.~\ref{sec:energy_attractor} introduces the energy-density attractor used in the phenomenological applications of Sec.~\ref{sec:phenomenology}.

\subsection{From hydrodynamization to forward attraction}
\label{sec:hydrodynamization}

\subsubsection{Hydrodynamization before local equilibration}

One of the central lessons of modern high-energy nuclear phenomenology is that hydrodynamic models appears to describe the evolution of the QGP much earlier, and under much more extreme conditions, than one might have expected from the traditional interpretation of relativistic hydrodynamics. In the textbook view, hydrodynamics is an effective description of systems close to local thermal equilibrium. From this point of view, one might expect hydrodynamics to become applicable only after microscopic degrees of freedom have equilibrated locally, pressure anisotropies have become small, and dissipative corrections are parametrically suppressed.

The phenomenology of heavy-ion collisions suggests a more subtle picture. Hydrodynamic simulations successfully reproduce a wide range of collective observables even when initialized at very early times after the collision. At such times the system is rapidly expanding and pressure anisotropies can be large. This observation motivates the distinction between \emph{local equilibration} and \emph{hydrodynamization}. Local equilibration refers to the expectation value of the stress-energy tensor being well-described by ideal hydrodynamics, with small dissipative gradient corrections. In Bjorken flow, this requires isotropization (the approximate equality of the longitudinal and transverse pressures). Hydrodynamization is weaker: it means that the expectation value of the stress-energy tensor is already well described by hydrodynamic constitutive relations, even if the viscous corrections to the ideal flow behavior are not small, and the state is not locally equilibrated in the stronger sense.

A particularly clean arena in which to sharpen this distinction is provided by holography \cite{Maldacena:1997re,Gubser:1998bc,Witten:1998qj} (see \cite{Casalderrey-Solana:2011dxg} for a review with a focus on heavy-ion physics). Holography provides a first-principles description of the far-from-equilibrium dynamics of certain strongly coupled non-Abelian gauge theories in the limit of a large number of colors, by mapping them to a dual classical gravity problem. In the canonical example of $\mathcal{N}=4$ supersymmetric Yang--Mills (SYM) theory, the relevant dual problem is the evolution of a five-dimensional asymptotically anti-de Sitter (AdS) geometry governed by Einstein equations with a negative cosmological constant. This problem can be solved numerically using standard methods in general relativity.

The strategy is conceptually simple. One specifies a far-from-equilibrium initial state in the field theory, represented holographically by suitable initial data for the bulk geometry. One then solves the classical gravitational initial value problem and extracts the boundary stress-energy tensor $\langle T^{\mu\nu}\rangle$ by holographic renormalization. From this microscopic stress-energy tensor, one defines the local energy density $\varepsilon$ and fluid velocity $u^\mu$ through the Landau-frame condition
\begin{equation}
\langle T^\mu_{\nu} \rangle u^\nu=\varepsilon u^\mu,
\end{equation}
with $u^\mu u_\mu=1$. These quantities are then inserted into the hydrodynamic constitutive relations. Hydrodynamization is said to occur when the microscopic stress-energy tensor obtained from the gravity computation is accurately reproduced by the hydrodynamic stress-energy tensor constructed from the same $\varepsilon$ and $u^\mu$, within some specified tolerance. 

This procedure allows one to quantify precisely the statement that hydrodynamics can become valid before local equilibration. In many holographic examples, the hydrodynamic prediction for the stress-energy tensor becomes accurate while leading-order viscous corrections to the ideal flow behavior are still sizable. 

The earliest and most studied example is boost-invariant Bjorken flow. Bjorken flow assumes translational and rotational invariance in the transverse plane, together with boost invariance along the collision axis. These symmetries reduce the dynamics to a function of proper time $\tau$, making the problem effectively (0+1)-dimensional and computationally tractable. Holographic simulations of Bjorken flow showed that the stress-energy tensor can become accurately described by viscous hydrodynamics at times when the pressure anisotropy is still large \cite{Chesler:2009cy,Heller:2011ju,Jankowski:2014lna}. This provided a sharp demonstration of hydrodynamization without isotropization.

Over time, this conclusion was tested in increasingly less symmetric setups. Holographic collisions of planar shock waves provided a closer analogue of the longitudinal dynamics in heavy-ion collisions \cite{Chesler:2010bi,Casalderrey-Solana:2013aba,Casalderrey-Solana:2013sxa,Chesler:2015fpa}. Later studies considered localized shocks \cite{Chesler:2015wra,Chesler:2015bba,Chesler:2016ceu}, allowing to consider off-center collisions with nontrivial transverse dynamics. These calculations moved holography closer to the physical situation of colliding nuclei, while still retaining theoretical control. 

As an example, let us focus on Chesler's analysis of hydrodynamization in small systems \cite{Chesler:2016ceu}. In this work, collisions of localized shock waves produce a localized droplet of plasma whose transverse size is of order the microscopic scale, schematically
\begin{equation}
R \sim \frac{1}{T_{\rm eff}},
\end{equation}
where $T_{\rm eff}$ is the effective temperature associated with the local energy density $\varepsilon$. The resulting system has large gradients and develops significant transverse flow. Nevertheless, after the collision its evolution is well described by viscous hydrodynamics.
The result demonstrates that, at strong coupling, hydrodynamic behavior can emerge even in plasma droplets whose size is only marginally larger than the microscopic relaxation scale. This supports the broader phenomenological idea that the applicability of hydrodynamics is not controlled by proximity to local equilibrium. 

The take-home message is therefore the following. Hydrodynamization is a dynamical phenomenon distinct from local equilibration and isotropization. Holography provides a framework in which this distinction can be demonstrated explicitly from first principles: one solves the microscopic dynamics, extracts the stress-energy tensor, and verifies that hydrodynamics describes it even if dissipative corrections to ideal flow behavior are not small. This motivates a deeper question: why does a low-order hydrodynamic description work so well when the assumptions behind a naive gradient expansion appear to be violated?

\subsubsection{The nature of the gradient expansion}

Hydrodynamization before local equilibration calls for an explanation. A natural first possibility would be that the hydrodynamic gradient expansion is much better behaved than one might have expected: perhaps it converges rapidly, so that terms beyond first order are already negligible around the hydrodynamization time. 

The study of holographic Bjorken flow showed that convergence cannot provide a general explanation. In Ref.~\cite{Heller:2013fn}, the hydrodynamic gradient expansion of the stress-energy tensor of large $N$, strongly coupled $\mathcal{N}=4$ SYM undergoing Bjorken expansion was computed through 240 orders. Its coefficients were found to grow factorially, providing strong evidence that the series has zero radius of convergence. Subsequent work established the general divergence of the gradient expansion in Bjorken flow across MIS-type theories \cite{Heller:2015dha,Basar:2015ava,Aniceto:2015mto} and kinetic theory \cite{Heller:2016rtz, Heller:2018qvh}, modulo a counterexample found in \cite{Denicol:2019lio} which concerns a MIS description of an ultrarelativistic gas of hard spheres. Further analysis in general longitudinal flows \cite{Heller:2021oxl} demonstrated that the gradient expansion is also factorially divergent in the model considered by \cite{Denicol:2019lio} once boost invariance is relaxed.

The overarching conclusion of the aforementioned studies is that, in general, the hydrodynamic gradient expansion is an asymptotic expansion with vanishing radius of convergence, just like familiar perturbative expansions in quantum mechanics and QFT. A low order, and more generally an optimally truncated, asymptotic series can nevertheless be highly accurate; what fails is the idea that the formal infinite series defines the full answer by ordinary summation.

This observation changes the conceptual framework. If the gradient expansion were convergent, hydrodynamics could in principle be defined by summing it. Once the series is known to be factorially divergent, the relevant questions are different: what information is encoded in its large-order behavior, how should the divergent series be resummed, what additional data are needed to remove the possible ambiguities of that resummation, and how is the resulting late-time representation related to the full far-from-equilibrium evolution?

\subsubsection{Transseries, resurgence, and forward attraction}

A natural framework for addressing these questions is provided by transseries and resurgence (see Ref.~\cite{Aniceto:2018bis} for a review). To present the general idea in a flow-agnostic manner, let us introduce a formal bookkeeping parameter $\epsilon$ that counts gradients and tends to zero in the hydrodynamic limit. Schematically, the stress-energy tensor can be organized in a transseries of the form
\begin{equation}\label{transseries_general}
\begin{aligned}
T_{\mu\nu}(\epsilon)
={}&T_{\mu\nu}^{(0)}(\epsilon)
+\sum_a \sigma_a e^{-\frac{\chi_a}{\epsilon}}
T_{\mu\nu}^{(a)}(\epsilon)\
&+\sum_{a,b}\sigma_a\sigma_b
e^{-\frac{\chi_a+\chi_b}{\epsilon}}
T_{\mu\nu}^{(ab)}(\epsilon)
+\cdots .
\end{aligned}
\end{equation}
Here, $T_{\mu\nu}^{(0)}$ denotes the hydrodynamic gradient expansion, while the expressions multiplying each exponential are themselves asymptotic expansions in $\epsilon$; these may also contain characteristic powers of $\epsilon$ and, in resonant cases, logarithms. The fundamental sectors labeled by $a$ are typically associated with non-hydrodynamic excitations, while the mixed sectors describe their nonlinear interactions. Once the hydrodynamic data have been fixed, the transseries parameters $\sigma_a$ encode the remaining dependence on the initial state that is visible within this asymptotic description. 

In this context, resurgence means that the asymptotic expansions associated with the different sectors are not independent. Singularities of the Borel transform of one sector encode the low-order fluctuations around other sectors. In particular, the large-order coefficients of the hydrodynamic gradient expansion contain information about the non-hydrodynamic sectors, with the dominant large-order behavior being controlled by the Borel singularity associated with the smallest $|\chi_a|$. If a Borel singularity lies on the chosen summation ray, the corresponding lateral Borel sums differ by an exponentially small ambiguity. Across the associated Stokes ray, the transseries parameters undergo compensating Stokes jumps, so that a consistent resummation of the full transseries can be made free of resummation ambiguities and compatible with the relevant reality conditions.

The transseries ansatz \eqref{transseries_general} encodes the dynamical hierarchy underlying the late-time emergence of hydrodynamics. The hydrodynamic sector depends algebraically on the gradient parameter $\epsilon$, whereas the transient sectors are exponentially suppressed. These exponential factors encode the dynamical decay of the non-hydrodynamic excitations. Consequently, solutions that share the same hydrodynamic data but differ in their transient amplitudes become exponentially close at late times. The imprint of the corresponding non-hydrodynamic initial data on $T_{\mu\nu}$ therefore becomes progressively smaller, and the solutions approach a common asymptotic equivalence class.

For Bjorken flow, the starting point was the holographic discovery of factorial growth of the gradient expansion and the identification of its leading Borel singularities with non-hydrodynamic quasinormal modes \cite{Heller:2013fn}. Explicit resurgent and transseries analyses in MIS-like theories were developed in~\cite{Heller:2015dha,Basar:2015ava,Aniceto:2015mto}. In kinetic theory, the divergence and non-hydrodynamic origin of the gradient expansion were studied in~\cite{Denicol:2016bjh,Heller:2016rtz}, while the structure of the transient sectors, their dependence on initial data, and multiparameter transseries were investigated further in~\cite{Heller:2018qvh,Behtash:2019txb,Blaizot:2020gql,Behtash:2020vqk}. Holographic studies include~\cite{Casalderrey-Solana:2017zyh,Spalinski:2017mel,Spalinski:2018mqg,Aniceto:2018uik}. Extensions beyond transversely homogeneous Bjorken flow include Gubser flow \cite{Behtash:2017wqg,Denicol:2018pak}, and longitudinal flows in which boost invariance is not imposed \cite{Heller:2021oxl,Heller:2021yjh}.

These considerations provide the first notion of a hydrodynamic attractor that we will encounter in this review: a forward attractor. It characterizes a family of solutions that are distinct on an initial Cauchy slice but converge toward one another at sufficiently late times. More precisely, the transseries describes the forward contraction of solutions that differ in their transient amplitudes $\sigma_a$ while sharing the same hydrodynamic data. It does not, by itself, select a unique preferred trajectory within the resulting asymptotic family: any two such solutions differ only by exponentially suppressed contributions and therefore provide equivalent representatives of the late-time behavior.

\subsubsection{Attractorization need not imply hydrodynamization}

The forward hydrodynamic attractor just described is not the most general form of attractor behavior, because its characterization relies specifically on the late-time behavior of the observable at hand. As we will see, universality in non-equilibrium dynamics need not arise solely through the decay of non-hydrodynamic excitations: other mechanisms can lead to the effective loss of the information contained in the initial state and drive the system into a simpler regime. It is therefore essential to distinguish attractorization from hydrodynamization, even though the modern study of the former notion was historically motivated by the latter.

In this review, we use \emph{attractorization} to denote the loss of sensitivity to some directions in the space of initial data and the confinement of the evolution near an attracting structure of lower dimensionality. This definition does not assume that the surviving dynamics is already hydrodynamic. At this stage, hydrodynamic and non-hydrodynamic contributions may still be comparable and need not admit a clean separation. Attractorization is also necessarily relative to the chosen state space and set of observables. For example, the collapse of different stress-energy tensor profiles onto a common evolution does not imply that every microscopic observable has lost sensitivity to the same initial state information.

By contrast, \emph{hydrodynamization} means that the surviving evolution is accurately described, to a specified tolerance, by hydrodynamic constitutive relations truncated at low or optimal order. This typically occurs once non-hydrodynamic contributions have become sufficiently suppressed. However, hydrodynamization requires neither pressure isotropy nor local equilibration, and viscous corrections may remain sizable.

Attractorization may therefore precede hydrodynamization. As illustrated by the canonical example of Bjorken flow in conformal BRSSS theory discussed below, rapid longitudinal expansion can initially collapse a family of states onto a universal far-from-equilibrium evolution in which hydrodynamic and non-hydrodynamic information remain simultaneously relevant. Only at a later stage does the evolution become dominated by hydrodynamic degrees of freedom.

\subsection{Hydrodynamic attractors in conformal MIS Bjorken flow}
\label{sec:BRSSS_attractors}

A useful, and historically crucial, case study illustrating the key points raised in the previous subsection is conformal MIS theory specialized to Bjorken flow and supplemented by the nonlinear second-order structure of BRSSS theory~\cite{Baier:2007ix}. We will refer to this model as conformal BRSSS theory. It provides an all-orders completion of conformal relativistic hydrodynamics whose gradient expansion agrees with the most general constitutive relations up to second order. At higher orders, however, its transport coefficients are fixed by the small set of parameters defining the model and therefore do not take their most general possible values. This system was analyzed in Ref.~\cite{Heller:2015dha}, which provided the modern formulation of the hydrodynamic attractor and established its explicit connection with transseries and resurgence.

For conformal Bjorken flow (see App.~\ref{app}), hydrodynamics predicts that, in the late-time limit $\tau \to \infty$,  
\begin{equation}\label{Bjorken_cooling}
T(\tau)=\frac{\Lambda}{(\Lambda\tau)^{1/3}}
\left[1+\mathcal{O}\!\left(\frac{1}{(\Lambda\tau)^\frac{2}{3}}\right)\right],
\end{equation}
where $T(\tau)$ is the effective temperature defined from the local energy density and $\Lambda$ is an integration constant fixed by the initial state. The leading power $T\propto\tau^{-1/3}$ is the conformal ideal fluid result; viscous effects generate the gradient corrections displayed in Eq.~\eqref{Bjorken_cooling}.

It is convenient to characterize the stress-energy tensor by the dimensionless pressure anisotropy
\begin{equation}\label{A_def}
\mathcal{A}
=\frac{\mathcal{P}_T-\mathcal{P}_L}{\mathcal{P}},
\qquad
\mathcal{P}=\frac{\varepsilon}{3},
\end{equation}
where $\mathcal{P}_T$, $\mathcal{P}_L$, and $\mathcal{P}$ are the transverse, longitudinal, and equilibrium pressures, respectively. The natural dimensionless clock is
\begin{equation}
w=\tau T(\tau).
\end{equation}
The expansion and the conformal MIS relaxation time are
\begin{equation}
\theta\equiv\nabla_\mu u^\mu=\frac{1}{\tau},
\qquad
\tau_\pi=\frac{C_{\tau_\pi}}{T(\tau)},
\qquad C_{\tau_\pi}>0.
\end{equation}
Consequently, the relevant dimensionless measure of gradient strength (the Knudsen number) is
\begin{equation}
\mathrm{Kn}\equiv \frac{\theta}{T}
=\frac{1}{w}.
\end{equation}
The hydrodynamic expansion is thus a large-$w$ expansion. 

In conformal BRSSS theory, for Bjorken flow the evolution of the pressure anisotropy \eqref{A_def} decouples from the energy density and obeys a single nonlinear first-order ODE
\begin{equation}\label{A_ODE}
C_{\tau_\pi}\left(1+\frac{\mathcal{A}}{12}\right)
\mathcal{A}'
+\left(\frac{C_{\tau_\pi}}{3w}
+\frac{C_{\lambda_1}}{8C_\eta}\right)\mathcal{A}^2
-\frac{3}{2}\left(\frac{8C_\eta}{w}-\mathcal{A}\right)=0,
\end{equation}
where a prime denotes a derivative with respect to $w$, and we have defined 
\begin{equation}
\eta=C_\eta s,
\qquad
\lambda_1=C_{\lambda_1}\frac{\eta}{T}.
\end{equation}
The minimal MIS model is recovered by setting $C_{\lambda_1}=0$. 

\subsubsection{Transseries and the gradient expansion}

The formal large-$w$ solution of Eq.~\eqref{A_ODE} is
\begin{equation}\label{A_exp}
\mathcal{A}_{\mathrm{hyd}}(w)
=\sum_{n=1}^{\infty}\frac{a_n}{w^n}
=\frac{8C_\eta}{w}
+\frac{16C_\eta(C_{\tau_\pi}-C_{\lambda_1})}{3w^2}
+\mathcal{O}(w^{-3}).
\end{equation}
In a general hydrodynamic effective theory, $a_n$ depends on the constitutive relations up to $n$th order in gradients. In the BRSSS completion, all $a_n$ are instead generated recursively from the finite set of parameters appearing in Eq.~\eqref{A_ODE}. They are independent of the non-hydrodynamic initial data, and found to grow factorially at large order,
\begin{equation}\label{Lipatov}
a_n=
\mathcal{S}\,\frac{\Gamma(n+\alpha)}{\chi^{\,n+\alpha}}
\left[1+\mathcal{O}\!\left(n^{-1}\right)\right],
\qquad n\to\infty,
\end{equation}
implying that \eqref{A_exp} is an asymptotic series with vanishing radius of convergence. The quantities $\chi$ and $\alpha$ take the values 
\begin{equation}\label{chi_alpha}
\chi=\frac{3}{2C_{\tau_\pi}},
\qquad
\alpha=\frac{C_\eta-2C_{\lambda_1}}{C_{\tau_\pi}}.
\end{equation}
The same result follows by linearizing Eq.~\eqref{A_ODE} about the large-$w$ hydrodynamic solution \eqref{A_exp}. The full one-parameter family of late-time solutions has the transseries representation
\begin{equation}\label{transseries_BRSSS}
\mathcal{A}(w;\sigma) = \sum_{m=0}^{\infty}
\sigma^m e^{-m\chi\,w}\Phi^{(m)}(w),
\qquad
\Phi^{(m)}(w) =w^{m\alpha}\sum_{n=0}^{\infty}\frac{a_n^{(m)}}{w^n},
\end{equation}
The sector $m=0$ is the hydrodynamic gradient expansion. The sector $m=1$ describes the single non-hydrodynamic mode present in this model, while the sectors with $m>1$ are generated by nonlinearities. Equations~\eqref{Lipatov}-\eqref{chi_alpha} display the most basic feature of resurgence: the weight $\chi$ governs both the large-order behavior of the hydrodynamic gradient expansion and the beyond all-orders transient contributions. 

Because the leading Borel singularity lies on the positive real axis, the Borel resummation of the hydrodynamic sector is ambiguous. This ambiguity is canceled by corresponding ambiguities in the transient sectors. For a chosen prescription, this cancellation fixes the ambiguity-dependent imaginary part of the transseries parameter $\sigma$; its remaining real part is the integration constant labeling physical solutions and is fixed by the initial data. At this stage the transseries therefore represents a family of solutions, not a uniquely selected curve.

It also displays their late-time contraction. For two solutions labeled by $\sigma_1$ and $\sigma_2$,
\begin{equation}\label{late_contraction}
\mathcal{A}_{\sigma_1}(w)-\mathcal{A}_{\sigma_2}(w)
\sim(\sigma_1-\sigma_2)e^{-\chi w}w^\alpha
\left[a_0^{(1)}+\mathcal{O}(w^{-1})\right].
\end{equation}
The separation between them is controlled by decay of the non-hydrodynamic mode. The mechanism behind late-time collapse is therefore already built into the transseries ansatz: an algebraically evolving common sector is separated from the exponentially damped dependence on the initial state. 

\subsubsection{Late-time forward attraction}

Let $U(w,w_0)$ denote the process that evolves an initial value at $w_0$ to a later value at $w$. A reference solution $\mathcal{A}_{\mathrm{ref}}(w)$ is forward attracting for a basin $B$ if, at fixed $w_0$,
\begin{equation}\label{forward_definition}
\sup_{\mathcal{A}_0\in B}
\left|
U(w,w_0)\mathcal{A}_0-\mathcal{A}_{\mathrm{ref}}(w)
\right|
\longrightarrow0,
\qquad w\to\infty.
\end{equation}
More generally, an invariant family of compact sets satisfying the analogous limit is called a forward attractor; the single solution $\mathcal{A}_{\mathrm{ref}}(w)$ is the case relevant to this reduced equation. Equation~\eqref{late_contraction} implies this property for solutions sharing the same late-time regime. Forward attraction, however, is nonselective here. Since any two such solutions collapse asymptotically, any one of them can be used as the reference trajectory in Eq.~\eqref{forward_definition}. In this precise sense, every regular member of the transseries family is a valid forward attractor. Forward attraction captures their late-time collapse, but it does not privilege the solution with $\sigma=0$, a particular resummation prescription, or any other unique representative~\cite{Heller:2020anv}.

\subsubsection{Pullback attraction}

A distinguished curve is selected by a different limiting procedure. The point $w=0$ is a singular endpoint of Eq.~\eqref{A_ODE}, not an ordinary initial time surface. As $w\to0$, a generic solution can either diverge as $w^{-4}$ or approach one of the following two constant values:
\begin{equation}
\mathcal{A}_{0,\pm}
=\pm 6\sqrt{\frac{C_\eta}{C_{\tau_\pi}}}.
\end{equation}
For $C_\eta,C_{\tau_\pi}>0$, the regular branch approaching the positive value is stable under evolution toward increasing $w$. Linearizing the difference between nearby solutions about this branch and taking the small-$w$ limit gives
\begin{equation}\label{early_contraction}
\delta\mathcal{A}(w)
\simeq \delta\mathcal{A}(w_0)
\left(\frac{w_0}{w}\right)^{\gamma_+},
\qquad
\gamma_+=\frac{8\mathcal{A}_{0,+}}{12+\mathcal{A}_{0,+}}>0.
\end{equation}
For a fixed observation time $w>0$, pullback attraction is defined by sending the initialization time to the singular endpoint:
\begin{equation}\label{pullback_definition}
\sup_{\mathcal{A}_0\in B}
\left|
U(w,w_0)\mathcal{A}_0-\mathcal{A}_\star(w)
\right|
\longrightarrow0,
\qquad w_0\to0^+,
\end{equation}
for every compact set $B$ contained in the basin of this stable branch. The regular global solution satisfying
\begin{equation}
\lim_{w\to0^+}\mathcal{A}_\star(w)=\mathcal{A}_{0,+}
\end{equation}
is therefore the pullback attractor. It approaches the hydrodynamic asymptotics at large $w$ and is the distinguished representative conventionally called the hydrodynamic attractor in Ref.~\cite{Heller:2015dha}. This curve is also forward attracting, but it is pullback selection, rather than forward collapse, that singles it out. Once the curve has been selected, its transseries parameter can be identified. In the resummation convention of Ref.~\cite{Heller:2015dha}, the pullback-selected solution corresponds to a particular value $\sigma_\star$, whose real part need not vanish. The leading transient displacement from the selected curve is consequently proportional to $\sigma-\sigma_\star$.

\begin{figure}[h!]
    \centering
    \includegraphics[width=0.65\linewidth]{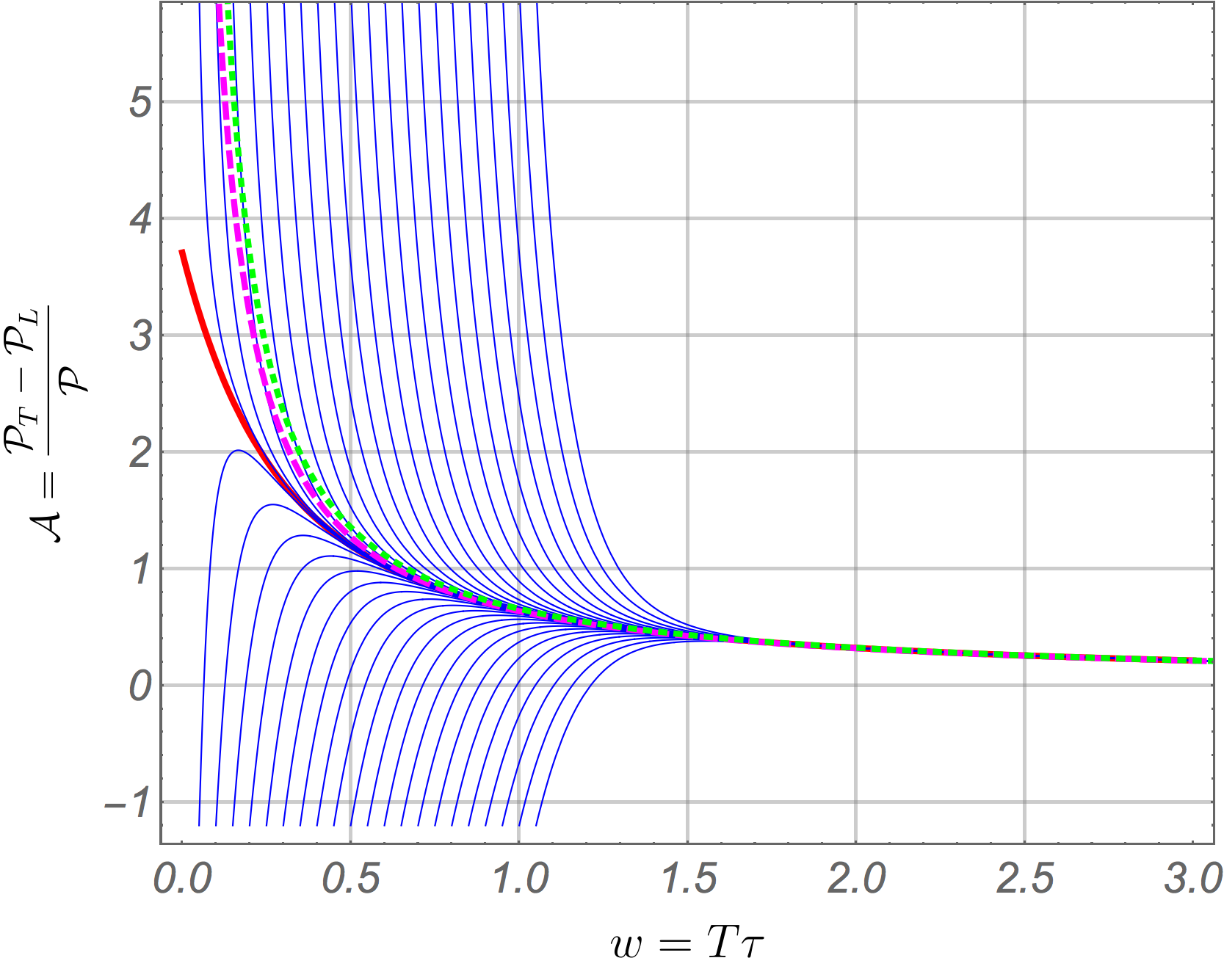}
    \caption{Numerical solutions with different initial values of the pressure anisotropy $\mathcal{A}$ approach a common evolution. The red curve is the stable regular solution $\mathcal{A}_\star(w)$ selected by the pullback limit, while the dashed magenta and the green dotted curves are approximations to the attractor given by the hydrodynamic gradient expansion truncated at first and second order respectively. The values of the parameters in the model have been chosen to match those of $\mathcal{N}=4$ SUSY YM theory in the holographic limit. Adapted from Ref.~\cite{Heller:2015dha}.}
    \label{fig:BRSSS_attractor}
\end{figure}

\subsubsection{Expansion-driven versus relaxation-driven attraction}

The forward and pullback limits also expose two different mechanisms. At small $w$, Eq.~\eqref{early_contraction} gives a power-law approach to the early-time branch of $\mathcal{A}_\star$. Since $\mathrm{Kn}=w^{-1}\gg1$ in this regime, the longitudinal expansion is faster than microscopic relaxation; the leading $1/w$ terms in Eq.~\eqref{A_ODE}, rather than the late-time exponential transient, drive the collapse. This is the expansion-dominated, early-time mechanism emphasized in Ref.~\cite{Kurkela:2019set}, and it can produce attractorization before hydrodynamization. At large $w$, by contrast, microscopic relaxation is fast compared with the longitudinal expansion, and the remaining separation between solutions is the exponentially suppressed non-hydrodynamic contribution in Eq.~\eqref{late_contraction}. It is important to keep in mind that in Bjorken flow, these are two regimes of approach to the same distinguished solution, not two independently selected attractor curves. However, it is possible that a system exhibits one of the two mechanisms but not the other, so it does make sense to distinguish them. In this spirit, we refer to attraction driven by the early-time expansion as an `expansion attractor'.

\subsubsection{Numerical attractor and transseries resummation}

The numerical analysis of Ref.~\cite{Heller:2015dha} makes these statements concrete. Solutions initialized at different values of $w_0$ and with widely different pressure anisotropies rapidly converge toward a common curve, as shown in Fig.~\ref{fig:BRSSS_attractor}. Initializing progressively closer to $w=0$ selects the stable regular solution $\mathcal{A}_\star(w)$, in accordance with the pullback limit. At large $w$, the same curve approaches the hydrodynamic gradient expansion. At smaller $w$, however, the first- and second-order hydrodynamic truncations can differ visibly from it, with the size and duration of that difference depending on the theory parameters. Thus the collapse of solutions and the accuracy of low-order hydrodynamics are empirically distinct even in this simple model.

Ref.~\cite{Heller:2015dha} also compared the numerically selected curve with the generalized Borel resummation of the transseries. After the remaining real integration constant was fixed to $\sigma_\star$, the resummed result tracked $\mathcal{A}_\star(w)$. This agreement shows that the late-time asymptotic data contain enough information to reconstruct the distinguished global solution over a substantially wider range of $w$. It does not alter the logical order of the construction: the numerical pullback criterion selects the curve, and matching to that curve fixes the real transseries parameter.

\subsubsection{Lessons from conformal BRSSS theory}

The conformal BRSSS example therefore distinguishes clearly several notions. The gradient expansion gives the common algebraic late-time asymptotics. The transseries and its resummation provide a nonperturbative completion of those asymptotics into a one-parameter family of solutions and makes their late-time collapse manifest through non-hydrodynamic mode decay. Forward attraction alone does not distinguish one member of this family. Pullback attraction selects the stable regular solution $\mathcal{A}_\star(w)$. Finally, convergence toward that solution need not coincide with the onset of hydrodynamics. Rapid expansion can first erase sensitivity to some features of the initial data and organize the evolution onto a universal far-from-equilibrium structure while hydrodynamic and non-hydrodynamic contributions remain comparably important. Subsequent decay of the transient sector then produces hydrodynamization: the motion on that structure becomes governed by hydrodynamic degrees of freedom and approaches the ordinary late-time gradient expansion. Pressure isotropization and local equilibration can occur later still.

\subsection{Hydrodynamic attractors in kinetic theory}
\label{sec:kt_attractors}

The fact that an attractor for the pressure anisotropy exists also in kinetic theory with an RTA collision kernel was discovered not long after its conceptualization in both the conformal~\cite{Heller:2016rtz} and non-conformal~\cite{Romatschke:2017acs} cases. However, when breaking conformal symmetry via the introduction of a particle mass~\cite{Chattopadhyay:2021ive}, neither shear nor the now nonzero bulk pressure by themselves exhibit attractor behavior, though the longitudinal pressure still does. QCD effective kinetic theory also exhibits an attractor at zero density and fixed coupling strength~\cite{Almaalol:2020rnu}. However, while in RTA evolution curves in terms of the scaled time variable $\tilde{w}$ converge also for different coupling strength, this is not the case in QCD EKT~\cite{Kurkela:2018oqw}. The coupling strength dependence was examined in detail only recently~\cite{Boguslavski:2023jvg} and it was discovered that the degree of collapse across different couplings depends on the observable and the time scaling. Extrapolation at fixed time scaling scheme defines distinct weak-coupling and large-coupling limiting curves, the `limiting attractors'. When comparing time evolution with different baryon densities~\cite{Du:2020zqg}, convergence occurs only at late times $\tau\gtrsim\frac{4\pi\eta}{\varepsilon+\mathcal{P}}$.

But the phase space distribution $f$ carries far more information than just the pressures, and the behavior manifests in all of it. In order to organize the information, usually one considers moments of $f$:
\begin{align}
    \mathcal{M}^{nm}[f]=\int \frac{\d^3p}{(2\pi)^3p^0}~(p\cdot u)^np_L^{2m} f\,.
\end{align}
For example, $n=\mathcal{M}^{10}$ is the number density, $\varepsilon=\mathcal{M}^{20}$ is the energy density, and $\P_L=\mathcal{M}^{01}$ is the longitudinal pressure. The factor 2 for the exponent $m$ is due to the fact that the system is assumed to be even in $p_L$.
It was found in numerical time evolution of both RTA~\cite{Strickland:2018ayk} and QCD EKT~\cite{Almaalol:2020rnu} that all of these moments exhibit convergence to their own attractor curve when normalized to their equilibrium value. However, the speed of convergence can differ strongly for different moments. Part of the reason is that the evolution couples the moments, typically in such a way that moments of higher order in the longitudinal momentum need to converge to their attractor values before lower order ones can. Another reason is that the expansion attractor at early times does not act in the entirety of state space like the late-time attractor does, as we will explain now.

The Boltzmann equation in Bjorken flow can be written in co-moving coordinates as
\begin{align}
    p^\tau\partial_\tau f-\frac{p^\tau p_L}{\tau}\partial_{p_L}f=C[f]\,,
\end{align}
where $p_L$ is the longitudinal momentum variable (see Sec.~\ref{sec:kinetic_theory}) and the term containing $\partial_{p_L}$ describes the longitudinal expansion. At late times $\tau\to\infty$, this term becomes negligible compared to interaction terms and the collision kernel will bring the distribution to local equilibrium, $f\to f_{\rm eq}$, which is fully determined by equilibrium quantities. Thus, the late-time fixed point of all moments of the distribution function is given by their equilibrium values. 

On the other hand, at early times, the expansion term dominates over the collision kernel, and the particles are effectively free-streaming. In the comoving coordinates, this causes $p_L$ to decrease, and the early-time fixed point takes the form
\begin{align}
    f(\tau,p_L,p_\perp)=\frac{1}{\tau}\delta(p_L)g(p_\perp)\,,
\end{align}
where $g(p_\perp)$ relates to the energy density as
\begin{align}
    \varepsilon=\int \frac{\d^3p}{(2\pi)^3}~ p f=\frac{1}{\tau}\int\frac{\d^2p_\perp}{(2\pi)^3}~ p_\perp g
\end{align}
and possibly to other conserved densities but is otherwise unconstrained and constitutes a degeneracy of the expansion attractor. In fact, the transverse momentum distribution remains unchanged in the early-time limit.

\subsection{Hydrodynamic attractors in holography}
\label{sec:holography_attractors}

Holography provides first-principles access to the real-time dynamics of strongly coupled QFTs. It is therefore a natural setting in which to investigate hydrodynamic attractors. Studies of holographic Bjorken flow provide strong numerical evidence for late-time forward attraction: when expressed in the relevant set of dimensionless variables, broad classes of initial states approach a common evolution whose late-time asymptotics is described by the hydrodynamic gradient expansion and its resummations \cite{Heller:2011ju,Heller:2013fn,Spalinski:2018mqg}.

The existence of an early-time attractor is substantially more subtle. In contrast to MIS-type theories, and just like in kinetic theory, the holographic initial value problem has an infinite-dimensional state space: its initial data include functions of the holographic radial coordinate. Consequently, agreement of a small number of boundary observables at one time need not imply similar subsequent evolution.

\begin{figure}[h!]
    \centering
    \includegraphics[width=\linewidth]{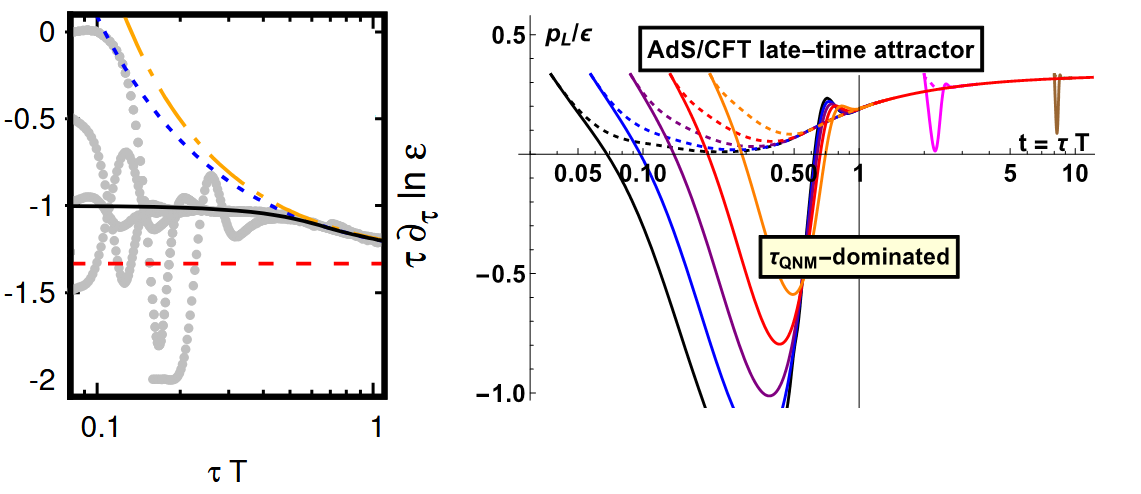}
    \caption{Early- and late-time behavior in holographic Bjorken flow. \emph{Left:} adapted version of the AdS/CFT panel of Fig.~1 in Ref.~\cite{Romatschke:2017vte}. Grey curves show numerical histories obtained from different bulk initial states, the solid black curve is the smooth attractor candidate, and the broken colored curves denote successive hydrodynamic approximations. The candidate approaches $\tau\partial_\tau\log\varepsilon=-1$ toward small $w=\tau T$. \emph{Right:} the holographic panel of Fig.~1 in Ref.~\cite{Kurkela:2019set}. Different colors denote different initialization times, while dashed and solid curves correspond, respectively, to the UV-localized and IR-extended families of bulk profiles; the red curve denotes the late-time attractor. Although the two profiles are matched in their initial $P_L/\varepsilon$ and its first time derivative, their evolutions remain distinct until $w=O(1)$. The UV-localized histories pass close to $P_L/\varepsilon=0$ when initialized sufficiently early. The observables in the two panels are related by $\tau\partial_\tau\log\varepsilon=-1-P_L/\varepsilon$, so the limiting value $-1$ in the left panel corresponds to zero longitudinal pressure in the right panel. 
    }
    \label{fig:attractors_holography}
\end{figure}

Two important points of reference concerning this question are Refs.~\cite{Romatschke:2017vte} and \cite{Kurkela:2019set}. Ref.~\cite{Romatschke:2017vte} studied numerical solutions describing boost-invariant evolution in strongly coupled $\mathcal N=4$ SYM theory. Among the solutions displayed in Fig.~1 of that work (reproduced here in the left plot of Fig.~\ref{fig:attractors_holography}), a smooth curve without apparent quasinormal mode oscillations was identified as an attractor candidate. Its backward extrapolation obeyed
\begin{equation}
\lim_{\tau\to0}
\frac{\partial\log\varepsilon}{\partial\log\tau}=-1 .
\end{equation}
Thus $\varepsilon\sim\tau^{-1}$, which, by Bjorken-flow energy conservation, is equivalent to $P_L/\varepsilon\to0$. This observation should not be interpreted as establishing a universal early-time attractor. The numerical solutions were initialized at finite proper time, and the displayed ensemble was only a low-dimensional projection of the infinite-dimensional space of holographic initial data. Moreover, Ref.~\cite{Romatschke:2017vte} explicitly noted a potential tension with the analysis of Ref.~\cite{Beuf:2009cx}, according to which regular bulk solutions admitting a power-series expansion around $\tau=0$ do not have a singular $\varepsilon\sim\tau^{-1}$ limit. The proposed curve is therefore best regarded as a numerical candidate selected by its smooth evolution and limiting slope, rather than as a solution whose uniqueness or regularity at $\tau=0$ has been established. The other displayed solutions approach it only after oscillatory non-hydrodynamic contributions have decayed.

Ref.~\cite{Kurkela:2019set} tested the sensitivity to genuinely different bulk initial data more directly. The authors considered two families of radial profiles, localized respectively near the AdS boundary and more broadly throughout the bulk. The profiles were chosen so that both the initial value of $P_L/\varepsilon$ and its first proper-time derivative agreed. They nevertheless differed in the additional information carried by the bulk radial profile, or equivalently by higher-order time derivatives of the boundary stress-energy tensor.

The two families did not collapse onto a common curve on a timescale set by the initialization time. Instead, convergence to common behavior occurred on essentially the same timescale $w=\tau T(\tau) = O(1)$ as hydrodynamization. Differences between solutions displayed quasinormal-mode-like oscillatory decay, with particularly pronounced excursions among the solutions supported deeper in the bulk. Thus the calculation provides direct counter-evidence to a universal prehydrodynamic curve determined by the boundary stress-energy tensor alone. Within the set of initial data examined there, only the late-time attractor was universal.

This conclusion does not exclude more restricted forms of early-time universality. The lower panel of Fig.~1 in Ref.~\cite{Kurkela:2019set} (reproduced here in the right plot of Fig.~\ref{fig:attractors_holography}) is consistent with the UV-supported initial data  being driven close to $P_L/\varepsilon=0$ when initialized at progressively earlier times, before they join the hydrodynamic evolution. These observations are suggestive of prehydrodynamic attractorization for UV-supported initial data, but they do not by themselves establish an attractor: doing so would require varying a broader class of initial data profiles and demonstrating quantitative loss of sensitivity to those variations.

Preliminary work in progress involving three of us~\cite{hha:2026} suggests a possible mechanism for such restricted behavior. In the boost-invariant and transversely homogeneous sector, infinitesimal perturbations of the bulk geometry dual to the vacuum can be described by a single gauge-invariant master field. Our analysis indicates that, for initial data that remain finite and nonzero at the Poincaré horizon, the late-time behavior of the boundary stress-energy tensor takes the form 
\begin{equation}
\lim_{\tau \to \infty} \frac{\partial \log \varepsilon}{\partial \log \tau} = - 1,
\qquad
\lim_{\tau\to\infty} \frac{P_L}{\varepsilon}=0 .
\end{equation}
This motivates the following working conjecture. Sufficiently weak initial data supported near the AdS boundary may possess an intermediate time interval during which their boundary response is governed predominantly by the linearized vacuum geometry. If subleading contributions decay before the evolution probes the finite-area apparent horizon present in the full bulk geometry or backreaction becomes important, the boundary stress-energy tensor will be driven toward $P_L/\varepsilon=0$. At later times, this approximation breaks down, and the full bulk geometry and dynamics drive the system toward hydrodynamization. IR-supported data need not pass through this intermediate regime. The conjecture predicts that the near-zero-$P_L$ interval should become more pronounced as the initial data are made weaker and more sharply UV-localized. Testing these predictions in nonlinear evolutions is part of the ongoing work. Because the vacuum has vanishing energy density, the power-law behavior found in the linearized problem is controlled by the expansion itself. In this sense, if prehydrodynamic attraction within the relevant class of initial data is established, this behavior could be described as an expansion attractor.

More broadly, the distinction between universality over the full space of initial data and attraction within physically selected families may be phenomenologically important. Heavy-ion collisions do not prepare arbitrary states. If holography is to be used as a model, they correspond to a subset of states prepared through high-energy shock-wave collisions. The physically relevant question is therefore not only whether a universal early-time attractor exists for every mathematically admissible initial state, but also whether the  relevant set of initial states lies in a restricted basin that attractorizes before hydrodynamizing. A systematic understanding of this question remains to be developed.

\subsection{The attractor for energy density}\label{sec:energy_attractor}

\begin{figure}
    \centering
    \includegraphics[width=0.5\linewidth]{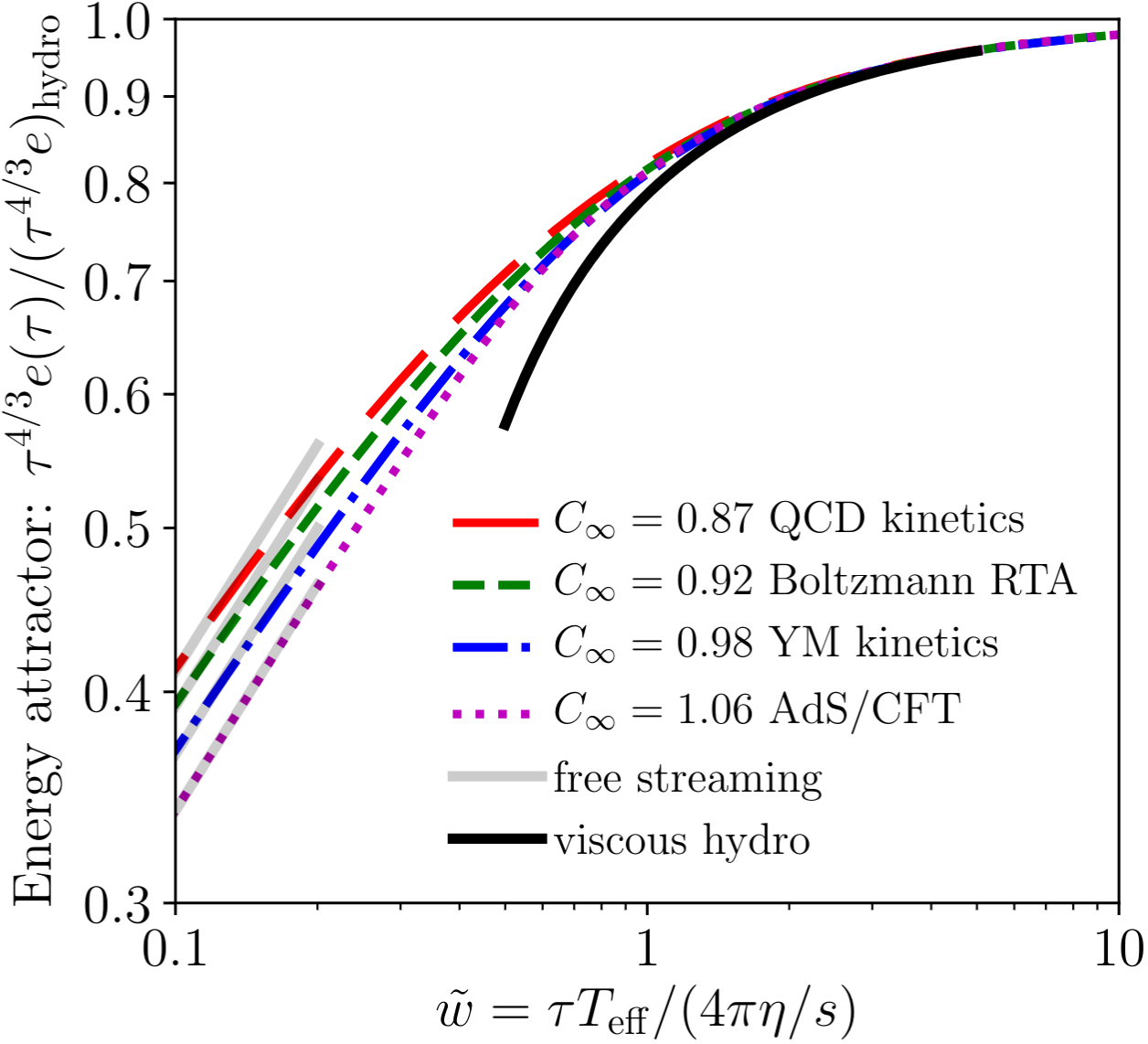}
    \caption{Evolution of the energy attractor $\mathcal{E}(\tilde{w})$ in different dynamical descriptions, which feature different total amount of cooling and therefore different values of $C_\infty$. Taken from~\cite{Giacalone:2019ldn}.}
    \label{fig:energy_attractor}
\end{figure}

 In Bjorken flow, the evolution of energy density is tied to that of longitudinal pressure according to
\begin{align}
    \tau\partial_\tau \varepsilon=-\varepsilon-\mathcal{P}_L\,.
\end{align}
Thus, if the pressure anisotropy evolves on a universal curve, so must also the energy density, up to a multiplicative constant, and the quantity $\mathcal{E}$ defined by
\begin{align}
    \mathcal{E}=\frac{\varepsilon\tau^{4/3}}{\underset{\tau\to\infty}{\lim} \varepsilon\tau^{4/3}}
\end{align}
will exhibit attractor behavior in a similar way to the pressure anisotropy, as was first pointed out in~\cite{Kurkela:2018vqr}. The factor of $\tau^{4/3}$ reflects the time dependence of energy density in the late-time ideal conformal hydrodynamic limit, where $\mathcal{P}_L=\mathcal{P}=\varepsilon/3$ and $\varepsilon\sim\tau^{-4/3}$. In the opposite free-streaming limit of kinetic theory at early times, $\mathcal{P}_L=0$ and $\varepsilon\sim\tau^{-1}$. In terms of the attractor scaled time variable $\tilde{w}=\frac{\tau T}{4\pi\eta/s}$, this means 
\begin{align}
    \mathcal{E}(\tilde{w}\gg1)&=1\,,\\
    \mathcal{E}(\tilde{w}\ll1)&=C_\infty^{-1}\tilde{w}^{4/9}\,.
\end{align}
Fig.~\ref{fig:energy_attractor} shows the evolution of this attractor curve for several different dynamical descriptions. As it turns out, the differences between different descriptions are not too large, with $C_\infty$ ranging between $0.87$ and $1.06$.

\section{Toward a general definition}
\label{sec:general_definition}

This section moves beyond the special description available in conformal Bjorken flow and presents two complementary frameworks for characterizing attractor behavior in more general systems. In Sec.~\ref{sec:state_space}, we explain how attractorization can be interpreted as the contraction of ensembles of states toward persistent lower-dimensional structures, explaining how the familiar BRSSS curve arises as a projection of an attracting hypersurface, and discussing both diagnostics of effective dimensional reduction and extensions to higher-dimensional theories. Within this discussion, \ref{sec:classical_YM} provides an example in which expansion produces early-time memory loss that is partly restored later on without subsequent equilibration-driven attraction. Sec.~\ref{sec:AH} then presents the complementary description provided by adiabatic hydrodynamization, in which gaps in the instantaneous spectrum and suppressed nonadiabatic mixing progressively isolate a low-lying sector. It also clarifies the facilitating (but nonessential) role of scaling evolution and relates this instantaneous-mode picture to the relaxation described by quasinormal modes.

\subsection{Dimensional reduction in state space}
\label{sec:state_space}

The original representation of the Bjorken-flow attractor in conformal BRSSS theory as a single curve $\mathcal{A}_\star(w)$ was exceptionally economical, but it relied on two special features: a preferred dimensionless clock, $w=\tau T$, and a partial decoupling which made the equation for $\mathcal{A}(w)$ close on itself. Neither feature need be available in a generic flow or microscopic theory. More importantly, the curve by itself obscures which initial data have actually been erased and which remain as coordinates along the attracting evolution. The state-space formulation restores this information and provides a language which does not require identifying in advance a single observable that displays universal
behavior~\cite{Heller:2020anv,Spalinski:2025ngd}.

Let $\mathcal{P}_t$ denote the state space of the theory at time $t$. The key idea behind the state-space picture is to consider an ensemble of initial states $B_{t_0} \subset \mathcal{P}_{t_0}$ and focus on its image $B_t \subset \mathcal{P}_t$ under time evolution. Dissipative effects may contract this ensemble much more strongly in some directions than in others. The evolved cloud then develops a hierarchy of widths and becomes concentrated, to a prescribed accuracy, near a lower-dimensional region. In this sense, attractorization can be viewed as dynamical reduction of the \emph{effective} dimensionality of an ensemble in state space \cite{Heller:2020anv}. 

This formulation does not rely on taking either an
asymptotically early or an asymptotically late limit. As a diagnostic of information loss, it is therefore logically independent of the definitions of pullback and forward attraction introduced above.
Dimensional reduction alone does not, however, identify an attractor: one must also verify that the resulting low-dimensional region persists
under time evolution and attracts neighboring states. In conformal MIS theory, the region identified in this way coincides with the lift to
state space of the pullback-selected solution
$\mathcal{A}_\star(w)$.

\begin{figure}[h!]
    \centering
    \begin{minipage}[h!]{0.47\textwidth}
        \vspace{0pt}
        \centering
        \includegraphics[width=.9\linewidth]{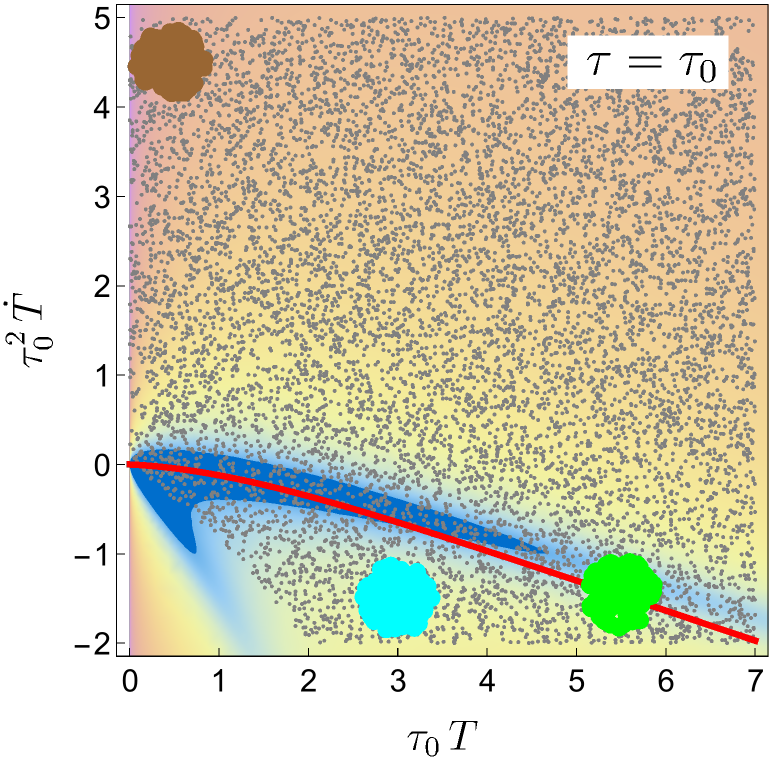}
    \end{minipage}
    \begin{minipage}[h!]{0.47\textwidth}
        \vspace{0pt}
        \centering
        \includegraphics[width=.9\linewidth]{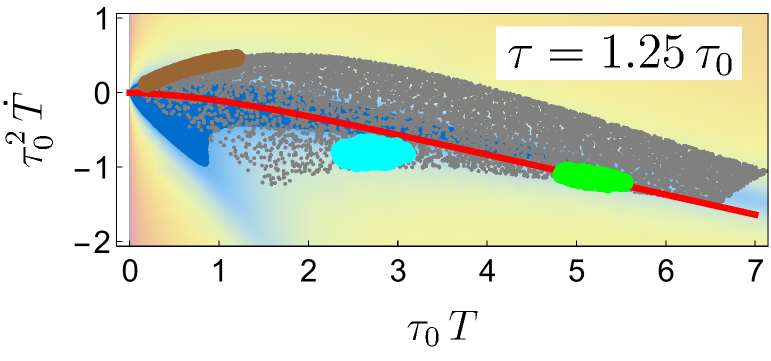}  
        \includegraphics[width=.9\linewidth]{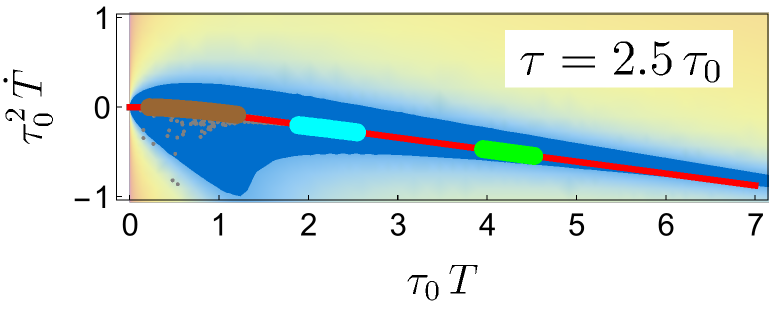}        
    \end{minipage}
    \caption{Three snapshots of the evolution of an ensemble of initial states
    in conformal MIS theory, adapted from
    Ref.~\cite{Heller:2020anv}. The red curve is the section $\mathcal{M}_\star(\tau)$ of the attracting hypersurface on the
    corresponding constant-$\tau$ slice. The colored ensembles illustrate local dimensional reduction: they first become narrow, effectively one-dimensional clouds, subsequently approach the attracting section, and finally evolve along it. The background
    color represents the magnitude of the state-space velocity, with bluer regions corresponding to slower evolution. The plots use
    $C_\eta=0.75$ and $C_{\tau_\pi}=1$.
    }
    \label{fig:BRSSS_state-space}
\end{figure}

To see this explicitly, consider again Bjorken flow in conformal MIS theory. On each constant-$\tau$ slice, the reduced state space is two-dimensional and may be parametrized by $(T,\dot T)$, or
equivalently by $(T,\mathcal{A})$. A two-dimensional initial region $\mathcal{B}_{\tau_0}$ is represented by a cloud of states on the slice
$\tau=\tau_0$. Evolving each point in this cloud reveals three conceptually distinct stages
\cite{Heller:2020anv}. First, the cloud becomes effectively one-dimensional, possibly while still far from the attractor locus. The resulting narrow cloud subsequently approaches this locus and
finally evolves along it toward the locally equilibrated hydrodynamic regime. The intersection of the attracting hypersurface with a
constant-$\tau$ slice is the curve
\begin{equation}\label{MIS_attractor_section}
\mathcal{M}_\star(\tau)=
\left\{
(T,\dot T):
6+18\tau\frac{\dot T}{T}
=\mathcal{A}_\star(\tau T)
\right\}.
\end{equation}
The complete two-dimensional attracting hypersurface in the extended state space parametrized by $(\tau, T, \dot{T})$ is therefore
\begin{equation}
    \widehat{\mathcal{M}}_\star
    =
    \bigcup_{\tau>0}
    \{\tau\}\times\mathcal{M}_\star(\tau).
\end{equation}
The three stages of the evolution are illustrated in
Fig.~\ref{fig:BRSSS_state-space}.

The direction tangent to $\mathcal{M}_\star(\tau)$ contains the initial state information which survives within the attracting family.
Near $\tau=0$, imposing $\mathcal{A}\rightarrow\mathcal{A}_{0,+}$ selects the regular branch but leaves one dimensionful amplitude, conventionally denoted by $\mu$,
undetermined \cite{An:2023yfq}. At late times, the same one-dimensional freedom may instead be parametrized by the hydrodynamic scale $\Lambda$; the evolution determines a map between $\mu$ and $\Lambda$. The direction transverse to $\mathcal{M}_\star(\tau)$ measures departure from the attracting family. In the regime controlled by the late-time transseries \eqref{transseries_BRSSS}, this departure is parametrized by $\sigma-\sigma_\star$. Thus a generic two-dimensional cloud contracts predominantly in the direction carrying transient information while remaining extended along the direction which labels different hydrodynamic histories. This identification with $\Lambda$ and $\sigma-\sigma_\star$ is asymptotic: outside the late-time transseries
regime, the tangent and transverse directions remain geometrically well-defined, but need not admit a controlled separation into hydrodynamic and non-hydrodynamic data.

The familiar $(w,\mathcal{A})$ plot is an efficient projection of this geometry. On every slice with $\tau>0$, the transformation
\begin{equation}
    (T,\dot T)\longleftrightarrow(w,\mathcal{A})
\end{equation}
is invertible. In conformal MIS theory, scale invariance and the partial decoupling of the equation for $\mathcal{A}(w)$ imply that the
attracting sections take the identical form
\begin{equation}
\mathcal{M}_\star(\tau) = \left\{(w,\mathcal{A}): \mathcal{A}=\mathcal{A}_\star(w)
\right\}
\end{equation}
on every proper-time slice. In the coordinates
$(\tau,w,\mathcal{A})$, the full attracting hypersurface is consequently a generalized cylinder defined by the $\tau$-independent relation
$\mathcal{A}=\mathcal{A}_\star(w)$. Projecting away the explicit time coordinate produces the familiar single curve $\mathcal{A}_\star(w)$ \cite{Spalinski:2025ngd}.

The scale $\Lambda$ has not disappeared from the physical solutions. At fixed $\tau$, it parametrizes position along the projected curve,
while along a physical trajectory it determines the temperature history $T(\tau)$. What has become universal is the relation between the two dimensionless variables $w$ and $\mathcal{A}$, not the complete history of every state. This also reconciles the uniqueness of the pullback-selected curve $\mathcal{A}_\star(w)$ with the existence of a
one-parameter family of physical histories on the attracting hypersurface.

The projection is a convenience rather than the defining feature of the attractor. In a generic non-conformal theory, the sections $\mathcal{M}_\star(\tau)$ will change shape from one proper-time slice to another, and there need not exist variables which turn them into a single time-independent curve. Ref.~\cite{Spalinski:2025ngd} exhibited this explicitly in a simple MIS model in which conformal symmetry is broken through the equation of state, while bulk pressure is neglected.
Although the evolution equations no longer partially decouple, regularity at $\tau=0$ selects a one-parameter family of solutions whose
constant-$\tau$ sections attract generic states. This example shows  that conformal symmetry is responsible for the particularly simple
projection onto a single curve, but is not necessary for early-time attraction in this class of models. 

\begin{figure}[h!]
    \centering
    \includegraphics[width=0.47\linewidth]{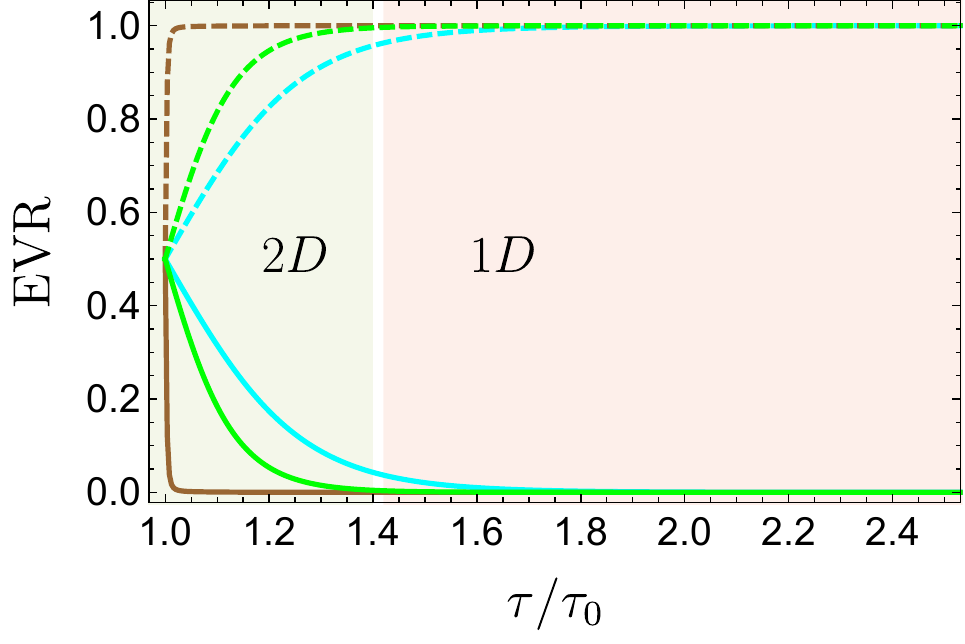}    \includegraphics[width=0.47\linewidth]{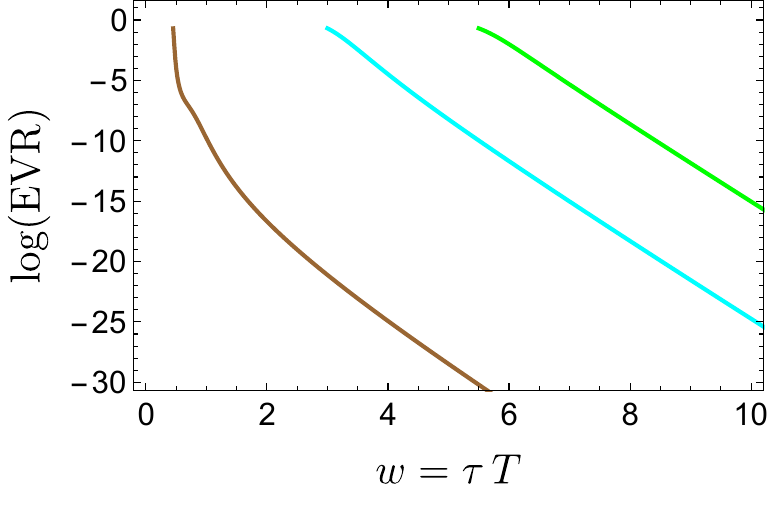}
    \caption{PCA of the three regions marked in brown, cyan, and green in Fig.~\ref{fig:BRSSS_state-space}, adapted from Ref.~\cite{Heller:2020anv}. For each region, the analysis is performed on an initially circular cloud of radius $10^{-4}$, in the dimensionless coordinates $(\tau_0T,\tau_0^2\dot T)$, centered within the corresponding colored ensemble. Left: evolution of the leading (dashed) and subleading (solid) EVR as function of $\tau/\tau_0$. The two components initially carry approximately equal variance; effective dimensional reduction is signaled by the leading EVR approaching unity and the subleading EVR approaching zero. Right: logarithm of the subleading EVR as a function of $w=\tau T$. Its asymptotically linear behavior shows that the transverse variance decays exponentially, with a rate compatible with twice the decay rate of the transient contribution to $\mathcal{A}(w)$. The early-time behavior of the brown cloud departs from this exponential law and is interpreted as expansion-driven.}
    \label{fig:BRSSS_EVR}
\end{figure}

The effective dimensionality of an evolving cloud can be quantified using Principal Component Analysis (PCA). For a sufficiently small cloud in a two-dimensional state space, let $\lambda_1\geq\lambda_2$ be the eigenvalues of its covariance matrix. The explained-variance ratios (EVR)
\begin{equation}
    r_i=\frac{\lambda_i}{\lambda_1+\lambda_2}
\end{equation}
measure the relative squared widths along the two principal directions, and reduction to one effective dimension is signaled by $r_2\ll r_1$. In Ref.~\cite{Heller:2020anv}, PCA was applied to small
initially circular clouds centered on the brown, cyan, and green ensembles shown in Fig.~\ref{fig:BRSSS_state-space}. The onset of
dimensional reduction was found to depend on the location of the initial cloud. At sufficiently large $w$, the subleading EVR decays exponentially with a rate compatible with twice the exponential decay rate of the MIS transient mode (see Fig.~\ref{fig:BRSSS_EVR}). This factor of two has a simple origin: the transverse displacement is linear in the transient amplitude, whereas its contribution to the covariance is quadratic. The early contraction of the brown cloud, which begins at the smallest value of $\tau_0T$, does not follow this late-time law and was interpreted as being driven primarily by the longitudinal expansion rather than by transient-mode decay \cite{Heller:2020anv}.

PCA itself is not invariant under nonlinear changes of state-space coordinates or independent rescalings of the variables. Moreover, a global PCA of a cloud extended along a curved manifold can interpret curvature as additional dimensionality. PCA is therefore most reliable as a local diagnostic applied to clouds which are sufficiently small, in the chosen coordinates, compared with the curvature scale of the attracting manifold.

Finally, let us emphasize that the main advantage of the state-space approach becomes manifest in higher-dimensional systems. In the Heller--Janik--Spaliński--Witaszczyk (HJSW) model \cite{Heller:2014wfa}, whose constant-$\tau$ state space is three-dimensional, the evolution displays
a first reduction from three effective dimensions to two, attributed to the expansion, followed by an oscillatory stage reflecting the complex non-hydrodynamic modes and a final reduction to one dimension. The same framework was also applied exploratorily to a five-dimensional family
of initial conditions embedded in a sixteen-dimensional moment representation of RTA kinetic theory~\cite{Heller:2020anv}. Subsequently, PCA was used in Yang--Mills kinetic theory to isolate one
surviving component associated with the overall energy scale and a subleading component carrying correlated deviations of the pressure anisotropy, screening mass, and scattering rate
\cite{Du:2022bel}. These examples illustrate the real payoff of the state-space picture: it remains meaningful when no single curve in a preferred pair of variables is available.

\subsubsection{Expansion attractor in classical Yang--Mills theory}
\label{sec:classical_YM}

In an effort to explore the general characteristics of attractors, Ref.~\cite{Werthmann:2026zso} studied the behavior in the gauge fields of classical Yang--Mills theory under Bjorken-like symmetries. This is also motivated by the fact that the earliest stages of a heavy ion collision feature high gluon occupation numbers and may be described by an ensemble of independent classical trajectories of the strong Yang--Mills fields in a framework called `Glasma'. However, the aim of Ref.~\cite{Werthmann:2026zso} was to find a simple example of a dynamical description that features expansion-driven but no equilibration-driven memory loss. Therefore, a minimal setup was chosen where not ensembles but individual trajectories of classical Yang--Mills dynamics are considered.

In the early-time regime of this setup, geometric terms from the longitudinal expansion dominate over the non-Abelian interaction terms, such that the evolution equations decouple and can be solved analytically. These solutions exhibit constant longitudinal chromoelectric and -magnetic fields, while transverse fields drop $\propto\tau^{-1}$. However, the transverse chromomagnetic fields, which in this setup are purely non-Abelian, transition to a linear growth in time on a timescale that is determined by the initial conditions. 

This behavior is also reflected in the pressure anisotropy, which depends quadratically on the field strengths. At early times, domination of longitudinal fields means $\mathcal{P}_L/\varepsilon\approx-1$, and the decay of transverse fields causes convergence towards this value as $\propto\tau^{-2}$. After the transition, $\mathcal{P}_L/\varepsilon+1$ starts to grow quadratically, and if the expansion dominated period is sufficiently extended, may even saturate to $\mathcal{P}_L/\varepsilon=1$, signaling domination of transverse fields. When interactions take over on the timescale $\tau_{\rm int}\sim Q_s^{-1}\sim [gA(\tau_0)]^{-1}$, the dynamics transitions to oscillatory behavior with no preferred value of the pressure anisotropy. Thus, this system features memory loss at very early times, which is partially restored while expansion still drives the dynamics, but the non-Abelian transverse fields dominate pressure contributions, while at late times information is scrambled but neither lost nor further restored.

In this system, the pressure anisotropy provides information only on the dominant fields. A full grasp of the behavior of different field components can only be obtained in the state-space picture, which here is spanned by all chromomagnetic and -electric field components. Indeed, clouds of trajectories will initially have constant extent in the directions of longitudinal fields but contract in directions of transverse fields. After the aforementioned transition, transverse chromomagnetic field components grow again, while the cloud extent in transverse chromoelectric fields keeps shrinking, which is not visible in just the pressure anisotropy. The framework further confirms that during interaction domination, the cloud rotates and oscillates, but no continuous trend of contraction or growth is observed.

\subsection{Adiabatic hydrodynamization}
\label{sec:AH}

The adiabatic hydrodynamization framework provides an overarching perspective on the emergence of hydrodynamics and hydrodynamic attractors. As we will see, its conceptual foundations differ from those of the state-space picture. Indeed, the two approaches are complementary, as they provide different languages to understand the same far-from-equilibrium dynamics, with the common emphasis on the effective loss of initial-state information as the dynamics becomes dominated by a lower-dimensional sector.

The basic idea underlying adiabatic hydrodynamization is to recast the evolution problem, possibly after a suitable time-dependent redefinition of variables, as a pseudo-Schr\"odinger equation,
\begin{equation}\label{AH_IVP}
\partial_t\psi=-H_{\rm eff}[\psi;t]\psi,
\end{equation}
and to analyze the instantaneous spectrum of $H_{\rm eff}$. The effective Hamiltonian is generally non-Hermitian and may depend on $\psi$; hence, although $H_{\rm eff}[\psi;t]$ acts linearly when $\psi$ and $t$ are held fixed, the evolution it generates can be nonlinear. Moreover, the evolution is generically non-unitary. Over a short interval $\delta t$, if the generator and its eigenbasis are treated as frozen, an
instantaneous right eigenmode with eigenvalue $E_n(t)$ acquires a
factor $e^{-E_n(t)\delta t}$, rather than the phase
$e^{-iE_n(t)\delta t}$. Consequently, when mixing between modes is suppressed, the real parts of the eigenvalue differences control their instantaneous relative damping, while their time integrals control the accumulated damping. The key word of `adiabaticity' in this framework refers to the suppression of transitions between instantaneous eigenmodes induced by the evolution of the eigenbasis.

Suppose that a band of low-lying modes is separated from the rest of the spectrum by a finite positive gap in the real parts of the instantaneous eigenvalues. If the evolution is adiabatic, higher modes are not efficiently repopulated as they decay. A broad class of initial states is therefore rapidly driven onto the low-lying sector, which acts as an attractor. The initial state information carried by the higher modes is lost first, while some residual memory may remain in the particular combination of modes within the low-lying band. In the kinetic theory examples studied so far, the subsequent spectral evolution opens further gaps within this band, eventually isolating a unique lowest mode. Hence, just like the state-space picture, adiabatic hydrodynamization describes a progressive reduction of the effective number of degrees of freedom: first onto a multi-modal prehydrodynamic attractor and ultimately onto a single dominant mode. Crucially, this last mode is not a static final state. It continues to evolve adiabatically and becomes well-described by hydrodynamics at late times. 

While the concept itself sounds quite general, it has so far been successfully applied only to a small class of systems. This is because constructing a meaningful effective Hamiltonian that fulfills the adiabaticity requirement is usually not straightforward. However, it is simplified in systems where a large part of the dynamics can be absorbed into a choice of frame, that is, into a time-dependent redefinition of variables which renders the dominant profile nearly stationary in the adapted variables. This is why, although scaling is not part of the definition of adiabatic hydrodynamization~\cite{Brewer:2019oha}, it played a central technical role in several of its early applications. When a system exhibits self-similar or prescaling evolution, suitable time-dependent rescalings can absorb the dominant time dependence of the distribution function and, for an exact scaling solution, render the rescaled profile time-independent. In favorable cases, the resulting adapted frame also makes the relevant low-lying eigenmodes of $H_{\rm eff}$ stationary or slowly varying, rendering the adiabatic structure manifest and simplifying the spectral analysis~\cite{Brewer:2022vkq,Rajagopal:2024lou}.

This technical convenience should not, however, be confused with a necessary condition for adiabatic hydrodynamization. Its essential ingredients are instead the existence, in a suitably adapted frame, of a low-lying sector separated from faster modes by a gap in the real parts of the instantaneous eigenvalues, together with the suppression of nonadiabatic mixing between these sectors. Neither requires the surviving modes to obey a self-similar or prescaling form. Indeed, adiabatic hydrodynamization has been used to describe continuously the passage between distinct scaling regimes across an intermediate evolution that is not itself scaling~\cite{Rajagopal:2024lou}, and, more recently, to identify prehydrodynamic attractor behavior in simplified boost-invariant gluon kinetic theories that exhibit no prethermal scaling~\cite{Rajagopal:2025nca}. Conversely, scaling is not sufficient for attraction: free streaming provides a scaling solution whose rescaled profile retains dependence on the initial conditions and therefore does not constitute an attractor~\cite{Rajagopal:2025nca}. Scaling can thus provide a particularly natural choice of variables and substantially simplify the implementation of adiabatic hydrodynamization, but the organizing mechanism proposed by the framework (spectral separation, relative decay of the higher modes, and adiabatic following of the surviving low-lying sector) is not intrinsically tied to scaling.

The idea of classifying decaying dynamics around a reference state is reminiscent of the concept of quasinormal modes, which describe the linearized evolution around the reference state by a set of complex coefficients in exponential time evolution, $\propto e^{i\Omega t}$, $\rm{Im}~\Omega > 0$. The imaginary part of these coefficients describes the decay, while the real part describes oscillatory behavior. This concept extends to scaling solutions~\cite{DeLescluze:2025jqx}, proving that deviations from these solutions relax in a classifiable manner, with the distinction that, in the absence of a reference timescale, the decay is not exponential but a power law, $\propto t^{i\Omega}$. The connection to the picture of an instantaneous spectrum of decaying eigenstates in adiabatic hydrodynamization has recently been made explicit~\cite{DeLescluze:2025gaa} by constructing an exact correspondence between these eigenstates and linear and nonlinear combinations of quasinormal modes close to the scaling solution in an example case. The key distinction of the two concepts is that quasinormal modes can describe relaxation dynamics close to the asymptotic reference state, while adiabatic hydrodynamization considers relaxation towards the instantaneous ground state, which itself evolves towards this reference state.

\section{Phenomenological predictions for nuclear collisions from the attractor}
\label{sec:phenomenology}

Estimates of the quantitative effects of many important phenomena in heavy ion collisions can be derived from a toy model description in Bjorken flow. These usually provide a good idea of the order of magnitude and the parametric dependencies of the considered effects. However, more accurate predictions need to take into account the transverse structure of the collision system. Moreover, observables that relate to details of this transverse structure, such as transverse flow, cannot be estimated in pure Bjorken flow at all.

While including a transverse structure introduces further complexity to the system, knowledge of the attractor curve can simplify the description of longitudinal dynamics. Naturally, making use of attractor behavior comes with the assumption that the system is close to conformal. Also, much of the simplification comes from the fact that the initial rapid convergence onto the attractor is neglected, and the system is assumed to be on its attractor at all times. In this section, we will present several models that provide predictions for heavy ion phenomenology with transverse profiles by making use of the Bjorken flow attractor.

\subsection{Local Bjorken flow}\label{sec:local_bjorken_flow}

The underlying idea for many of these models is transverse locality at early times. As the colliding nuclei carry high longitudinal momenta, which the medium created in the collision inherits, boost-invariant heavy ion collision feature involved longitudinal dynamics from the initial stage on. In contrast, transverse dynamics are only set in motion over time in response to the transverse gradients that are present in the system. This means that on timescales much smaller than the typical inverse size of transverse gradients of e.g. the energy density $\varepsilon$, $\tau\ll``\langle\nabla_\perp \rangle^{-1}$'', neighboring points in the transverse plane do not influence each other yet and thus follow purely longitudinal dynamics, i.e. Bjorken flow, which in a (close to) conformal system means that attractor behavior will manifest. Further assuming that the initial period of decay onto the attractor can be neglected, at any point in time the local quantities exactly match the corresponding attractor configuration. After mapping out the attractor curves, computing longitudinal dynamics then reduces to simple evaluation of these curves.

\subsubsection{Predicting final state charged particle multiplicity}\label{sec:mult_est}

The simplest example of such a model relies on the attractor curve of energy density to predict the multiplicity of particles produced in a heavy ion collision from its initial energy density profile. To understand how this works, we first introduce the energy attractor.

The model assumes that the system starts in the early-time limit on the curve $\E(\tilde{w})$ and describes the final state by its late-time limit, see Sec.~\ref{sec:energy_attractor}. Since these two regimes are connected by a universal time evolution, together with the dependence of $\tilde{w}$ on the energy density via temperature, $\varepsilon=aT^4$, this means 
\begin{align}
    (\varepsilon\tau^{4/3})_{\rm late}=C_\infty(4\pi\eta/s)^{4/9}a^{1/9}(\varepsilon\tau)_{\rm early}^{8/9}\,.
\end{align}
Again, the factor $\tau$ in $(\varepsilon\tau)_{\rm early}$ reflects the early-time free-streaming behavior in kinetic theory. This formula now allows to compute the final energy density from the initial one in one step. Note that the relation between the two is not a linear one due to the fact that more initial energy density leads to more interactions and therefore more cooling. In this model, different dynamical models yield slightly different entropy production only due to the slightly different values of $C_\infty$.

In the late-time equilibrated case one can make use of the thermodynamic relations $Ts = \varepsilon+\mathcal{P}$ and $\varepsilon=3\mathcal{P}$ as well as the empirical fact that the multiplicity of final state particles is proportional to the total entropy on the freeze-out surface -- $N_{\rm ch}/S\approx0.133$ ~\cite{Hanus:2019fnc} -- to arrive at an estimate for said multiplicity. Together with all the parametric dependencies in the proportionality constants, this estimation formula reads
\begin{align}
    \frac{\d N_{\rm ch}}{\d \eta}=\frac{4}{3}\frac{N_{\rm ch}}{S}C_{\infty}^{3/4}(4\pi\eta/s)^{1/3}a^{1/3}\int \d^2x_\perp (\varepsilon\tau)_0^{2/3}\,.\label{eq:multiplicity_estimate}
\end{align}
Let us summarize the model simplifications that allow to derive this formula. It is assumed that the system follows local Bjorken flow and is conformal, starts its evolution in free-streaming and fully equilibrates exactly on the local Bjorken flow attractor at all points in the transverse plane. Under these assumptions, an estimate of the total entropy is computed. This means that entropy production due to transverse expansion is neglected while entropy production due to longitudinal dynamics is overestimated, as the real system would not fully equilibrate. As demonstrated in Ref.~\cite{Giacalone:2019ldn}, the resulting prediction assuming constant $N_{\rm ch}/S$ is in good agreement with LHC data.

The formula~\eqref{eq:multiplicity_estimate} has recently been fine-tuned by comparing with simulation results from a setup using K\o MP\o ST for pre-equilibrium, MUSIC for the hydrodynamic stage and SMASH as a hadronic afterburner~\cite{Andronic:2025ylc}. It was found that an even more accurate estimate can be obtained by multiplying~\eqref{eq:multiplicity_estimate} by another $\eta/s$-dependent factor
\begin{align}
    C\left(\frac{\eta}{s}\right)=a+b\sqrt{\frac{\eta}{s}}+c\frac{\eta}{s}\,,
\end{align}
where $a\approx0.94$ $b\approx -0.18$ and $c\approx -0.07$ slightly depend on the timescale of switching from pre-equilibrium to hydrodynamics.

\subsubsection{Predicting evolution of transverse energy}

In a similar way to energy density, an attractor curve can also be defined for transverse energy density, which in kinetic theory is the integrated transverse momentum of particles
\begin{align}
    \frac{\d E_\perp}{\d^2x_\perp\d \eta}=\tau\int\frac{\d^3p}{(2\pi)^3} p_\perp f\,.
\end{align}
By writing it as a sum, the total transverse energy can also be extracted from the experimentally measured final state particle spectrum. Thus, physics lessons can be learned from studying it in theoretical models and comparing to data. This quantity is constant in free-streaming, where the momenta do not change, and drops as $\tau^{-1/3}$ in equilibrated Bjorken flow. 

\begin{figure}
    \centering
    \includegraphics[width=0.5\linewidth]{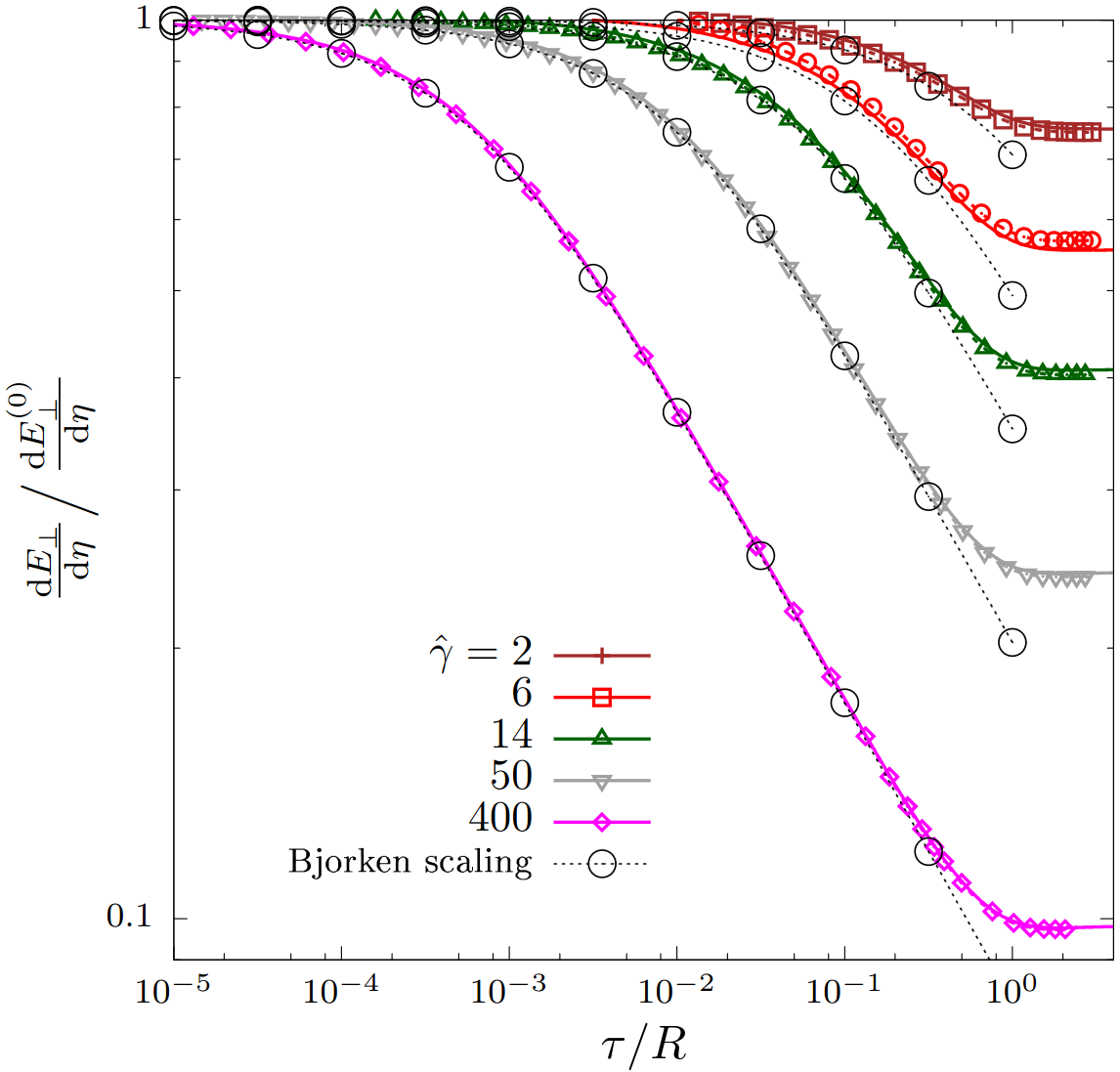}
    \caption{Time evolution of total transverse energy in 2+1D simulation of a heavy ion collision based on kinetic theory in conformal relaxation time approximation compared to the estimate from a transverse collection of attractor curves. For a larger number of interactions -- as quantified by the opacity $\hat{\gamma}$, the transition to the equilibrium power law $\propto\tau^{-1/3}$ occurs faster and the system cools more in total. Taken from~\cite{Ambrus:2021fej}.}
    \label{fig:transverse_energy}
\end{figure}

The attractor for the transverse energy density can be expressed via the quantity~\cite{Ambrus:2021fej}
\begin{align}
    f_{E_\perp}=\frac{\frac{\d E_\perp}{\d^2x_\perp\d\eta}\tau^{1/3}}{\underset{\tau\to\infty}{\lim}\varepsilon\tau^{4/3}}\,.
\end{align}
The normalization by the late-time limit of $\varepsilon\tau^{4/3}$ ensures that this quantity agrees with $\mathcal{E}$ at early times, where vanishing longitudinal pressure implies that all energy of massless particles is in their transverse momentum and $\frac{\d E_\perp}{\d^2x_\perp\d \eta}=\tau\varepsilon$. However, this also means that the late-time limit of $f_{E_\perp}$ will differ from $1$:
\begin{align}
    f_{E_\perp}(\tilde{w}\ll 1) &= C_\infty^{-1}\tilde{w}^{4/9}=\mathcal{E}(\tilde{w}\ll1)\,,\\
    f_{E_\perp}(\tilde{w}\gg 1) &=\frac{\pi}{4}\,.
\end{align}
Ref.~\cite{Ambrus:2021fej} mainly considered 2+1D simulations of heavy ion collisions in kinetic theory in conformal relaxation time approximation, including full transverse dynamics, but also compared these 2+1D results to a transverse integral of local attractor curves. An important step in this integration is to realize that points in the transverse plane that have higher temperatures will also evolve on their local attractor curves faster, as $\tilde{w}$ depends also on temperature. In order to determine the evolution of the local attractor scaling variable $\tilde{w}(\tau,x_\perp)$, it is necessary to also track the evolution of energy density. Thus, both $f_{E_\perp}(\tilde{w})$ and $\mathcal{E}(\tilde{w})$ need to be mapped out in pure Bjorken flow before performing this calculation. The comparison between the prediction from integrated local Bjorken flow and the 2+1D simulation results as shown in Fig.~\ref{fig:transverse_energy} proves that the former is accurate until the onset of transverse expansion at $\tau\sim R$. This is the first explicit verification that describing the system as a transverse collection of local Bjorken flows indeed yields accurate predictions.

\subsubsection{Inhomogeneous cooling}\label{sec:inhom_cooling}

An important consequence of the purely longitudinal dynamics at early times, as already alluded to in the previous subsection, is the fact that hotter regions in the transverse plane will evolve on the local attractor curve faster than colder ones, which will have effects on the transverse geometry of the system. This is relevant in particular for the dynamical flow response of the medium to this transverse geometry. During transverse expansion, anisotropies in the initial geometry will cause an anisotropic expansion of the medium, which is reflected in an anisotropic distribution in the transverse momentum of final state particles. This is measured in experiment via the Fourier coefficients of the azimuthal dependence in this distribution, the so-called flow harmonics $v_n$, which are some of the main observables for probing the dynamical properties of the medium created in a heavy ion collision.

For lower harmonics $n\le 3$, the leading contribution is a linear response to the eccentricities $\epsilon_n$ in the initial state, which are weighted radial means of the Fourier components in the azimuthal angle, i.e. the position space angle in the transverse plane. The proportionality constants $\kappa_{nn}$ in $v_n=\kappa_{nn}\epsilon_n$ are called flow response coefficients and vary significantly depending on the size, energy and to some extent geometry of the system~\cite{Drescher:2007cd,Kurkela:2019kip,Borghini:2022iym,Ambrus:2024hks}. However, within centrality classes, i.e. groups of similar collision events, they vary only little~\cite{Niemi:2015qia}. This is true even when comparing two different but similar collision systems, like Au+Au and U+U, which makes it possible to draw conclusions on the initial geometry by comparing the measured flow harmonics in these systems without considering the effect of dynamical response~\cite{STAR:2024wgy}.

However, what is important for the dynamical buildup of momentum anisotropies during transverse expansion are not the eccentricities in the state of the system right after collision, but at the onset of transverse expansion, which may be altered via the preceding longitudinal dynamics. These longitudinal dynamics follow Bjorken flow, which means that the evolution of energy density can locally be described via the energy attractor curve $\mathcal{E}$ as in Sec.~\ref{sec:mult_est}. Because the eccentricities are computed from the energy density profile, their evolution can also be determined from that of the energy density.

During periods where the entire system follows the same dynamics of the energy density, i.e. $\varepsilon\propto\tau^{-1}$ in free-streaming at early times or $\varepsilon\propto\tau^{-4/3}$ in ideal hydrodynamics at late times, only the global scale changes and eccentricities will be unchanged. However, during the local Bjorken flow evolution hotter regions will transition to faster decay at an earlier point in time and therefore experience a larger total decrease of the energy density, changing the values of the eccentricities. The total change can be easily estimated in local Bjorken flow. In Sec.~\ref{sec:mult_est}, we have already discussed a relation between the early and late-time values of the local energy density on the attractor. There we considered kinetic theory, which fixes the early-time evolution. More generally~\cite{Ambrus:2022koq}, we can consider a system following unspecified dynamics whose early-time energy attractor follows a $\gamma$-power law in the scaled time variable $\tilde{w}$, i.e. $\mathcal{E_\gamma}(\tilde{w}\ll1)=C_{\infty,\gamma}^{-1}\tilde{w}^\gamma$. This exponent $\gamma$ is related to the ratio of longitudinal pressure to energy density in the early-time limit of the given theory as
\begin{align}
    \gamma=\frac{4/3-4(\mathcal{P}_L/\varepsilon)_0}{3-(\mathcal{P}_L/\varepsilon)_0}\,.
\end{align}
Now, the late-time energy density in local Bjorken flow relates to the early time one as $(\varepsilon\tau^{4/3})_{\gamma, \rm late}\propto \left(\varepsilon\tau^{(\frac{4}{3}-\gamma)/({1-\gamma/4})}\right)^{1-\gamma/4}_{\rm early}$, reflecting the early-time power law of the energy density in this theory, $\varepsilon\propto\tau^{(\gamma-\frac{4}{3})/(1-\gamma/4)}$. To be more precise, including all parametric dependencies, the relation is~\cite{Ambrus:2023oyk}
\begin{align}
    (\varepsilon\tau^{4/3})_{\gamma, \rm late}=C_{\infty}(4\pi\eta/s)^\gamma a^{\gamma/4} \left(\varepsilon\tau^{(\frac{4}{3}-\gamma)/{1-\gamma/4}}\right)^{1-\gamma/4}_{\rm early}\,.\label{eq:inhom_cooling}
\end{align}
Since the eccentricities are defined as certain weighted means of the energy density profile, their late-time limit under local Bjorken flow can thus simply be predicted by computing these same means of the profile of $\varepsilon^{1-\gamma/4}$ instead. This means that different dynamical descriptions will yield different changes in the eccentricities. Sec.~\ref{sec:scaled_hydro} deals with a prescription of how to combat these differences.

\begin{figure}
    \centering
    \includegraphics[width=0.5\linewidth]{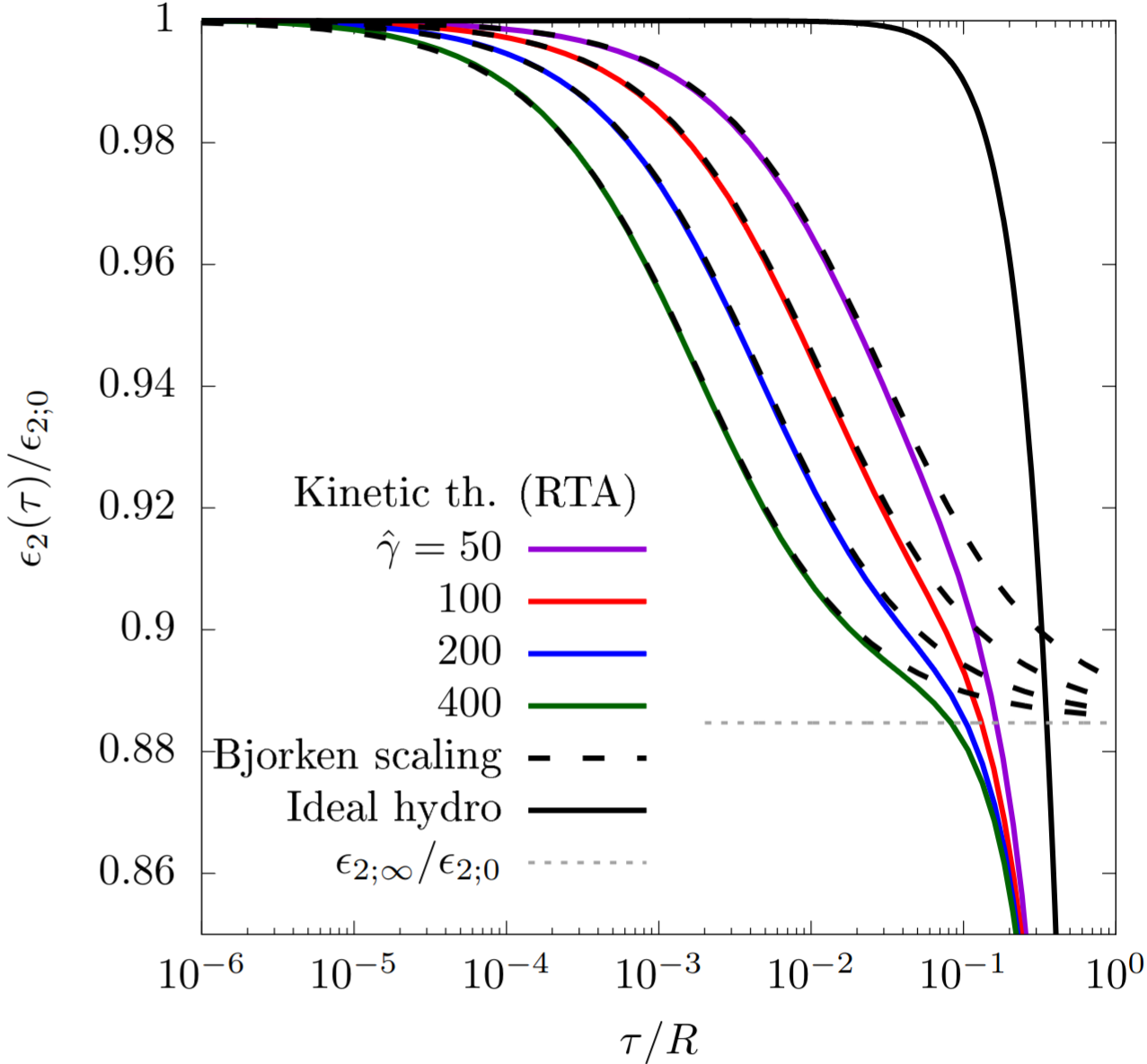}
    \caption{Inhomogeneous cooling causing a decrease of the eccentricity $\epsilon_2$ before the onset of transverse expansion. The change occurs exclusively in the transition from global $\propto\tau^{-1}$ to global $\propto\tau^{-4/3}$-behavior in the energy density, which happens earlier for larger number of interactions (opacity $\hat{\gamma}$). The total amount of decrease in $\epsilon_2$ depends on this only because the onset of transverse expansion interrupts the process at different times. Taken from~\cite{Ambrus:2021fej}.  }
    \label{fig:inhomogeneous_cooling}
\end{figure}

In Ref.~\cite{Ambrus:2021fej}, it was found that flow response coefficients in the large interaction rate limit of kinetic theory starting in the free-streaming regime do not converge to the ideal hydrodynamic response coefficient as they should, if the response is computed relative to the initial eccentricities. Note that in ideal hydrodynamics, the Bjorken flow evolution of energy density follows the same power law at all times and no inhomogeneous cooling occurs. For kinetic theory, predictions of longitudinal cooling from local Bjorken flow are in good agreement with 2+1D simulations before the onset of transverse expansion, as can be seen in Fig.~\ref{fig:inhomogeneous_cooling}. The prediction of the total change of eccentricities in the case of local Bjorken flow allowed to lift the discrepancy of response coefficients in ideal hydrodynamics and kinetic theory at large interaction rate.

\subsubsection{Scaled initial condition of MIS hydrodynamics}\label{sec:scaled_hydro}

Inhomogeneous cooling affects how kinetic theory compares to hydrodynamics not only in the ideal but also in the viscous case. MIS hydrodynamics does feature an attractor that transitions from an early time to a late-time behavior and thus does exhibit inhomogeneous cooling. However, the MIS attractor differs from the kinetic theory one at early times. In particular, the longitudinal pressure in the very early-time limit of the MIS attractor is negative, causing a decrease of the energy density at a rate slower than $\propto\tau^{-1}$. In fact, this problem has two facets: the inhomogeneous cooling rate changes the eccentricities, but there is also a change in the total energy scale due to the different early-time power law. At higher energies, the interaction rates during transverse expansion will be higher and thus it will feature larger flow response coefficients, which interferes with a meaningful comparison of MIS to kinetic theory. The idea of counteracting this by a global rescaling of the initial energy density was put forward in Ref.~\cite{Kurkela:2020wwb}.

In order to perform more detailed comparisons to simulation results in kinetic theory and arrive at a practical answer for the range of applicability of simulations based on MIS hydrodynamics, one would like to match the state of the system as described in MIS as closely as possible to the one in kinetic theory around the onset of transverse expansion. Using the local Bjorken flow model, as already described in Sec.~\ref{sec:inhom_cooling}, it is possible to predict the evolution of the energy density profile in both dynamical descriptions. Thus, we may combine both predictions and locally scale the initial condition of hydrodynamics to account for the difference in inhomogeneous cooling~\cite{Ambrus:2022koq}. 

The total effect of cooling on the local energy density in a theory with an early-time attractor $\mathcal{E}\propto\tilde{w}^\gamma$ is given by Eq.~\eqref{eq:inhom_cooling}. In kinetic theory $\gamma=4/9$ and in ideal hydrodynamics $\gamma=0$, while in MIS with RTA transport coefficients one finds $\gamma=(8\sqrt{5}-4)/19$ in the BRSSS scheme and $\gamma=(\sqrt{505}-13)/18$ in the DNMR scheme. Comparing the case of kinetic theory to that of arbitrary $\gamma$, we can compute the initial energy density $\varepsilon_{0,\gamma}$ that a dynamical description has to be initialized with in order to match $(\varepsilon\tau^{4/3})_{\rm late}$ with the kinetic theory case at the same shear viscosity and initialized at the same time $\tau_0$ as
\begin{align}
    \varepsilon_{0,\gamma}=\left[\left(\frac{4\pi\eta/s}{\tau_0}a^{1/4}\right)^{\frac{1}{2}-\frac{9\gamma}{8}}\left(\frac{C_{\infty,\rm kin.th.}}{C_{\infty,\gamma}}\right)^{9/8}\varepsilon_{0,\rm kin.th.}\right]^{\frac{8/9}{1-\gamma/4}}.\label{eq:scaled_hydro}
\end{align}
The values for the attractor normalization constants in hydrodynamics are $C_{\infty}\approx0.80$ for MIS with RTA transport coefficients, $C_{\infty,\rm kin.th}\approx0.88$ for RTA kinetic theory  and $C_{\infty}=1$ for ideal hydro. Indeed, this same formula applies also for ideal hydrodynamics. Here, $\eta/s$ describes only the kinetic theory, but this scaling factor makes the coefficient of the $\propto \tau^{-4/3}$ power law in ideal hydrodynamics match with the corresponding late-time behavior in kinetic theory. Note that technically the global scale of the initial energy density does not matter in conformal ideal hydrodynamics, because flow response coefficients will be unchanged. The scale does somewhat matter for a meaningful definition of dimensionful observables, like the large interaction rate limit of the final state transverse energy, which will scale $\propto(\eta/s)^{4/9}$. However, this factor can be applied in hindsight to the ideal hydrodynamic result when initialized according to Eq.~\eqref{eq:scaled_hydro} at any reference value of $\eta/s$.

This scheme of adjusting the initial condition of hydrodynamics was employed under the name ``scaled hydrodynamics'' in several works comparing the 2+1D evolution in simulations of heavy ion collisions based on hydrodynamics and kinetic theory~\cite{Ambrus:2022koq,Ambrus:2022qya,Ambrus:2024hks,Ambrus:2024eqa}.

\subsubsection{Pre-flow from local Bjorken flow and conservation laws}\label{sec:pre-flow}

Since the conservation equations of energy and momentum, $\nabla_\mu T^{\mu\nu}=0$, already constrain how energy and momentum must propagate, it is possible to derive expressions that estimate the buildup of flow velocities at early times of a heavy ion collision when assuming a diagonal energy-momentum tensor at initial time and linearizing in the time evolution. When this was first discussed, the pressure anisotropy was assumed to be constant in order to avoid having to compute its evolution~\cite{Vredevoogd:2008id}. Under these assumptions, the early-time buildup of transverse flow velocities ($i=x,y$) can be computed as
\begin{align}
    u^i=-\frac{\tau}{3-\mathcal{P}_L/\varepsilon}\frac{\partial_i\varepsilon_0}{\varepsilon_0}\,.\label{eq:preflow_simple}
\end{align}
This idea can be combined with the knowledge of how the system evolves on the hydrodynamic attractor~\cite{Ambrus:2022koq}, allowing to lift the assumption of a constant pressure anisotropy. Both the evolution of energy density and that of the pressure anisotropy enter at multiple points, such that the formula becomes rather involved. Defining $f_\pi(\tilde{w})=P_L/\varepsilon-1/3$, it states:
\begin{align}
    u^i=-\frac{\tau^{1/3}}{3-(\mathcal{P}_L/\varepsilon)_0}\frac{\partial_i\varepsilon_0}{\varepsilon_0}\frac{2}{\mathcal{E}(1-\frac{3}{8}f_\pi)}\int_{\tau_0}^{\tau}\d\bar{\tau}\frac{\frac{1}{3}-\frac{1}{2}f_\pi-\frac{\bar{\tilde{w}}f_\pi'}{8}}{1-\frac{\bar{\tilde{w}}}{4\mathcal{E}}\mathcal{E}'}\bar{\tau}^{-1/3}\mathcal{E}\label{eq:preflow_complicated}
\end{align}
Note that here the attractor functions depend on time via the scaling variable $\tilde{w}$ (or $\bar{\tilde{w}}\propto\bar{\tau}T(\bar{\tau})$ under the integral) and the prime denotes differentiation with respect to the scaling variable. It also depends on temperature, bringing additional dependencies on time and the transverse position. However, Eq.~\eqref{eq:preflow_complicated} is local in transverse space. For $\tau_0\to0$, and in both the early and very late time (meaning the integrand is dominated by the late-time case) limits of $f_\pi$ and $\mathcal{E}$,  Eq.~\eqref{eq:preflow_complicated} reduces to the simpler result in Eq.~\eqref{eq:preflow_simple}.

\subsection{Perturbations on top of Bjorken flow}

The models outlined in the following describe perturbations on top of some form of local or global Bjorken flow. This Bjorken flow background is described by the attractor, and the perturbations on top, which constitute deviations from transverse homogeneity, are propagated in some approximate model, assuming that they do not influence the evolution of this background itself.

\subsubsection{K\o MP\o ST: Perturbation of local average Bjorken flow}

A big problem in simulation setups for heavy ion collisions is that hydrodynamic models may not be applicable to very early times, while models that are applicable require a prohibitive amount of computational power to employ in 2+1D or even 3+1D. The idea of the kinetic theory based pre-equilibrium evolution code K\o MP\o ST~\cite{Kurkela:2018wud,Kurkela:2018vqr,Kurkela:2018gitrep} is to expand the regime where one can profit from the simplicity of Bjorken flow by adding transverse perturbations. Of course, since transverse inhomogeneity is locally only propagated to linear order, this model still has a limited range of applicability in time.

Initially, K\o MP\o ST implemented the dynamics of Yang--Mills effective kinetic theory, but the concept is not restricted to this model and could be extended to any dynamical description of heavy ion collisions. In fact, the accompanying GitHub repository also includes a version of the code that can simulate the dynamics of RTA kinetic theory. This version has been tested against 2+1D simulations and was found to yield accurate predictions on the timescales it is supposed to apply to for all components of the energy-momentum tensor except the ones related to transverse anisotropy, $T^{xx}-T^{yy}$ and $T^{xy}$~\cite{Ambrus:2022koq}. K\o MP\o ST has also been extended to a version called ShinyK\o MP\o ST, which uses universal scaling functions for the emission spectra to predict event by event the production of photons~\cite{Garcia-Montero:2023lrd} and dileptons~\cite{Garcia-Montero:2024lbl} during the pre-equilibrium stage. Furthermore, by including longitudinal perturbations to averaged Bjorken flow, K\o MP\o ST can be extended to 3+1D~\cite{Du:2025hyk}.

Starting from an initial energy density profile given at a proper time $\tau_{\rm EKT}\sim0.2\,\mathrm{fm}/c$, chosen according to estimates of when kinetic theory should start to be applicable, K\o MP\o ST computes in one timestep the profiles of the components of the energy-momentum tensor at time $\tau_{\rm hydro}$, where a hydrodynamic model can be employed for the subsequent evolution. For each point $x_0$ at $\tau=\tau_{\rm hydro}\sim1\,\mathrm{fm}/c$, the energy-momentum tensor is split into an average Bjorken flow background at the given point and a perturbation:
\begin{align}
    T^{\mu\nu}(\tau,x)=\bar{T}^{\mu\nu}_{x_0}(\tau)+\delta T_{x_0}^{\mu\nu}(\tau,x)\,.\label{eq:kompost_splitting}
\end{align}
At each $x_0$, the background $\bar{T}^{\mu\nu}_{x_0}(\tau_{\rm EKT})$ is computed by folding the profiles at $\tau=\tau_{\rm EKT}$ with a Gaussian of a width proportional to $\Delta\tau=\tau_{\rm hydro}-\tau_{\rm EKT}$. This background is the part that is assumed to evolve according to the hydrodynamic attractor. Note that this means that the energy-momentum tensor at the same point $x$ at $\tau_{\rm EKT}$ will be split in different ways when viewed from different $x_0$ at $\tau_{\rm hydro}$, so the splitting procedure has to be repeated for each point. This idea is similar but due to the averaging step distinct from the concept of local Bjorken flow that was described in Sec.~\ref{sec:local_bjorken_flow}. Technically, it is only the energy density that has to be averaged at $\tau=\tau_{\rm EKT}$. Due to the assumption of a conformal system on its attractor, the background pressures can readily be determined from the background energy density at any time. Another slight difference is the fact that in K\o MP\o ST attractors are mapped as a function not of $\tilde{w}=\frac{\tau T}{4\pi\eta/s}$ but of the scaled time variable $\frac{\tau T_{\rm id}}{\eta/s}$. This now depends on the instantaneous ideal temperature $T_{\rm id}$, determined as a backward continuation of the late-time $\tau^{-1/3}$-trend. The energy attractor provides a one-to-one correspondence between this ideal and the actual (effective) temperature of the system, meaning that the two parametrizations are equivalent. However, this makes it necessary to determine the initial ideal temperature scale via an iterative solution to a self-consistency equation.

The perturbations are normalized to the background and propagated to linear order using Green's functions as follows.
\begin{align}
    \frac{\delta T^{\mu\nu}(\tau_{\rm hydro},x_0)}{\bar{T}^{\tau\tau}_{x_0}(\tau_{\rm hydro})}=\int\d^2x~G^{\mu\nu}_{\alpha\beta}(|x-x_0|,\tau,\tau_0)\frac{\delta T^{\alpha\beta}_{x_0}(\tau_{\rm EKT},x)}{\bar{T}^{\tau\tau}_{x_0}(\tau_{\rm EKT)}}\,,\label{eq:kompost_perturbation}
\end{align}
where the Green's functions are obtained from numerical solutions of linearized kinetic theory in a Bjorken flow background and tabulated. Due to transverse homogeneity and isotropy of this background, individual components depend on $x$ and $x_0$ only via the magnitude of their difference. The direction of $x-x_0$ is reflected in the tensorial structure. Causality implies that the Green's functions are nonzero only for $|x-x_0|<\tau_{\rm hydro}-\tau_{\rm EKT}$. In order to keep the number of necessary evaluation points for the Green's function tabulation low, they are computed in terms of scaled time $\frac{\tau T_{\rm id}}{\eta/s}$ and in Fourier space for the transverse coordinate. In terms of these coordinates, the regime of nontrivial dynamics always falls in the same coordinate range. Furthermore, the large $|k|$ tail is regulated by multiplying with a Gaussian. At each point $x$ in the transverse plane, the code will transform the relevant Green's function for the corresponding scaled time to position space and compute the perturbation $\delta T^{\mu\nu}$ according to Eq.~\eqref{eq:kompost_perturbation}. The total energy-momentum tensor is of course reconstructed as the sum of the background and this perturbation.

\subsubsection{Perturbation of global Bjorken flow}

Another idea for making use of the predictive power of the attractor in Bjorken flow in an approximate description of the dynamics of a 2+1D system is to describe perturbations to a global Bjorken flow. Specifically, instead of solving the full nonlinear MIS equations, this Ansatz describes the evolution according to the linearization of these equations around the Bjorken flow background~\cite{An:2023yfq}, meaning that the energy-momentum tensor is split in a similar way to Eq.~\eqref{eq:kompost_splitting}, but with one global background $\bar{T}^{\mu\nu}(\tau)$ instead of an $x_0$-dependent one. The linearized equations are solved in Fourier space, where they turn into ordinary differential equations.

Part of the reason for this endeavor was to determine where the information on transverse details in a heavy ion collision such as flow anisotropies comes from, given that the Bjorken flow attractor suppresses memory of the initial state. It was found that at asymptotically late times, the solution for the spatially inhomogeneous perturbations can be written as a transseries with the background attractor being stable against these perturbations, and that all the transverse plane information is encoded in the coefficients of this transseries expansion.

This model has also been used to make a range of predictions for common observables in heavy ion collisions, including transverse momentum spectra $\frac{\d N}{p_\perp\d p_\perp \d y}$ and elliptic flow $v_2$, which showed the same qualitative behavior as experimental data. Though the accuracy of the linearization scheme, which may be a rough approximation especially in the outskirts of the system, has not been tested so far.

\subsubsection{Jet quenching in local Bjorken flow}

The evolution of an initially highly energetic parton passing through the strongly interacting medium created in a heavy ion collision can give insights into this medium's properties. Over time, the parton will split into less energetic partons. On the theory side, the collection of all partons originating from one initial hard parton is called a jet. In experiment, jets have to be reconstructed as localized regions of high total energy of measured particles. The typical quantities describing the interaction of a jet with its medium are the energy loss to the medium and the broadening of the transverse momentum distribution of the partons in the jet. Both are related to the jet quenching parameter $\hat{q}$, which is defined as the change in the mean squared transverse momentum per path length and quantifies the strength of the interaction with the medium.

Jet quenching has been described in out-of-equilibrium theories, specifically in the Glasma~\cite{Ipp:2020nfu,Avramescu:2023qvv}, i.e. the very early stage described by classical-statistical dynamics of Yang--Mills fields, and in kinetic theory~\cite{Boguslavski:2024ezg,Boguslavski:2024jwr}, and the effects in these stages were found to be sizable. However, event-by-event descriptions typically cannot make use of these more expensive descriptions and therefore use formulae that were derived for a homogeneous isotropic medium in equilibrium and are then extrapolated beyond this strict limit. Various models exist, which employ varying levels of detail in the description of the underlying medium.

Ref.~\cite{Pablos:2025cli} describes jet quenching by computing jet evolution on top of an event-by-event background medium that was evolved in hydrodynamics. In an effort to describe at least part of the effects of the early-time stage, these medium evolutions were locally extended to early times using the hydrodynamic attractor curve for the energy density, see Sec.~\ref{sec:mult_est}. Assuming local Bjorken flow, knowledge of $\mathcal{E}(\tilde{w})$ allows one to extend the evolution of temperature to a fixed onset time of jet interactions that is earlier than the starting time of the hydrodynamic evolution. What is left to determine is the local multiplicative constant, i.e. the limiting late-time energy $(\varepsilon\tau^{4/3})_{\rm late}$. In this work, it was expressed via the temperature, and was estimated from the local temperature $T$ at the start of the hydrodynamic evolution by using the Navier--Stokes correction as
\begin{align}
    (\tau^{1/3}T)_{\rm late}=\tau^{1/3}\left(T+\frac{2}{3}\frac{\eta/s}{\tau_{\rm hyd}}\right)\,.
\end{align}
Similarly, the evolution of flow velocities was extended by using pre-flow formulae~\cite{Vredevoogd:2008id} in the vein of what was discussed in Sec.~\ref{sec:pre-flow}.

This model description of jet quenching was able to achieve a simultaneous description of the experimentally measured jet yield ratio to the proton+proton case, $R_{AA}=N_{AA}/(\langle N_{\rm coll}\rangle N_{pp})$, and jet elliptic flow $v_2$ when starting the jet evolution at a time $\tau=0.2\,$fm/c. Previously, building on results from models that started both jet quenching and the hydrodynamic evolution at the same fixed time $\tau_0$~\cite{Andres:2019eus,Stojku:2020wkh}, it was thought that these observables could only be reproduced simultaneously if this time was chosen to be quite late, $\tau\sim0.6-1\,$fm/c.

\subsection{Modifying hydrodynamic models}

Insights from the attractor behavior may be used to construct improved phenomenological hydrodynamic models by adjusting transport coefficients. A word of warning is in order that indeed such models are purely phenomenological as opposed to constructed consistently as effective field theories and their transport coefficients may not necessarily reflect the actual transport properties of the described medium.

It is also possible to use comparisons of attractors in advanced hydrodynamic theories to those in more microscopic models to argue for their improved accuracy over traditional hydrodynamic models.

\subsubsection{Gradient-dependent shear viscosity}

One may reinterpret the attractor in theories with transient behavior as a time- or gradient-strength-dependent shear viscosity from the point of view of non-transient hydrodynamics, which has been dubbed ``Borel-resummed'' viscosity $\eta_{\rm B}$~\cite{Romatschke:2017vte,Romatschke:2017ejr}. This amounts to a phenomenological definition of $\eta_{\rm B}$ as the ratio of instantaneous shear stress $\pi^{\mu\nu}$ to instantaneous shear strain $\sigma^{\mu\nu}$. As this is obtained by matching the attractor solution to the general evolution equation obtained for a purely hydrodynamic theory that allows additional dependencies in the shear viscosity, it really just amounts to ``plugging the attractor into hydro''.

\begin{figure}
    \centering
    \includegraphics[width=0.57\linewidth]{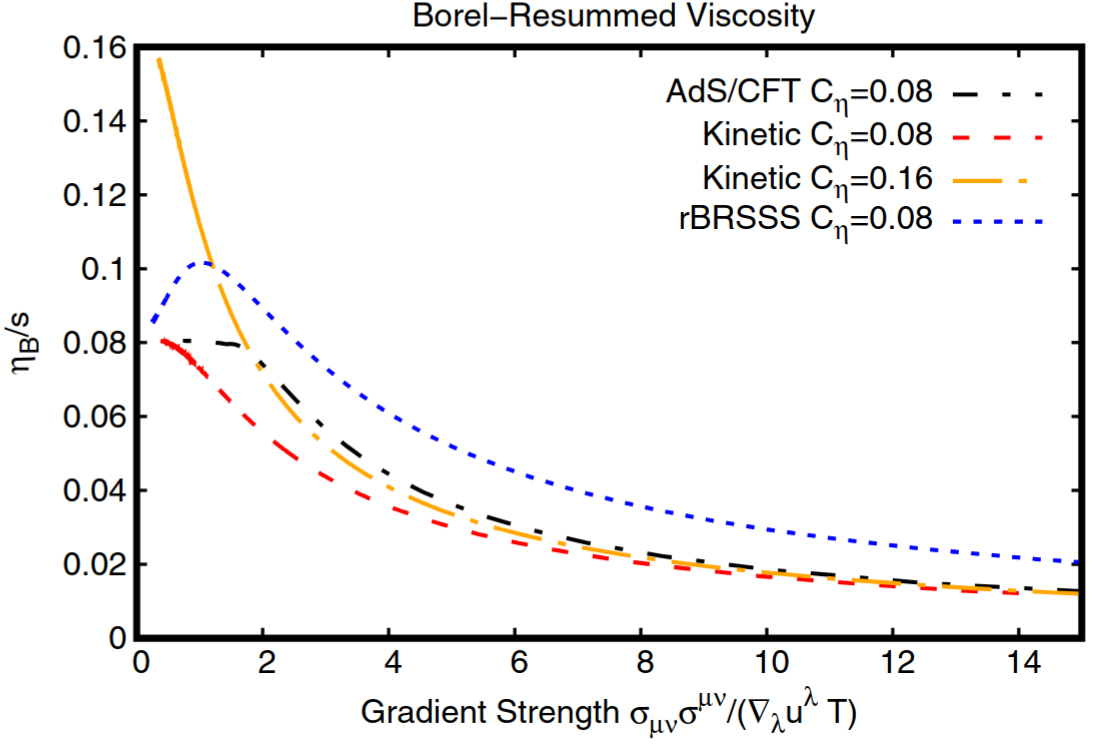}
    \includegraphics[width=0.42\linewidth]{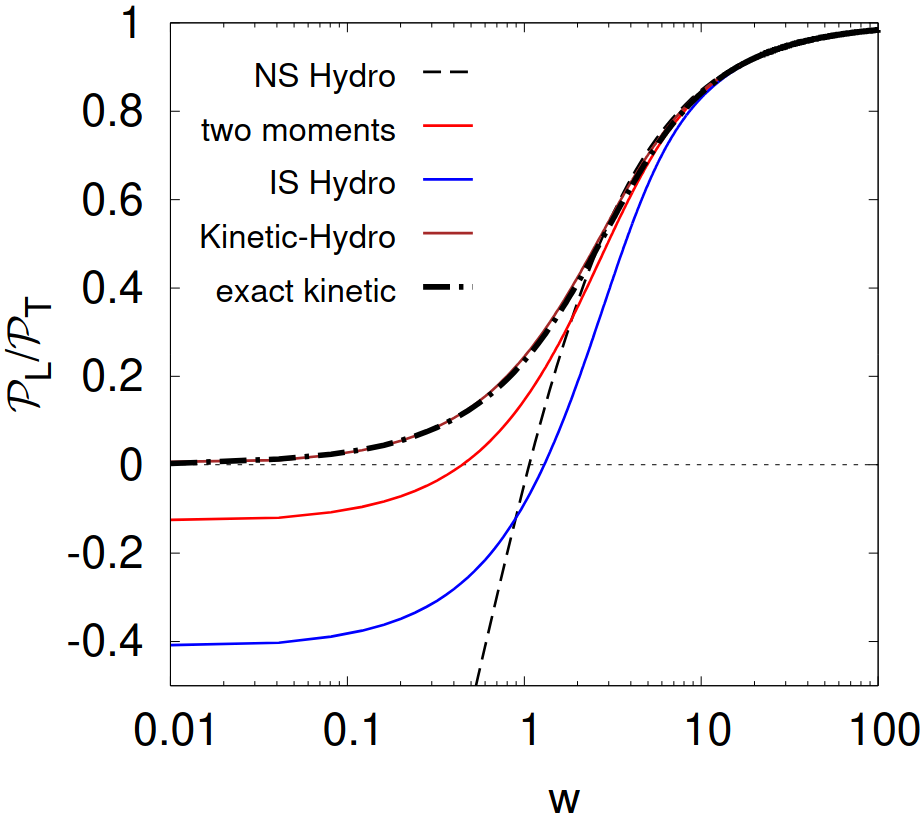}
    \caption{Left: Gradient dependent shear viscosity after reinterpreting the pressure anisotropy attractor of different dynamical models. Taken from~\cite{Romatschke:2017vte}. Right: Attractor curves for the pressure anisotropy in different dynamical descriptions and late-time Navier--Stokes limit. Attractor curves from hydrodynamics disagree with those from kinetic theory. Changing one transport coefficient yields agreement. Taken from~\cite{Blaizot:2021cdv}}
    \label{fig:changed_coefficients}
\end{figure}

The left plot of Fig.~\ref{fig:changed_coefficients} shows the Borel-resummed viscosities that one obtains in different dynamical theories. As expected, for small gradient strength or late times, the constant shear viscosity for which the attractor was obtained in the transient theory is recovered. However, away from this limit, the Borel-resummed viscosity will deviate, typically downwards. In the limit of large gradient strength, the Borel-resummed viscosity goes to zero.

The hydrodynamic model defined in this way is tuned to describe the attractor of a given transient theory and only this attractor. By construction, it assumes that the system is on this attractor from the start and does not describe the decay onto this attractor. Since the ``Borel-resummed'' viscosity is defined to reflect the longitudinal dynamics, it stands to reason that the propagation of transverse inhomogeneities will be described inaccurately.

\subsubsection{Modified second-order transport coefficients}

In an effort to describe the dynamics and the attractor of a reduced set of moments of the phase space distribution in RTA kinetic theory~\cite{Blaizot:2019scw,Blaizot:2020gql,Blaizot:2021cdv}, it was discovered that the attractor of MIS hydrodynamics can be made to closely agree with that of kinetic theory by simply adjusting the value of a second-order transport coefficient which dictates the early-time behavior~\cite{Blaizot:2021cdv}.

The transport coefficient $a_1$ in question is the one appearing in front of the following term for the evolution equation of shear stress $\pi=\pi^\eta_\eta$ in Bjorken flow:
\begin{align}
    \partial_\tau\pi=-\frac{a_1}{\tau}\pi+\dots\,.
\end{align}
In terms of a more general equation for $\pi^{\mu\nu}$, it may actually be the sum of two terms. For example, in DNMR hydrodynamics~\cite{Denicol:2012cn}, the two terms
\begin{align}
    \tau_R\dot{\pi}^{\mu\nu}=-\delta_{\pi\pi}\theta\pi^{\mu\nu}-\tau_{\pi\pi}\pi^{\langle\mu\lambda}\sigma^{\nu\rangle}_\lambda+\dots
\end{align}
will take the same form $\propto\pi/\tau$ in Bjorken flow and thus $a_1=\delta_{\pi\pi}/\tau_R+\tau_{\pi\pi}/(3\tau_R)=4/3+10/21=38/21$~\cite{Denicol:2017lxn}. Actually, the set of two coupled moments considered in Ref.~\cite{Blaizot:2021cdv} is by construction equivalent to DNMR hydrodynamics and thus also the attractor curves are identical, see the red curve in the right plot of Fig.~\ref{fig:changed_coefficients}.

The motivation for changing the value of $a_1$ comes from the limiting value of the attractor as $\tilde{w}\to0$. Assuming the proportionality constant in $\eta\propto \varepsilon\tau_R$ is fixed, this value only depends on $a_1$. Now in order to achieve zero longitudinal pressure in the early-time limit, one may choose $a_1=31/15$, which is how the brown curve in the right plot of Fig.~\ref{fig:changed_coefficients} was obtained. Evidently, by making this adjustment, we have brought the curve into good agreement with full kinetic theory, which is the black dashed curve. In full kinetic theory, the value of $a_1$ is indeed $38/21$, but there are additional terms in the evolution equation coming from the coupling of $\pi$ to higher order moments of the phase space distribution, which themselves take on constant ratios to the energy density in the early-time limit of the attractor~\cite{Aniceto:2024pyc}. At early times, the effect of these terms is equivalent to changing the value of $a_1$ to $31/15$.

This tells us that changing $a_1$ is actually a purely phenomenological model adjustment. It reproduces the effect of the coupling to higher order moments only in the early-time limit. The fact that the attractor agrees well also away from this limit is somewhat coincidental, partly owing to the fact that the $\frac{a_1}{\tau}\pi$-term becomes less important over time, and so do the couplings to higher order moments. This version of hydrodynamics no longer has the interpretation of a consistently derived effective field theory for the long wavelength sector close to equilibrium, because $a_1$ was adjusted to match far-from-equilibrium behavior in a specific flow. Its accuracy for other flows, for perturbations and for other observables must be assessed independently. In contrast, the canonical value $a_1=38/21$ was determined from perturbations to equilibrium~\cite{Denicol:2012cn}.

\subsubsection{Anisotropic hydrodynamics}\label{sec:ahydro}

Anisotropic hydrodynamics was proposed already before the discovery of the hydrodynamic attractor to more accurately describe early times of heavy ion collisions, which feature strong pressure anisotropies~\cite{Florkowski:2010cf,Martinez:2010sc,Martinez:2010sd}. The underlying idea is to construct hydrodynamics not as an expansion around equilibrium, but around an anisotropic reference state. For example, in terms of a (quasi-)particle distribution in momentum space, this reference state may be defined as
\begin{align}
    f(x,p,\tau)=f_{\rm eq}\left(\sqrt{p_\perp^2+(1+\xi)p_z^2}\Big/\Lambda_{\rm aniso}(x)\right)+\delta f(x,p,\tau)\,
\end{align}
and will introduce an additional dynamical parameter, here $\xi$. The momentum scale $\Lambda_{\rm aniso}$ of the deformed distribution is distinct from the Landau-matched temperature $T_{\rm eff}$. For the massless spheroidal ansatz 
\begin{align}
T_{\rm eff}=\Lambda_{\rm aniso}\left[ \frac{1}{2}\left(\frac{1}{1+\xi}+\frac{\mathrm{arctan\sqrt{\xi}}}{\sqrt\xi}\right) \right]^{1/4}\,,
\end{align}
 and the two coincide only for $\xi\to0$. The hydrodynamic equations derived from this will exhibit differing transverse and longitudinal pressures already on the ideal level and therefore need an additional time evolution equation for this extra degree of freedom.

Hydrodynamic attractors do come into play when comparing the behavior in anisotropic or traditional MIS hydrodynamics to kinetic theory. The attractor of traditional MIS will agree only at late times, but disagree at early times, in particular featuring negative longitudinal pressure. The attractor for anisotropic hydrodynamics can reproduce zero longitudinal pressure as one would find in kinetic theory~\cite{Strickland:2017kux}. Also at intermediate times, the attractor curve for the pressure anisotropy remains close to that of the appropriate kinetic theory~\cite{Strickland:2018ayk}, even when going beyond strict boost-invariance~\cite{Chen:2024grb}. Reconstructions of other moments of the phase space distribution function may strongly deviate~\cite{Strickland:2018ayk}, but this should be expected from an effective theory that aims only at describing the dynamics of the energy-momentum tensor. In fact, it has been found that viscous anisotropic hydrodynamics yields predictions for final state observables that remain accurate on a significantly wider range in interaction rates than traditional viscous hydrodynamics~\cite{Peng:2025gbj}.

\subsubsection{Attractodynamics}

Ref.~\cite{DuPlessis:2026qjy} proposes a way to exploit the universal dynamics around an attractor through a specific extension of hydrodynamics similar to the framework of anisotropic hydrodynamics, which the authors call `Attractodynamics'. It is constructed similarly, in a moment expansion in kinetic theory, but instead of an anisotropically deformed equilibrium distribution, this ansatz uses as a reference state the locally attractive distribution $f_{\mathcal{A}}$ with its logarithmic momentum dependence expanded in powers:
\begin{align}
    f(x,p,\tau)&=f_\mathcal{A}(x,p,\tau)+\delta f(x,p,\tau)\,,\\
    \ln f_\mathcal{A}(x,p,\tau)&\approx\alpha^{\mathcal{A}}+\beta^{\mathcal{A}}_\mu p^\mu+\frac{1}{2}p^\mu\mathcal{U}^{\mathcal{A}}_{\mu\nu}p^\nu+\dots\,.
\end{align}
$f_\mathcal{A}$ in general depends on the state of the system, as not all directions in phase space will contract. Thus, it constitutes an attractor surface in phase space, which is mapped by underlying attractor coordinates following their own evolution equations. The coefficients $\alpha^{\mathcal{A}}$, $\beta_\mu^{\mathcal{A}}$, $\mathcal{U}_{\mu\nu}^{\mathcal{A}}$ are also functions of these coordinates. The limit of conventional hydrodynamics for a classical fluid is restored when $\alpha^\mathcal{A}\to\mu/T$, $\beta_\mu^\mathcal{A}\to -u_\mu/T$ and $\mathcal{U_{\mu\nu}^\mathcal{A}}\to0$ and the same for higher orders. In general, though, these quantities are prescribed by the form of the attractive subspace in momentum space.

It is to be expected that a state in this subspace requires more parameters to be fully described than equilibrium, at minimum the second-order parameters $\mathcal{U}_{\mu\nu}^\mathcal{A}$, to describe for example a momentum anisotropy. $\mathcal{U}_{\mu\nu}^{\mathcal{A}}$ is symmetric, and its trace can through the on-shell condition $p^2=m^2$ be absorbed into $\alpha^{\mathcal{A}}$. Thus, at this order, in general ideal attractodynamics has nine additional degrees of freedom more than ideal hydrodynamics, for a total of 14. However, the independent attractor variables are those that parametrize the particular attracting family, and their number depends on the model and symmetry sector: the coefficients $\alpha^{\mathcal A}, \beta^{\mathcal A}_\mu, \mathcal U^{\mathcal A}_{\mu\nu}$ are functions of the attractor coordinates, not necessarily independent coordinates themselves. Their evolution is given by the dynamics tangential to the attractor surface.

The dynamics orthogonal to the attractor surface will decay towards it and can be viewed as transient viscous corrections. Projection is facilitated via appropriate matching conditions $\langle\cdot\rangle_{\delta f}=0$. As the ideal case is already comprised of 14 degrees of freedom, resulting in 14 such matching conditions, the usual 14-moment approximation 
\begin{align}
    N^\mu=\langle p^\mu\rangle_f\,, \qquad T^{\mu\nu}=\langle p^\mu p^\nu\rangle_f
\end{align}
with $\langle \mathcal{O} \rangle_f=\int\frac{d^3p}{(2\pi)^3p^0}\mathcal{O}f$ is insufficient to introduce viscous degrees of freedom. One has to introduce further moments, for example
\begin{align}
    I^{\mu\nu\lambda}=\langle p^\mu p^\nu p^\lambda\rangle_f\,.
\end{align}
The viscous corrections are accordingly fixed by the following ansatz for the deviation $\delta f$ from the attractor surface:
\begin{align}
    \delta f=f_\mathcal{A}\left(\alpha_\mu\frac{p^\mu}{p\cdot u}+\beta_{\mu\nu}\frac{p^\mu p^\nu}{p\cdot u}+w_{\mu\nu\lambda}\frac{p^\mu p^\nu p^\lambda}{p\cdot u}\right)\,.
\end{align}
Similar to MIS theory, they are modeled to follow their own transient relaxation equtions. Ref.~\cite{DuPlessis:2026qjy} goes on with a proof of principle of this framework in an example case where even the pre-thermal part of the attractor is exactly solvable. They find that while ideal attractodynamics shows deviations, viscous attractodynamics can quite precisely reproduce the exact solution. 

\section{Attractors in ultracold atomic gases}
\label{sec:cold_atoms}

\subsection{Motivation of studying attractors in cold atoms}
While it constitutes an important aspect of its dynamics, an experimental verification of attractor behavior in heavy ion collisions is obfuscated by the fact that it occurs early on in the evolution of the fireball, followed by an extended transversely expanding stage and particlization. Instead, one may replicate the behavior in a highly controlled laboratory quantum system that allows to mimic the dynamics of the QGP. Ultracold atomic gases\cite{bloch2008} provide exactly this platform. These systems allow to observe and discriminate
attractor behavior experimentally in real time, as the system can be prepared in a well-defined state and a snapshot of the momentum distribution can be taken at any time during its evolution. Furthermore, cold atomic gases have a simple, well-known Hamiltonian yet
are almost perfect fluids. 

One typical setup for transport
experiments uses a cloud of $N\sim10^6$ neutral $^6$Li atoms, which
follow fermionic statistics, confined in a harmonic or box trap
potential \cite{patel2020}.  The range $r_0\sim 1$nm of the
interatomic van-der-Waals interaction is much shorter than the particle spacing
$\ell\sim1\mu$m, realizing a very dilute gas.  At the same time, the
scattering length $a$ can be made arbitrarily large by tuning the
magnetic field near a Feshbach resonance \cite{chin2010}---a scattering resonance arising when the energy of two colliding atoms is tuned into degeneracy with a molecular bound state---allowing the interaction strength to be tuned in real time.  A two-component Fermi
gas---made of two hyperfine atomic states---with short-range
interactions tuned to a scattering resonance is called the
\emph{unitary Fermi gas}.  It is scale invariant because the particle
spacing $\ell$ is the only remaining length scale for $r_0\to0$,
$a\to\pm\infty$, and it has universal properties similar to those of
dilute neutron matter in neutron stars.  The unitary
Fermi gas exhibits a phase transition between a high-temperature
normal state and a low-temperature superfluid of fermion pairs.

Despite its diluteness, similarly to the QGP the unitary Fermi gas has very remarkable
transport properties of an almost perfect fluid: it has the lowest
shear viscosity to entropy ratio measured in a nonrelativistic fluid
\cite{cao2011, schaefer2009} and quantum limited sound diffusion
\cite{patel2020}.  Its slow, long-wavelength dynamics are very well described by collisional, Navier--Stokes hydrodynamics in the
normal state and superfluid hydrodynamics at low temperature.
Theoretical studies of the unitary Fermi gas have predicted the
equation of state and transport coefficients from the microscopic
Hamiltonian, in agreement with experiment \cite{vanhoucke2012, bluhm2017, frank2020}.  Thus, one can explicitly compute collective properties from first principles. This can help to understand the onset of hydrodynamics and what comes
before, mapping out the boundary of hydrodynamics
\cite{berges2026}.  The results will be representative for a
large universality class of strongly-correlated fermion systems with
short-range interaction, covering not just ultracold atoms but also
dilute neutron matter in neutron stars.  They can provide a guide to
find a universal mechanism of hydrodynamization common to very
different physical systems.

Despite the many similarities to the QGP fireball in heavy ion collisions, there are also obvious sharp distinctions. The biggest one is that the systems live on opposite ends on the spectrum of energy scales: while heavy ion collisions are ultrarelativistic and hot, atomic gases are deeply nonrelativistic and cold. Due to the mechanism of its creation, the QGP fireball is initially highly anisotropic, which is not easily replicated in cold atoms. Consequently, a more direct way to produce a controlled far-from-equilibrium state is to introduce a bulk strain to a homogeneous and isotropic system, in contrast to the boost invariant and transversely homogeneous heavy ion case, where attractor behavior is discussed in the shear sector. Thus, cold atoms not only provide a model system that allows to directly probe attractor behavior, but also demonstrate that the phenomenon is much more general than it may appear in the discussion of conformal Bjorken flow.

\subsection{Non-hydrodynamic behavior in cold atoms}

Early studies have sought to explain the emergence of hydrodynamics in
ultracold atom experiments by analyzing the decay of non-hydrodynamic
modes \cite{brewer2015}.  Specifically, when the quadrupole
mode of a trapped gas is excited, it performs damped oscillations with
the frequency determined by the equation of state and the damping by
shear viscosity; this is well described by hydrodynamics. At early
times immediately after the excitation, instead, a transient
non-hydrodynamic mode that decays without oscillation is predicted, but
could not be unambiguously confirmed with the experimental resolution at the time \cite{brewer2015}.

More recently, new ways to probe hydrodynamic attractors in ultracold
atomic gases have been proposed: by varying the scattering length
$a(t)$ over time (via the magnetic field, see above), the system is
brought out of local equilibrium and will relax by local dissipation
\cite{fujii2018}.  This relaxation will be visible in thermodynamic
and spectroscopic observables, most clearly in the \emph{contact
  density} $\mathcal C(t)$, which quantifies short-range pair
correlations and can be accurately measured \cite{tan2008large, mukherjee2019,
  jager2024, xie2026}. Its equilibrium contribution $C_{\rm eq}(t)$ depends on time through $\varepsilon(t)$ and $a(t)$, though for weak drives the dependence on $\varepsilon(t)$ can be neglected. A time lag between the change in $a(t)$ and
$\mathcal C(t)$ quantifies local dissipation, which is governed by the
bulk viscosity $\zeta$. Crucially, this protocol probes the
equivalent of relaxation to hydrodynamics in isotropic scaling flows in a
completely uniform gas with no moving parts and no spatial gradients. For slow drives that create a bulk strain rate $V_a(t) = -3\dot a(t)/a(t)$ much smaller than the bulk relaxation rate, $|V_a(t)|\ll\tau_\Pi^{-1}$, the response is well described by standard Navier--Stokes
hydrodynamics and is given as $\Pi_{\rm NS}(t)=-\zeta V_a(t)$.
Theoretically, the bulk viscosity has been computed by virial
expansion and Luttinger-Ward techniques \cite{nishida2019, enss2019bulk,
  hofmann2020}. Faster drives probe also
non-hydrodynamic evolution and the onset of hydrodynamics. In the scale invariant unitary Fermi gas, the bulk
relaxation time $\tau_\Pi\sim1/T$ is found to scale inversely with
temperature with typical values of $0.1$ms \cite{enss2019bulk, fujii2024,
enss2025quantum}.

\begin{figure}
    \centering
    \includegraphics[width=0.6\linewidth]{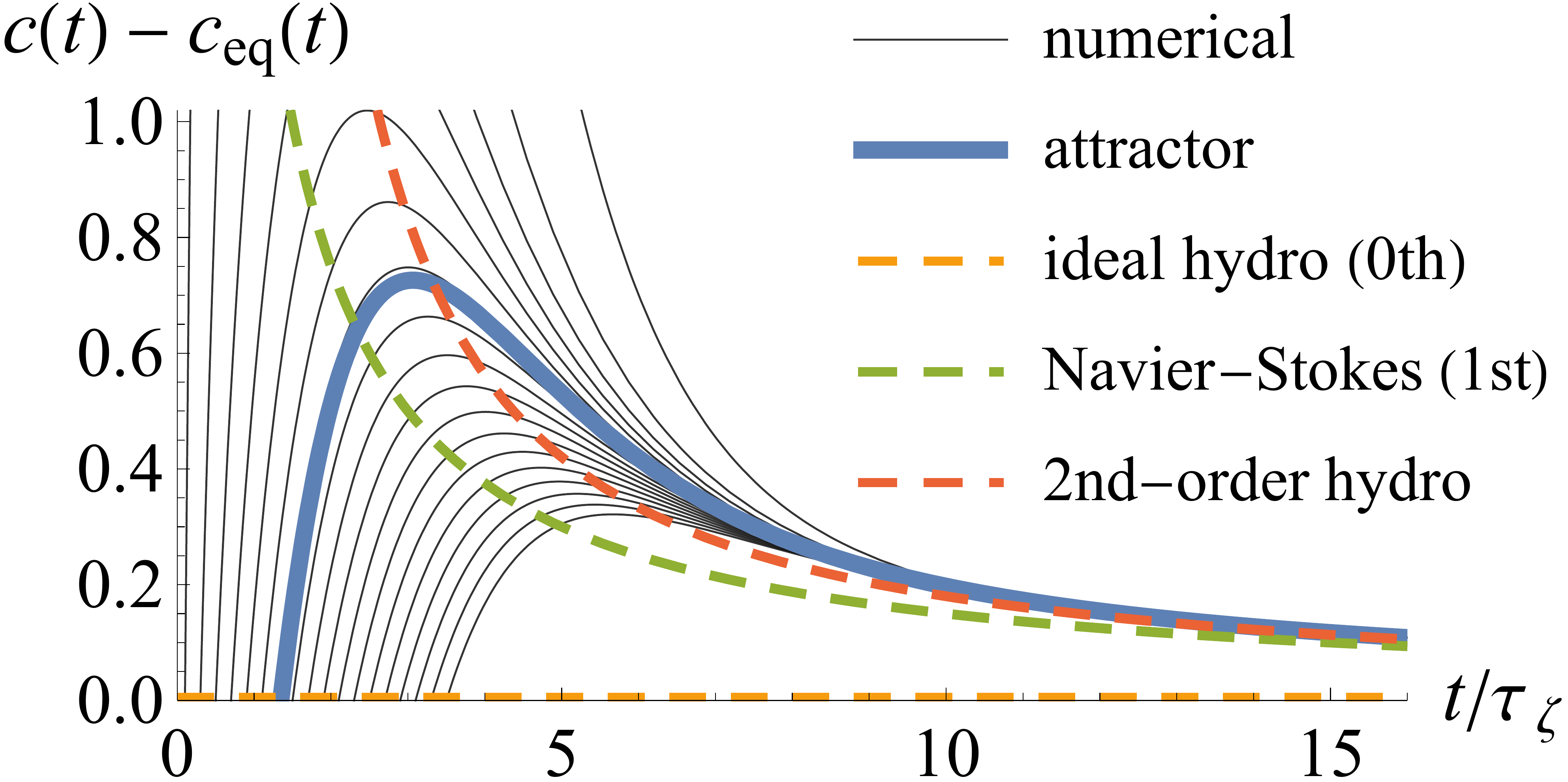}
    \hfill
    \includegraphics[width=0.35\linewidth]{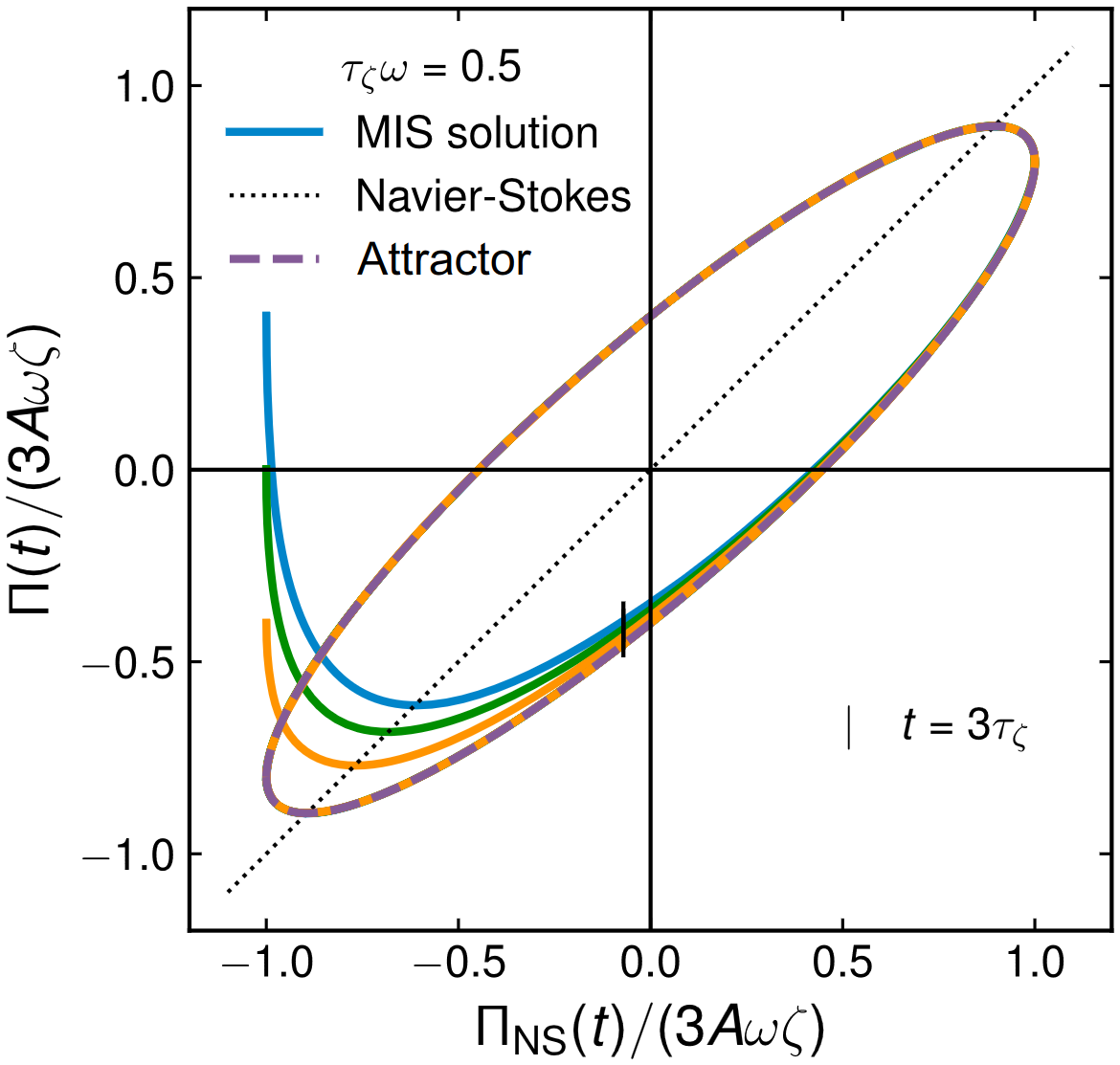}
    \caption{Hydrodynamic attractor in ultracold atoms. \textbf{Left:} For a power law drive, the normalized bulk pressure $c(t) - c_\text{eq}(t)= \Pi(t)/\chi$ exhibits different time evolutions for different initial conditions (thin lines), which quickly converge toward the attractor solution (thick blue line) and only later approach Navier–Stokes hydrodynamics (green dashed line). After \cite{fujii2024}, where $\tau_\zeta=\tau_\Pi$.
    \textbf{Right:} Cyclic attractor in a periodically driven system.  The normalized full bulk pressure $\Pi(t)$ (vertical axis) follows the Navier--Stokes approximation $\Pi_\text{NS}(t)$ (horizontal axis) with a time lag due to the relaxation time $\tau$, which is manifest in the elliptic shape.  Different initial conditions (colors) converge toward the same cycle.  After \cite{mazeliauskas2026}.}
    \label{fig:ca:attractor}
\end{figure}

\subsection{Late-time attractor}

\subsubsection{Monotonic drive}

There are two recent theoretical proposals to observe hydrodynamic
attractor behavior based on this scheme.  The first starts from an
interacting equilibrium state and then suddenly applies a monotonic
power-law drive of the scattering length $a(t)\sim t^\alpha$ toward
resonance \cite{fujii2024}. This creates Bjorken flow-like conditions, as the bulk strain rate takes the form $V_a(t)\sim t^{-1}$. A thermodynamic description would predict that
the contact $\mathcal C(t)=\mathcal C_\text{eq}[a(t)]$ follows the
drive instantaneously; this is the limit of slow drive or fast
relaxation. However, for faster changes in the scattering length the system is driven out of equilibrium and attains a non-equilibrium contribution to the instantaneous $\mathcal C(t)$, the so-called bulk
pressure
$\Pi(t) = (\mathcal C(t)-\mathcal C_\text{eq}[a(t)])/(12\pi ma(t))$
\cite{fujii2018}. However, even the bulk pressure itself is not determined by the drive instantaneously due to memory effects in the system. Modeling this effect by a single relaxation time $\tau_\Pi$ -- an ansatz that is compatible with the Drude peak in the low frequency behavior of the bulk viscosity, but whose validity at short times must still be assessed -- the bulk pressure relaxes to the bulk strain rate dictated by the drive on the timescale $\tau_\Pi$ as described by the MIS type equation
\begin{align}
  \label{eq:ca:attractor}
  \tau_\Pi\dot\Pi(t) = -\Pi(t)-\zeta\, V_a(t).
\end{align}
For the monotonic drive, the change of scattering length is
initially faster than the relaxation time $\tau_\Pi$ and $\Pi$ follows the drive excitation with a lag, which causes a deviation from the Navier--Stokes value. Figure~\ref{fig:ca:attractor} shows the
dimensionless bulk pressure vs time for different initial equilibrium
states \cite{fujii2024}.  When the drive starts, bulk pressure first
builds up (thin black lines) and then relaxes back to zero as the
unitary equilibrium state is approached for long times.  Notice that
the black lines converge toward the blue curve: this depicts the
(analytical) attractor solution of the MIS equation for the bulk
pressure $\Pi(t)$ driven by the bulk strain rate. The initial condition decays exponentially over the relaxation time
$\tau_\Pi$, which is a manifestation of the non-hydrodynamic mode of this
system.  The attractor solution agrees with Navier--Stokes
hydrodynamics (green dashed) for long times $t\gg\tau_\Pi$ but extends the
regime of fluid-like universal dynamics to earlier times $t\gtrsim\tau_\Pi$.

\subsubsection{Periodic drive}

A second proposal aims to make the beyond-Navier--Stokes behavior
visible for longer times \cite{mazeliauskas2026}.  It achieves this by
periodic drive of the scattering length
$a(t)=a_0+a_1\sin(\omega t) \Theta(t)$, with $a_1\ll a_0$, such that the dynamics can be linearized and transport coefficients assumed to be constant. The full bulk pressure is
again given by the solution of the MIS equation
\eqref{eq:ca:attractor}, while the Navier--Stokes prediction follows
with $\tau_\Pi\to0$.  Plotting the full bulk pressure $\Pi(t)$
parametrically against the Navier--Stokes approximation
$\Pi_\text{NS}(t)$, due to the time lag one finds for faster drive an elliptical trajectory
(see Fig.~\ref{fig:ca:attractor}): as the deviation from Navier--Stokes increases, the ellipse first opens more and then starts closing again. The trajectories for
different initial equilibrium states converge toward the same ellipse,
which is identified as a cyclic attractor curve.  For larger-amplitude
drives the nonlinear response was modeled in massive relativistic RTA kinetic theory, which in the non-relativistic limit provides a model description for the the normal gas. In this description, the time evolution still
shows convergence toward an attractor curve; this curve, however, no
longer closes because the system heats up under the drive.

This cyclic attractor is conceptually distinct from the monotonic one as it does not converge to Navier--Stokes at late times and therefore cannot be described through its late-time constitutive behavior. Rather, it probes the transient theory at different excitation frequencies, with a low-frequency hydrodynamic limit.

\begin{figure}[t]
    \centering
    \includegraphics[width=0.49\linewidth]{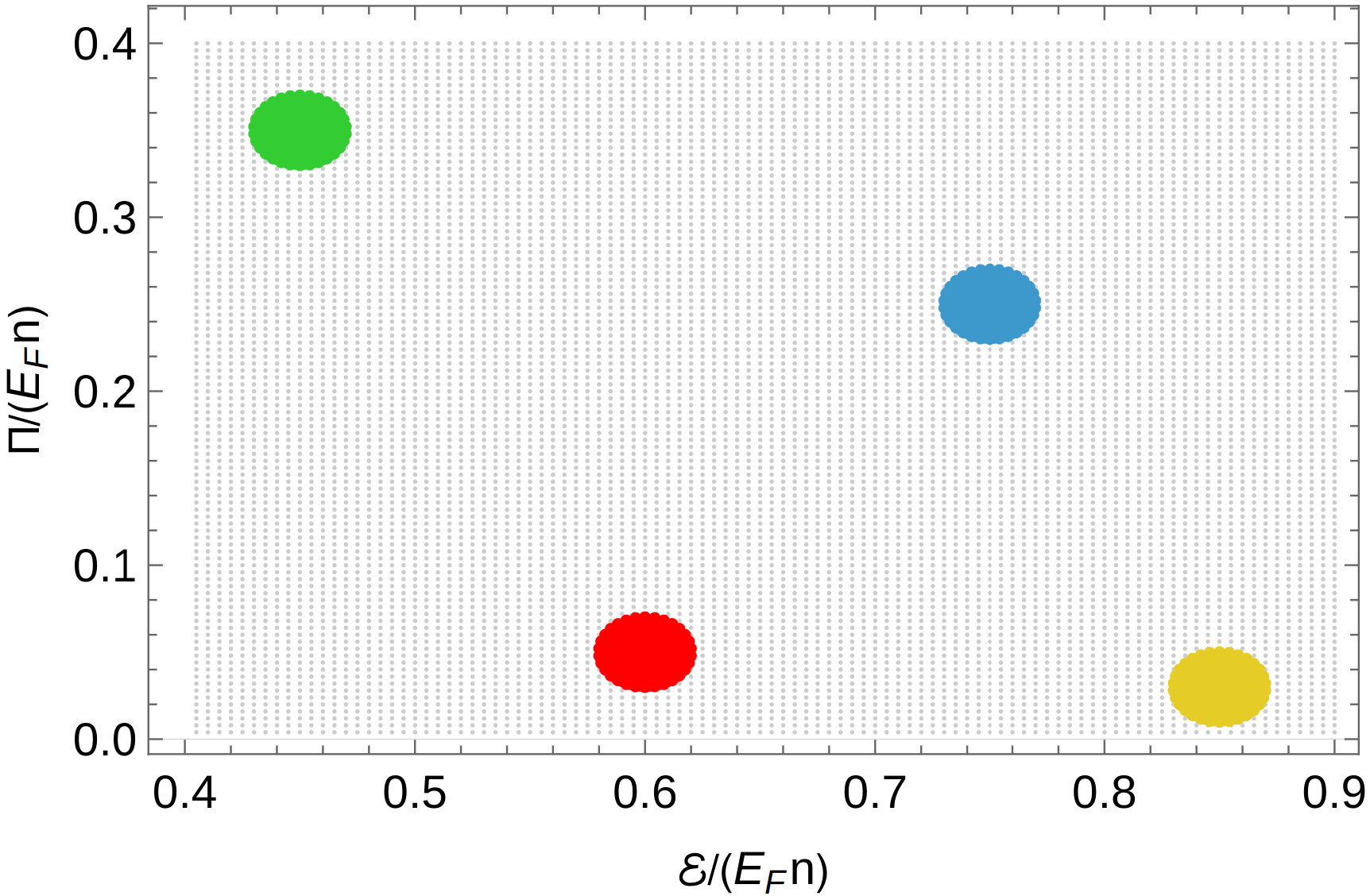}
    \includegraphics[width=0.49\linewidth]{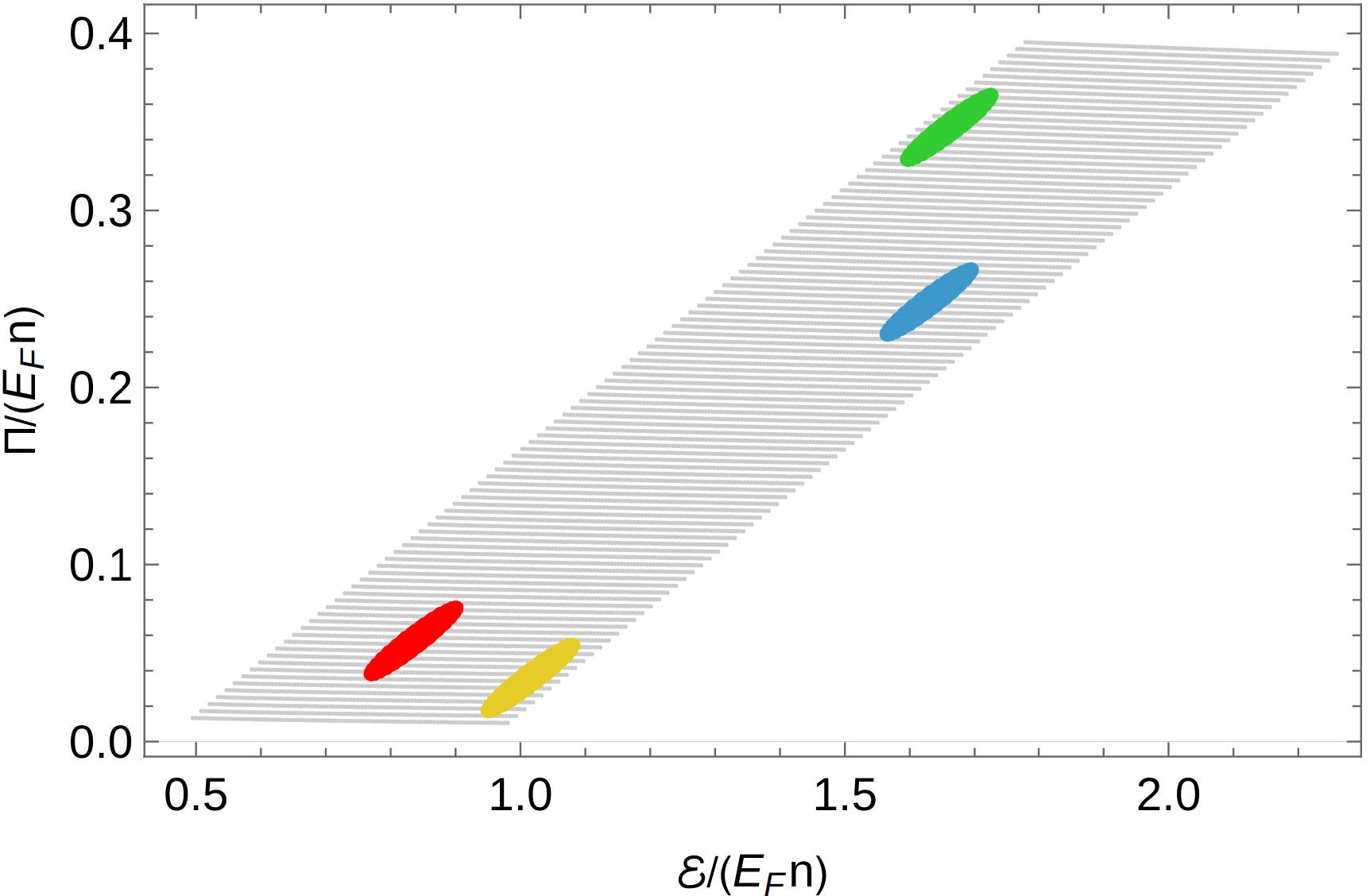}
    \includegraphics[width=0.49\linewidth]{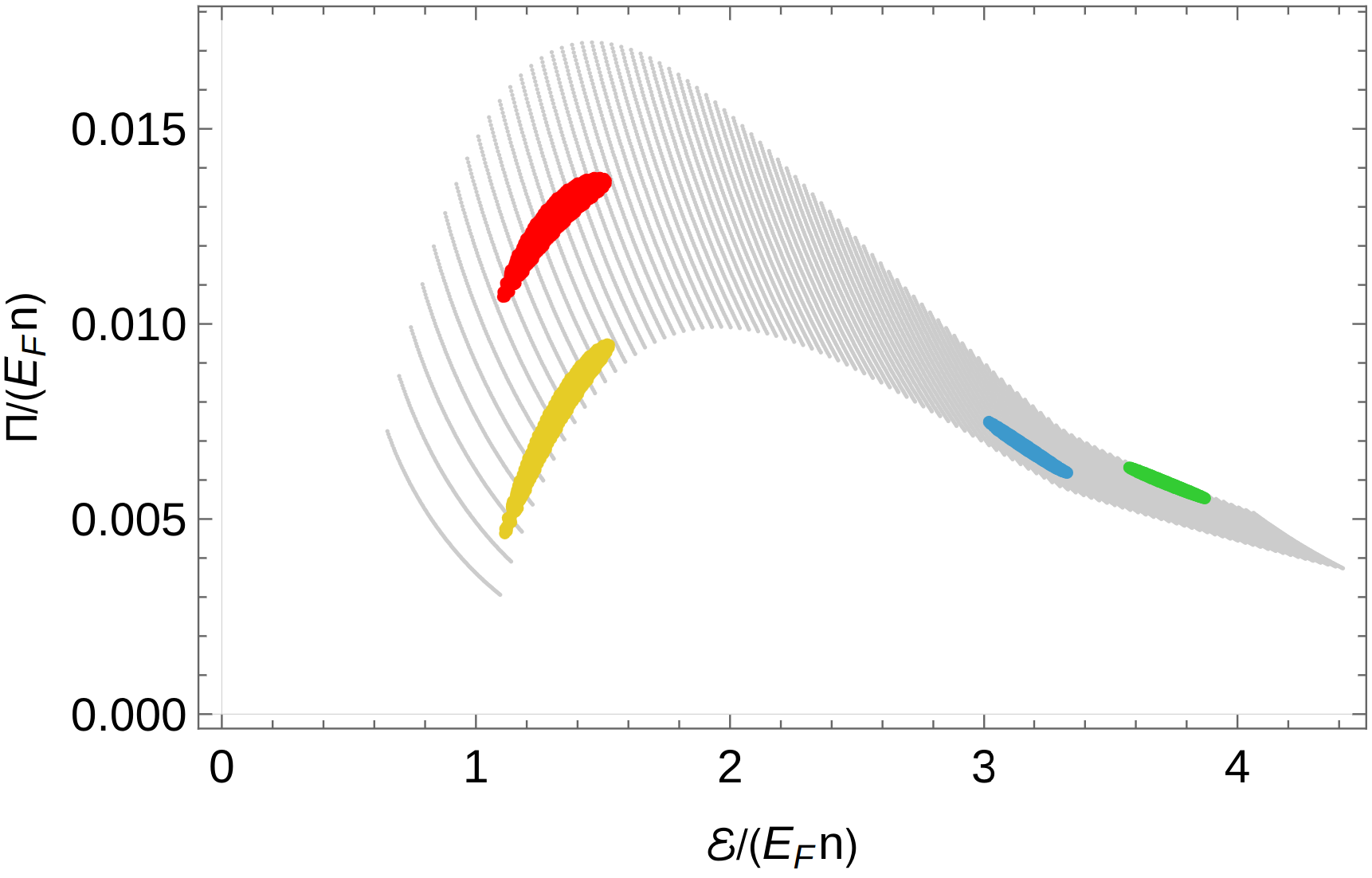}
    \caption{Early-time attractor in ultracold atoms. Starting from an initially rectangular (grey) and several initially spherical (colored) sets of initial conditions, the evolution contracts trajectories in the state space of normalized bulk pressure $\Pi(t)$ vs energy density $\mathcal E(t)$ (note difference in notation) toward a unique functional relationship, showing the loss of information. Shown are snapshots at $t=t_0=0.01T_F^{-1}$, $t=3t_0$ and $t=T_F^{-1}\sim\tau_\Pi$. After \cite{heller2025early}.}
    \label{fig:ca:shorttime}
\end{figure}

\newpage
\subsection{Early-time attractors}

\subsubsection{Expansion attractor}

While these attractors describe the universal approach toward a fluid
description (MIS/NS) at long times, a recent proposal studies the emergence of universality in the dynamics of many-body systems through strong expansion at short times $|V_a(t)|\gg\tau_\Pi^{-1}$, i.e. an instance of an expansion attractor~\cite{heller2025early}. An argument based on the fixed point structure known from Bjorken flow shows that this behavior is not evident from the dynamics of bulk pressure alone. Tracking the joint evolution of the bulk pressure and the energy
density $\partial_t \varepsilon= -(\mathcal C(t)/4\pi m) \partial_t
a^{-1}(t)$, which are coupled through the contact $C(t)$ and the temperature dependence of transport coefficients and $C_{\rm eq}$, the solutions for sets of
initial conditions $(\Pi(t_0)$, $\varepsilon(t_0))$ contract along one direction in this state space toward a
 functional relation between $\Pi(t)$ and $\varepsilon(t)$
already at short times $t\lesssim\tau_\Pi$, as shown in Fig.~\ref{fig:ca:shorttime}. In contrast to Bjorken flow, the form of this functional relation evolves with time and shows no early-time limiting fixed point. However, the contraction demonstrates that expansion can
provide a mechanism of information loss besides equilibration also in the context of cold atoms, which allows for an experimental verification. Ref.~\cite{heller2025early} also provides a concrete experimental protocol for the search for expansion-driven information loss.

\begin{figure}
\centering    \includegraphics[width=0.49\linewidth]{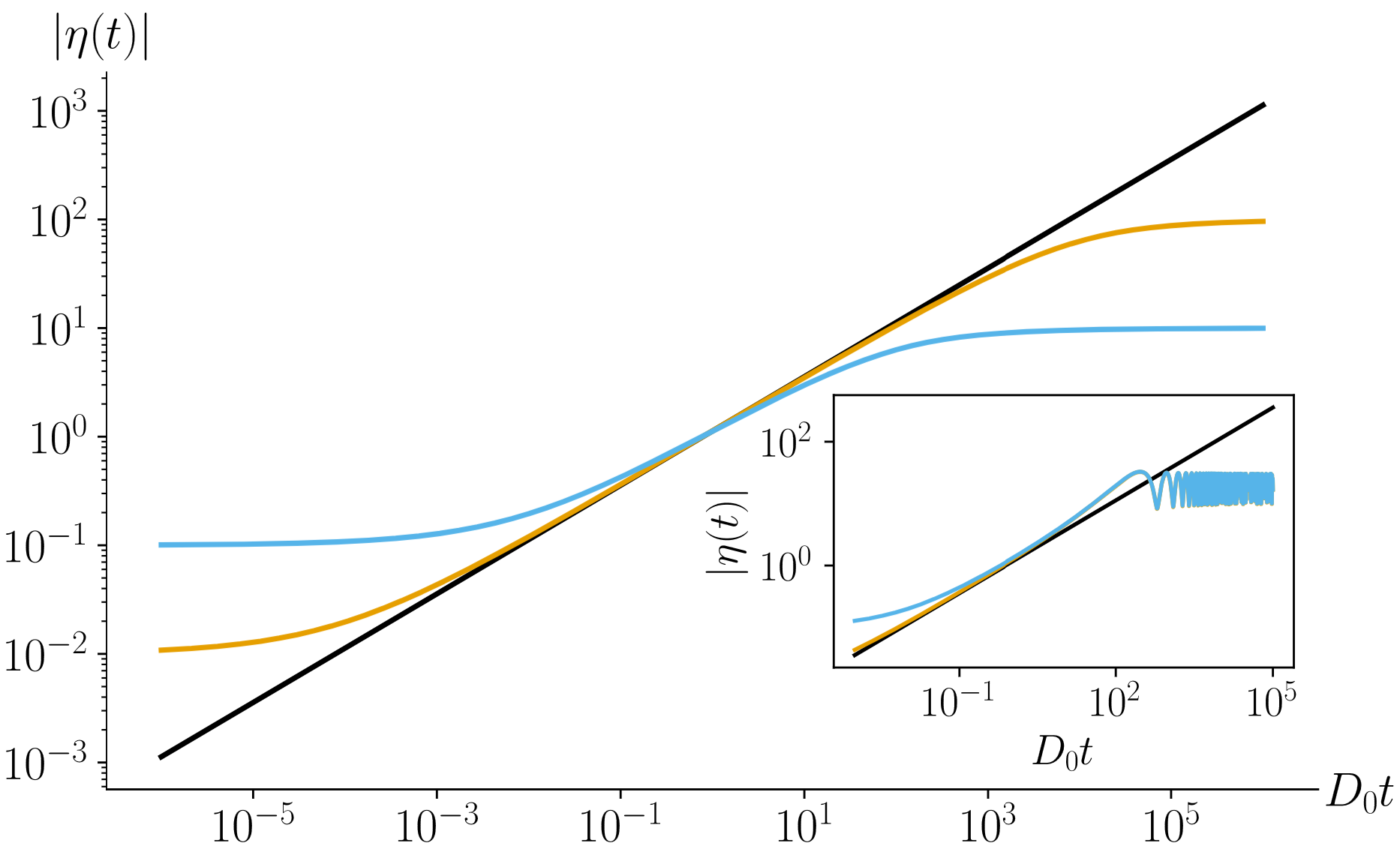}
\hfill
\includegraphics[width=0.49\linewidth]{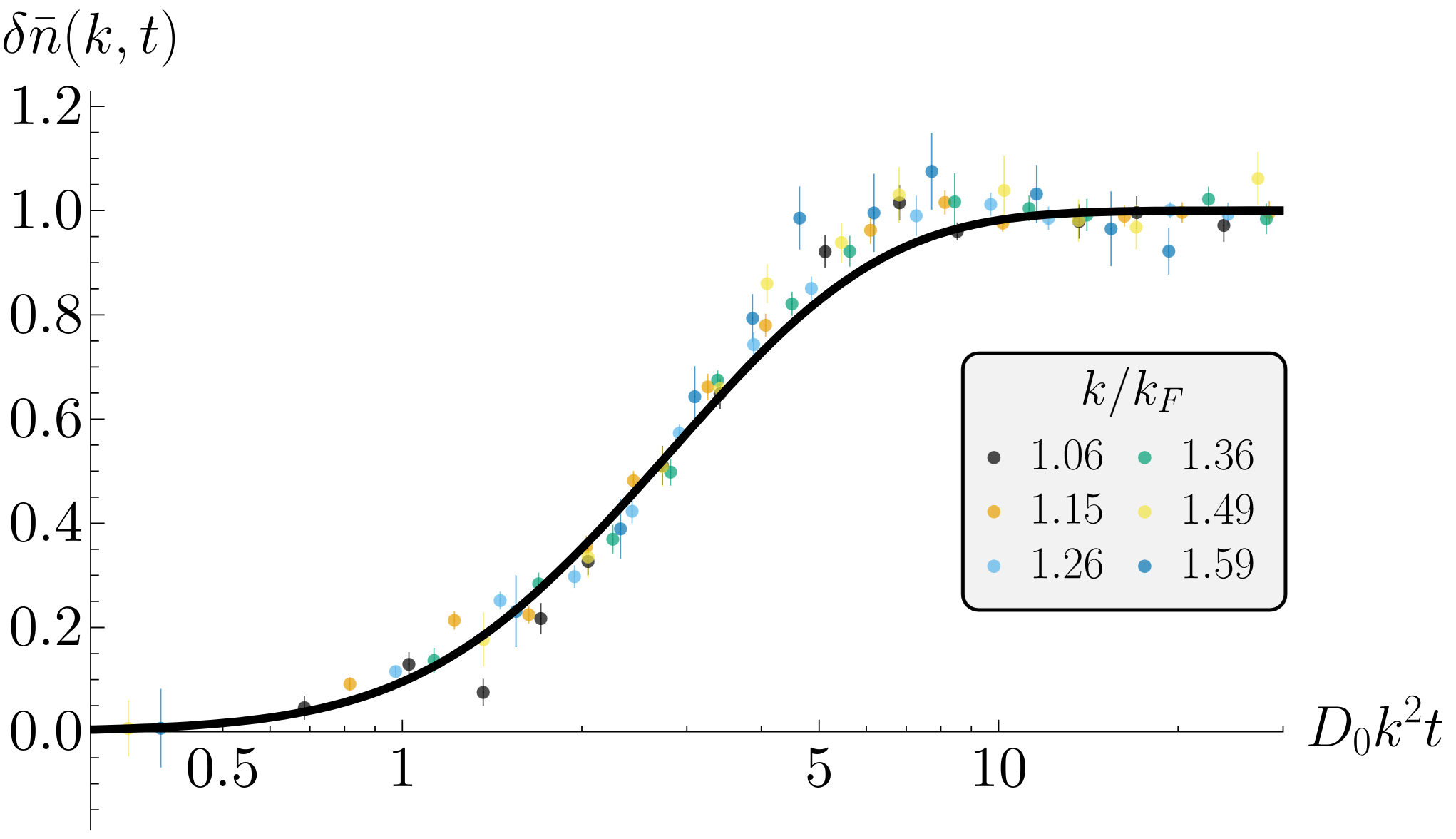}
    \caption{(Left) Growth of the pairing amplitude $\eta$ (notation exclusive to this Figure) after a sudden increase of interaction strength: initial convergence  onto few-body conformal attractor $\eta \sim (D_0t)^{1/2}$ (black line), before relevant perturbations lead away.  (Right) Momentum distribution change $\delta \bar n(k,t)$ measured in ultracold fermionic atoms after a sudden change of interaction, compared to the scaling prediction (solid line).  After \cite{sharell2026fractional}.}
    \label{fig:ca:shorttimesudden}
\end{figure}

\subsubsection{Few-body scaling and attraction}

For a sudden change of the interaction on time scales even shorter
than the many-body time scale, $t\lesssim\hbar/E_F$, the response of
the dilute gas is dominated by few-body physics.  The growth of local
correlations and the contact $\mathcal C(t)$ is fully described by the
two-body Schrödinger equation for a single pair.  In the local limit
of vanishing distance between the two particles, in contrast to hydrodynamic evolution equations it takes the form of a
\emph{fractional} differential equation, which can be solved
analytically for different drive protocols \cite{sharell2026fractional}.  For a
quench from weak to strong interaction, local correlations build
up and spread diffusively with the diffusion quantum $D_0=\hbar/m$, leading to a power-law loss of information about the initial interaction, as shown in Fig.~\ref{fig:ca:shorttimesudden} (left).

Adding a phenomenological relaxation time $\tau$ to describe decay of pair
coherence in the many-body medium, the fractional differential equation goes over into an MIS equation in the long-time limit: this model covers the full evolution from an early-time few-body conformal fixed point toward
many-body equilibration at late times.  Recent measurements of the time evolution of the momentum distribution of excitations $n(k,t)$ \cite{yi2025quantum} provide a new, fine-grained experimental observable. As shown in Fig.~\ref{fig:ca:shorttimesudden} (right), the large-momentum data show a collapse when plotted against $D_0k^2t$, consistent with the predicted conformal scaling, and the simple phenomenological model is able to capture the approach to a first quasi-stationary state \cite{sharell2026fractional}.

\section{Outlook}
\label{sec:outlook}

The developments covered in this chapter replace a single question---when does matter equilibrate?---with three distinct ones: which initial-state differences cease to matter, which variables (observables) remain necessary, and when does their evolution admit an accurate constitutive description? Progress now requires quantitative tests that distinguish these questions rather than infer one answer from another.

\paragraph{Testing attraction in ultracold gases.}
The proposed scattering-length protocols make it possible to formulate an operational test: prepare different states, apply the same subsequent drive, and measure whether their evolution contracts in a specified space of observables~\cite{fujii2024,mazeliauskas2026,heller2025early}. For the expansion-driven proposal, the relevant space includes both bulk pressure $\Pi$ and energy density $\varepsilon$, rather than either observable alone. Contact measurements \cite{xie2026}, together with the equilibrium contact evaluated at the instantaneous density, energy, and scattering length, provide access to $\Pi$ through the relation developed in~\cite{fujii2018}.

The practical constraints are part of the physics. The representative bulk-relaxation scale of order $0.1\,\mathrm{ms}$ discussed in this chapter requires appropriately fast control and readout, not merely a long observation window~\cite{enss2019bulk,fujii2024,enss2025quantum}. Box confinement or spatially resolved analysis can reduce the mismatch between a uniform-gas calculation and an inhomogeneous cloud~\cite{patel2020}. The normal-state description must be reassessed on crossing the superfluid transition, and periodic protocols must account for heating and the consequent drift of transport coefficients~\cite{enss2025quantum,mazeliauskas2026}.

Anisotropic cloud expansion and collective-mode measurements offer a complementary route through shear stress, reconnecting the attractor question with the long-standing viscosity comparison between atoms and nuclear matter~\cite{cao2011,brewer2015}. Interaction-driven elliptic flow in few-fermion systems further motivates tests of how collective behavior emerges as particle number grows, without identifying elliptic flow itself with constitutive validity~\cite{brandstetter2025,gao2026, berges2026}.

\paragraph{Connecting few-body correlations to many-body relaxation.}
The short-time fractional equation describes the local two-body problem, whereas its continuation toward an MIS-type equation introduces phenomenological decay of pair coherence~\cite{sharell2026fractional}. A microscopic derivation of that decay, its memory kernel, and its dependence on the many-body state would determine when this interpolation can become predictive. Measurements of momentum distributions constrain the early scaling regime and subsequent relaxation, but their collapse onto a scaling form does not by itself establish the proposed many-body attracting relation~\cite{yi2025quantum,sharell2026fractional}. Varying initial correlations as well as drive rate would help separate few-body scaling, expansion-driven contraction, and dissipative relaxation. The broader mesoscopic-gas program provides a setting in which to investigate how the relevant degrees of freedom change between few-body and collective dynamics~\cite{brandstetter2025,berges2026,gao2026}.

\paragraph{Establishing the scope of early-time holographic attraction.}
The preliminary work described in Sec.~\ref{sec:holography_attractors} raises a precise but unresolved possibility. It reports a late-time response $\varepsilon\propto\tau^{-1}$ and $\mathcal P_L/\varepsilon\to0$ in a linearized vacuum problem for a specified class of initial holographic data, finite and nonzero at the Poincar\'e horizon. The separate nonlinear conjecture is that sufficiently weak data localized near the boundary can pass through an intermediate interval governed by this response before backreaction or the finite-area horizon in the bulk geometry becomes important. Note that this late-time linearized statement also differs from a claim about regular nonlinear solutions at $\tau\to0$~\cite{Beuf:2009cx}.

A convincing test must vary amplitude, radial localization, and profile shape independently, demonstrate contraction across initial states, and identify the information that survives. A smooth trajectory or $\mathcal P_L\simeq0$ alone is insufficient~\cite{Kurkela:2019set}. It is equally important to determine whether states prepared by holographic shock collisions enter such a restricted basin~\cite{Grumiller:2008va,Chesler:2010bi,Casalderrey-Solana:2013aba,Casalderrey-Solana:2013sxa,Chesler:2015wra}, although it should be stressed that this requires going beyond strict boost invariance~\cite{Casalderrey-Solana:2013aba}. Universality for physically prepared states could be valuable even without universality over all admissible bulk initial data.

\paragraph{Beyond conformal Bjorken flow.}
Non-conformal matter, finite conserved-charge densities, and departures from boost invariance introduce additional scales and surviving variables~\cite{Romatschke:2017acs,Du:2020zqg,Chen:2024grb,Du:2025hyk}. The resulting attracting set may evolve in time without admitting a preferred single-curve representation~\cite{Spalinski:2025ngd}. The bulk channel provides a natural conceptual bridge to interaction-driven cold atoms, but its coupling to shear stress and conserved densities must be determined separately in each system.

An effective description must also accurately reproduce perturbations about its attracting background, not just that background itself. Kinetic linear-response evolution, perturbative transseries, anisotropic hydrodynamics, and attractodynamics provide complementary starting points~\cite{Kurkela:2018wud,An:2023yfq,Strickland:2017kux,DuPlessis:2026qjy}. Bjorken symmetry hides independent tensor structures and therefore cannot fix their coefficients or the response to general spatial gradients. In a relativistic completion, characteristic propagation, stability, and the initial-value problem must be checked on the non-equilibrium states to which the model will be applied, rather than inferred from an attractor fit~\cite{Israel:1979wp,Baier:2007ix,Romatschke:2017ejr}. This requirement applies in particular to phenomenological adjustments such as the coefficient $a_1=31/15$ discussed in Sec.~\ref{sec:phenomenology}; reproducing the early-time pressure ratio does not establish a controlled or causal extension of the microscopic constitutive theory~\cite{Blaizot:2021cdv}.

\paragraph{Connecting universal regimes and controlling errors.}
Adiabatic hydrodynamization and the relaxation spectrum of nonthermal fixed points suggest a description in terms of evolving slow sectors rather than a succession of unrelated universal curves~\cite{Brewer:2022vkq,Rajagopal:2024lou,DeLescluze:2025gaa,berges2025far}. Their relation to the stages of weak-coupling bottom-up evolution is particularly important. Coupling-dependent kinetic studies distinguish a bottom-up limiting attractor from a viscosity-scaled hydrodynamic one; the useful scaling depends on the observable and stage of evolution~\cite{Boguslavski:2023jvg}. This does not imply absence of attraction at finite coupling, nor does extrapolating a kinetic model to large coupling establish the holographic limit. The open question is how the surviving variables and relaxation mechanisms connect across these regimes.

Quantitative control requires more than identifying a spectral gap or an apparently thin state-space cloud. In conformal BRSSS theory, the early separation scales as $(w_0/w)^{\gamma_+}$, whereas the leading late displacement from the selected attractor is proportional to $(\sigma-\sigma_\star)e^{-\chi w}w^\alpha$, with the parameters defined in Sec.~\ref{sec:BRSSS_attractors}. These give concrete measures of initial-state sensitivity within their asymptotic regimes. Extending such estimates requires control of prefactors, nonlinear deviations, and the interval of validity. In adiabatic hydrodynamization, spectral separation must be assessed together with time-dependent mixing and the non-Hermitian nature of the generator; a gap alone is not an error bound~\cite{Brewer:2019oha,Rajagopal:2024lou,Rajagopal:2025nca}.

The reduced classical Yang--Mills example also cautions against assuming monotonic information loss~\cite{Werthmann:2026zso}. Whether its partial recovery of suppressed sensitivity persists in ensembles appropriate to the Glasma, when spatial fluctuations and possible instabilities are included, remains to be established. Comparing ensemble-averaged observables with the evolution of individual field configurations would identify which part of the contraction is dynamical and which follows from the chosen statistical description.

\paragraph{Discriminating early-time dynamics in nuclear collisions.}
Attractor-based multiplicity, cooling, and flow relations are most useful when their uncertainty can be separated from changes in the initial energy deposition and later transport~\cite{Giacalone:2019ldn,Ambrus:2021fej,Ambrus:2022koq}. A concrete next step is to compare pre-equilibrium prescriptions within a common Bayesian inference framework, varying their uncertainties alongside viscosity and initialization parameters rather than interpreting a fitted switching time as a universal hydrodynamization time. Consistent matching is essential: differences in early longitudinal cooling can otherwise masquerade as differences in subsequent hydrodynamic response~\cite{Kurkela:2020wwb,Ambrus:2022qya,Ambrus:2024eqa}. Small collision systems sharpen these tests because their shorter evolution leaves less room for the hydrodynamic stage to dominate~\cite{Kurkela:2019kip,Ambrus:2024hks}.

Photons, dileptons, and jets provide complementary sensitivity, but stress-energy universality does not automatically imply universal emission or scattering rates. Kinetic calculations of pre-equilibrium photon and dilepton production already exhibit useful scaling functions, demonstrating that universality can extend to such observables when established explicitly~\cite{Garcia-Montero:2023lrd,Garcia-Montero:2024lbl}. Momentum broadening and energy loss require their own microscopic input, particularly in an anisotropic plasma or the earlier gauge-field stage~\cite{Boguslavski:2024jwr,Ipp:2020nfu}. Combining these probes with bulk observables can test which features of the early evolution remain distinguishable after hydrodynamic matching~\cite{Pablos:2025cli}.

The objective is not to replace local equilibrium by another universal assumption. It is to determine when an economical description exists, what information it must retain, and how its accuracy can be tested. Nuclear collisions motivate this problem under extreme conditions; cold atoms provide complementary control over preparation and observation. Bringing constitutive accuracy, contraction of initial-state ensembles, and experimental sensitivity into quantitative agreement is the central task ahead.

\subsection*{Acknowledgements}
We are grateful to our collaborators and peers who have been sharing with us the joys of studying far-from-equilibrium dynamics. We would like to thank Xin An for feedback on the draft. MPH would like to acknowledge the wonderful hospitality of the Institute of Theoretical Physics and Astronomy, Vilnius University during the final stages of this chapter's preparation. This work has received funding from the European Research Council (ERC) under the European Union’s Horizon Europe research and innovation programme (grant number: 101089093 / project acronym: High-TheQ). Views and opinions expressed are however those of the authors only and do not necessarily reflect those of the European Union or the European Research Council. Neither the European Union nor the granting authority can be held responsible for them. This work is funded by
the Deutsche Forschungsgemeinschaft DFG (German Research Foundation) under Project-ID 273811115 – SFB
1225 ISOQUANT.

\subsection*{Declaration of generative AI and AI‑assisted technologies in the writing process}

During the preparation of this work, the authors used ChatGPT and Claude to help formulate parts of the abstract, introduction, and outlook; cross-check scientific statements; ensure consistency in grammatical conventions; and correct typographical errors. After using these tools, the authors reviewed and edited the content as needed and take full responsibility for the content of the publication.

\appendix

\section{Dictionary of foundational concepts}
\label{app}

\subsection{Heavy ion collisions}

In (ultra-)relativistic heavy ion collisions, nuclei are accelerated close to the speed of light and collided in order to create a small volume of strongly interacting matter at a high temperature. Under these conditions, strongly interacting particles deconfine and form a fluid-like phase of matter called the quark-gluon plasma (QGP), which at very high temperatures obeys conformal symmetry. It has also been observed to have a very small shear viscosity, close to a theoretical bound derived in the strongly interacting limit, and is thus an almost perfect fluid. The main goal of heavy ion collisions is to extract further details of the properties of this phase of matter.

However, this information is shrouded by the fact that the collisions themselves happen on scales of femtometers and yoctoseconds, which makes a direct measurement impossible. What is experimentally accessible are the spectra of final state particles reaching the detectors after a few meters of free flight and nanoseconds after the collision. These particles are again confined bound states under the strong interaction. The task of extracting this information therefore requires constructing detailed theoretical models of the dynamics in these collisions, which predict final state distributions that can be compared to experimental data.

\subsection{Bjorken flow}

Motivated by the fact that for nuclear collisions at very high energies, starting from their center of mass system small boosts towards one nucleus or the other are negligible compared to their rapidities, J.D. Bjorken proposed a highly symmetric model for the midrapidity region~\cite{Bjorken:1982qr}. Neglecting also the transverse structure, it assumes transverse homogeneity and isotropy on top of longitudinal boost invariance. In coordinates of the proper time $\tau$ and spacetime rapidity $\eta$ defined as
\begin{align}
    \tau=\sqrt{t^2-z^2}\,, \qquad \eta=\mathrm{artanh}(z/t)\,,
\end{align}
it becomes apparent that this description is $0+1$-dimensional, as $\tau$ is invariant and $\eta$ behaves additively under longitudinal boosts, so the assumption of boost invariance allows spacetime-dependent functions to depend only on proper time $\tau$. Furthermore, in the coordinate set $(\tau,x,y,\eta)$ the collective velocity $u^\mu$ and the energy-momentum tensor $T^{\mu\nu}$ take on simple tensorial structures:
\begin{align}
    u^\mu=(1,0,0,0)\,,\qquad T^{\mu\nu}=\mathrm{diag}(\varepsilon,\mathcal{P}_T,\mathcal{P}_T,\tau^{-2}\mathcal{P}_L)\,,
\end{align}
where $\varepsilon$ is the energy density, $\mathcal P_T$ is the transverse and $\mathcal P_L$ the longitudinal pressure.

\subsection{Conformal symmetry}

A conformal system exhibits scale invariance, meaning in particular that there are no dimensionful coupling constants. Conformality presumes proportionality relations between thermodynamic quantities: $\varepsilon\propto\mathcal{P}\propto T^4$. In terms of non-equilibrium quantities, it restricts bulk pressure to be zero and the specific shear viscosity to be constant. As a consequence, the energy-momentum tensor is traceless and in Bjorken flow $\varepsilon=2\mathcal{P}_T+\mathcal{P}_L$. It also implies that interaction timescales are proportional to $T^{-1}$, meaning that the variable
\begin{align}
    w=\tau T(\tau)
\end{align}
is a natural dimensionless clock. When comparing systems with different interaction strength, the proportionality constant differs between them. As interaction strength is tied to the size of dissipative effects in the non-equilibrium sector, it is then convenient to define the dimensionless time variable
\begin{align}
    \tilde{w}=\frac{\tau T(\tau)}{4\pi\eta/s}\,.
\end{align}

\subsection{Hydrodynamics}

Hydrodynamics is the effective field theory for long-lived, long wavelength excitations, which are typically associated with conserved currents such as energy, momentum, and possibly currents of conserved quantum numbers. Hydrodynamics then models these degrees of freedom in terms of macroscopic quantities: the energy density $\varepsilon$, the thermodynamic pressure $\mathcal{P}$, the number densities $n_i$ the collective velocity $u^\mu$ and gradients thereof. The hydrodynamic evolution equations are derived from the continuity equations for the conserved currents. 

The theory is organized in an expansion in the order of these gradients, starting with no gradients in ideal hydrodynamics and first order gradients in Navier--Stokes hydrodynamics. On top of this, hydrodynamics also requires closeness to a reference state of universal dynamics, typically equilibrium, which fixes the equation of state $\mathcal{P}(\varepsilon)$ and transport coefficients. The latter are constants multiplying the gradient terms appearing in the evolution equations. In Navier--Stokes hydrodynamics of energy and momentum, there are two such constants. The bulk viscosity $\zeta$ converts the expansion scalar into bulk pressure and the shear viscosity $\eta$ translates between tensorial shear strain rate and shear stress.

\subsection{MIS theory}

In relativistic hydrodynamics, the fact that changes in gradient terms instantaneously determine the stresses in the system may cause superluminal propagation, which leads to instabilities in the evolution. Müller~\cite{1967ZPhy..198..329M}, Israel~\cite{Israel:1976tn} and Stewart~\cite{Israel:1976efz,Israel:1979wp} formulated an ansatz to tackle this problem by promoting dissipative quantities to dynamical variables. However, even then, causality and well-posedness are conditional on constraints for the transport coefficients and non-equilibrium states~\cite{Bemfica:2020xym}.

In this theory, the values that these quantities take instantaneously in a purely hydrodynamic theory become sources for a relaxation equation of the form $\tau_X\dot{X}+{X}=\dots$ with new transport coefficients: the relaxation times $\tau_X$. This means that the theory now describes not only the long-lived hydrodynamic sector, but also a sector of transient dynamics. After a sufficiently long time without any significant external perturbations, transient dynamics will decay and the system will relax towards hydrodynamics.

\subsection{Kinetic theory}\label{sec:kinetic_theory}

Taking a weakly coupled perspective, kinetic theory describes a system as a statistical phase space distribution of (quasi-)particles,
\begin{align}
    f(t,x,p)=(2\pi)^3\frac{\d N}{\d^3x\d^3p}(t)\,.
\end{align}
For the sake of simplicity, we consider here any degeneracy factor for the particle type to be absorbed into $f$. This picture of course assumes that all excitations in the system are of this type and that interactions are short range, as potentials are neglected. Describing only the one-particle distribution given above also assumes that higher order correlations do not play a role. 

The time evolution of the phase space distribution is described by the Boltzmann equation
\begin{align}
    p^\mu\partial_\mu f=C[f]\,,
\end{align}
where the left hand side describes the motion of particles and the right hand side their interaction through a collision kernel, which can be modeled in different ways. If supplied with microscopic information on the scattering matrices, the underlying assumption is that scatterings happen independently of each other in a negligibly small spacetime volume, which is accurate only for sufficiently dilute systems. More macroscopic choices of the collision kernel constitute effective descriptions of the scatterings through for example  drag and diffusion in momentum space in the case of Fokker--Planck kinetic theory or relaxation to local equilibrium in the relaxation time approximation.

Boost invariance requires that the phase space distribution depends on the spacetime rapidity $\eta$ and the longitudinal momentum rapidity $y$ only through their difference. Assuming additionally a forward-backward symmetry makes the dependence on $y-\eta$ even. This may also be expressed through the co-moving momentum coordinates
\begin{align}
   p^\tau={m_\perp}{}\cosh(y-\eta)\,,\qquad p^\eta=\frac{m_\perp}{\tau}\sinh(y-\eta)\,, \qquad m_\perp=\sqrt{p_\perp^2+m^2}\,,
\end{align}
though it is more intuitive to work with co-moving physical momentum $p_L=\tau p^\eta$. In Bjorken flow, transverse homogeneity and isotropy further constrain the phase space distribution to depend only on $\tau$, $p_L$ and $p_\perp$.

\subsection{Holography}

Holography provides a dual description of certain strongly coupled non-Abelian gauge theories with a large number of local degrees of freedom (colors in the QCD sense) in terms of classical gravity in a higher-dimensional asymptotically anti-de Sitter spacetime. For the purposes of this review, its main advantage is that it makes the real-time dynamics of such theories accessible from first principles using well-established tools of general relativity. Thermal states of the field theory correspond to static black hole geometries with a planar horizon, known as black branes. Studying perturbations of these black branes with different characteristic amplitudes and length scales then provides access to linear response, hydrodynamics, and fully nonlinear far-from-equilibrium dynamics, depending on the chosen perturbation. These different regimes, which are notoriously difficult to tackle using Euclidean lattice methods, are mapped by holography onto classical gravitational initial value problems. The price is that theories admitting tractable classical gravitational duals are not real-world QCD: they differ from it in their large number of colors, symmetries, and matter content. Nevertheless, the expectation is that, for questions governed by generic features of strong coupling, such as the absence of quasiparticle excitations, holography can reveal correspondingly general lessons relevant to real-world phenomena. The hydrodynamization studies discussed in this review provide a prominent and insightful example of this strategy.

\subsection{Ultracold atomic gases}
Ultracold atomic gases typically consist of a million neutral atoms (alkali or alkaline-earth) confined into a cloud of micrometer size, much more dilute than air. The typical length scales (micrometers) and time scales (milliseconds) allow for real-time imaging with single-atom resolution. At the same time, the effective van-der-Waals interaction is experimentally tunable and has a range much shorter than the particle spacing, so the nonrelativistic microscopic Hamiltonian is precisely known and exhibits similar universal properties as dilute neutron matter. At large scattering length the atomic cloud behaves as a nearly perfect hydrodynamic fluid. This opens the possibility of real-time and -space probes of hydrodynamic attractors, which also serve as benchmarks for the development of non-equilibrium many-body quantum theory.

\bibliographystyle{nuclthpreprint}
\bibliography{literature_attractors_pheno} 

\end{document}